\documentclass[iop]{emulateapj}
\usepackage{lineno}
\usepackage{color}
\usepackage{amsmath}
\usepackage{hyperref}  
\usepackage{breakurl}  

\newcommand\ms{\ifmmode{\rm m\thinspace s^{-1}}\else m\thinspace s$^{-1}$\fi}
\newcommand\kms{\ifmmode{\rm km\thinspace s^{-1}}\else km\thinspace s$^{-1}$\fi}
\newcommand{\vni}[1]{\vskip 5pt\noindent{#1 --}}

\shortauthors{Torres et al.}
\shorttitle{Hyades Survey}

\begin{document}
\submitted{Accepted for publication in The Astrophysical Journal Supplement Series}

\title{Long-term Spectroscopic Survey of the Hyades Cluster: The Binary Population}

\author{Guillermo Torres}
\author{Robert P.\ Stefanik}
\author{David W.\ Latham}

\affiliation{Center for Astrophysics $\vert$ Harvard \& Smithsonian,
  60 Garden St., Cambridge, MA 02138, USA; gtorres@cfa.harvard.edu}

\begin{abstract}
We report the results of a radial velocity monitoring program in the
Hyades region, carried out at the Center for Astrophysics over a period of more
than 45~yr. Nearly 12,000 spectra were gathered for 625 stars brighter
than $V \approx 14.5$, of which 55\% are members or possible members
of the cluster.
New or updated spectroscopic orbital solutions are presented for more
than 100 members and non-members, including several triple systems.
In a few cases we incorporate available astrometry. The frequency
of binaries in the Hyades with periods up to $10^4$\,d is
determined to be $40 \pm 5$\%, after corrections for incompleteness.
This is marginally higher than in other open clusters. The orbital period
and eccentricity distributions are found to be similar to those of
solar-type binaries in the field. The mass ratio distribution is
essentially flat, or slightly rising toward mass ratios of unity.
We revisit the determination of the tidal circularization period,
obtaining a longer $P_{\rm circ}$ value of $5.9 \pm 1.1$\,d compared to
the previous estimate of 3.2\,d, still somewhat short of the value
expected if most or all of the action of tides happens during the
pre-main-sequence phase. We estimate a
line-of-sight velocity dispersion of $0.21 \pm 0.05~\kms$
within 5.5~pc of the cluster center (approximately the half-mass
radius) and a larger dispersion beyond that distance.
Our velocity
measurements are accurate enough to clearly reveal the signatures of
gravitational redshift and convective blueshift among the dwarfs
and giants in the Hyades.
\end{abstract}


\section{Introduction}
\label{sec:introduction}

The assemblage of stars we know today as the Hyades cluster has been recognized
since antiquity by its familiar `V' shape, formed by five member stars
brighter than the 4th magnitude, plus Aldebaran, which is not a member.
Because of its proximity ($\sim$47~pc) and moderately rich membership,
the Hyades has served for many decades
as an essential laboratory for stellar astrophysics. It has provided the basis for the construction of some of the most fundamental maps of astronomy, including the mass-luminosity relation and the location of the main sequence in the Hertzsprung-Russell diagram. Throughout much of the 20th century, the distance to the Hyades was an important initial stepping stone for the determination of the extragalactic distance scale.

In many respects, it is a well-studied cluster: its membership has been investigated in great detail \citep[e.g.,][]{Gaia:2018}, and its age ($\sim$700~Myr) and chemical
composition are known \citep[e.g.,][]{Gossage:2018, Dutra-Ferreira:2016}. We now have a fairly good picture of its internal structure and kinematics,
and we know of the presence of extended tidal tails on either side, resulting from the interaction with the Galactic potential \citep{Meingast:2019}. The activity level and rotational properties of many of its member stars have been examined \citep{Savanov:2018}, as well as their X-ray emission \citep{Freund:2020}. The cluster has also been a fertile hunting ground for exoplanets, and a number of them have already been found \citep[see, e.g.,][]{Quinn:2014, Mann:2018, Mayo:2023, Distler:2025}.

On the other hand, even though scores of spectroscopic binary and multiple systems have been identified in the Hyades, the statistical properties of this important segment of the population have yet to be investigated. This is somewhat surprising for such a well-known cluster. It stands in contrast with the status of several other classical open clusters including NGC~188 \citep{Geller:2013, Narayan:2026}, M67 \citep{Geller:2021}, NGC~7789 \citep{Nine:2020}, and the Pleiades \citep{Torres:2021}, for which definitive studies of their binary constituents have already uncovered their key properties. The present paper aims to remedy this situation.

Binary searches in the Hyades have been carried out both astrometrically \citep[e.g.,][]{Mason:1993, Patience:1998} and spectroscopically, with the latter method being the most effective for characterizing multiple systems in greater detail. The history of radial velocity (RV) observations in the Hyades is as old as the technique itself. The earliest measurements were made in the 1890s \citep[see][]{Campbell:1928}, and the first spectroscopic orbits for binaries in the cluster made their appearance in the literature shortly after (e.g., 63~Tau, \citealt{Jantzen:1913}; $\theta^2$~Tau, \citealt{Russell:1914, Plaskett:1915}). Since then, several dedicated
observing campaigns have been carried out
to measure RVs for stars in this region \citep[e.g.,][]{Wilson:1948, Kraft:1965, Detweiler:1984}, although in general those efforts were not sustained for more than a few years. An exception is the program conducted by Roger Griffin and his collaborators \citep{Griffin:1974}, which started in the early 1970s and ran for four decades, comprising some 350 stars. The last major data release from this effort was published in a lengthy 172-page monograph reporting about 3000 RV measurements \citep{Griffin:2012}.  Those measurements supported the determination of 52 newly derived spectroscopic binary orbits in the Hyades field, adding to the several dozen others published earlier by that team.

A few years after the start of Griffin's project, a similar but independent long-term RV monitoring program was initiated at the Center for Astrophysics (CfA) \citep{Stefanik:1985}, and has now been completed. It has run for more than 45~yr since the end of the 1970s, and includes nearly twice as many stars in the Hyades region. While there is considerable overlap with Griffin's study in terms of the scope, the target list, and the time coverage, our survey has shown that the harvest of spectroscopic binaries in the cluster had by no means been exhausted. 
We have now been able to determine the orbits of several dozen new systems, some with very long periods, but also some with short periods that had been missed. This has benefited in part from the complementary nature of the observations from the two surveys, which in many cases has allowed us to fill in the gaps in the orbital phase coverage. 
The enlarged sample of spectroscopic binaries provides the basis for a long overdue analysis of their properties, including the distribution of orbital periods, eccentricities, and mass ratios, as well as the overall binary frequency. All of these are important ingredients for population synthesis of clusters \citep[e.g.,][]{Hut:1992} and for the study of their dynamical evolution. 

The plan for the paper is as follows. Section~\ref{sec:sample} explains the selection of our sample of 625 targets in the Hyades region, and is followed in Sections~\ref{sec:observations} and \ref{sec:rvs} with a description of the spectroscopic observations and the determination of RVs for single-, double-, and triple-lined objects. The procedures for solving the spectroscopic orbits, which in many cases incorporated additional RVs from other sources, are presented in Section~\ref{sec:orbits}. 
Nearly three dozen of our binaries have orbital solutions independently reported by the Gaia mission. This section discusses those external solutions as well, and how they compare with ours. 
Section~\ref{sec:variability} explains how we decided when a star has a variable RV. Membership in the cluster is a key attribute, and our criteria for this are laid out in Section~\ref{sec:membership}. Aside from having variable RVs, other indicators of binarity relying on astrometry are discussed in Section~\ref{sec:binarity}. These are used later in the paper to produce cleaner samples for extracting the distributions of binary orbital elements. Section~\ref{sec:completeness} describes the determination of our detection sensitivity as a function of orbital period, eccentricity, and mass ratio. The following sections contain our main results concerning the binary frequency (Section~\ref{sec:frequency}) and the distribution of orbital properties (Section~\ref{sec:binaryproperties}), as well as a discussion of tidal circularization in the Hyades. Then, in Section~\ref{sec:dispersion}, we make a determination of the internal velocity dispersion in the cluster, and in the process we show that our RVs clearly reveal the effects of the gravitational redshift and convective blueshift. In Section~\ref{sec:rotation} we present tentative evidence of rotation or shear in the cluster, and discuss its interpretation. Final remarks are found in Section~\ref{sec:conclusions}. Three appendices provide additional information,
including (A) the discussion of more complicated orbital solutions, such as those
for triple systems or for objects that benefit from incorporating astrometric
observations; (B) notes of interest for selected targets; and (C) graphical
representations of our orbital solutions for the single- and double-lined
binaries from the main text.

\section{Sample}
\label{sec:sample}

The most conspicuous part of the Hyades cluster spans an area on the
sky more than 20\arcdeg\ across. Other members are scattered much
farther out.  This does not include stars that are in the tidal tails
of the cluster, revealed in recent years by the Gaia mission
\citep[see, e.g.,][]{Meingast:2019, Roser:2019, Risbud:2025}. Some of
those outliers may be as far as 800~pc, or more than 100\arcdeg\ away
from the center \citep{Jerabkova:2021}. The classical photographic
studies of cluster membership, on the other hand, focused mostly on
the brighter members in the central parts of the Hyades
\citep[e.g.,][]{vanBueren:1952, Hanson:1975, Pels:1975}, and relied
almost exclusively on astrometric or photometric information.  Those
were largely the lists of targets that this project drew from to begin
with, extended down to the magnitude limits accessible to our
facilities, $V \approx 14$--14.5, depending on the telescope. The goal
was to provide the missing radial velocities needed in many cases to
confirm membership, and also to identify binary star candidates
\citep{Stefanik:1985}.  The initial target list of some 200 objects
evolved with time to reach a current total of 625 stars, gradually
incorporating many more objects that had been claimed or suspected to
be members based on their proper motions or photometric properties,
some quite removed from the center \citep[e.g.,][among other
  sources]{Weis:1979, Weis:1983, Perryman:1998, Reid:2000,
  deBruijne:2001, Douglas:2014, Douglas:2016}. In many cases, known or
newly discovered visual companions of the main targets were included
in the sample as well, when bright enough and far enough away from the
brighter star to be observed separately.

\begin{figure}
\begin{center}
  \begin{tabular}{c}
  \hspace*{-7mm}\includegraphics[width=0.453\textwidth]{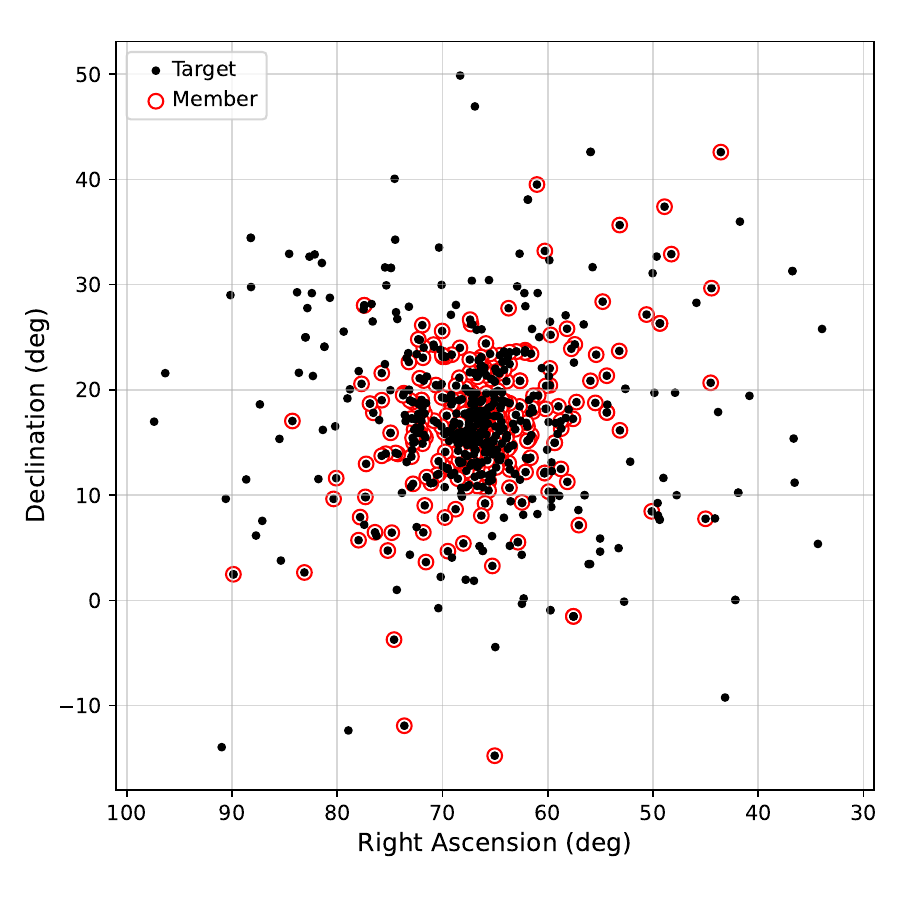} \\ [-3ex]
  \hspace*{-4mm}\includegraphics[width=0.5\textwidth]{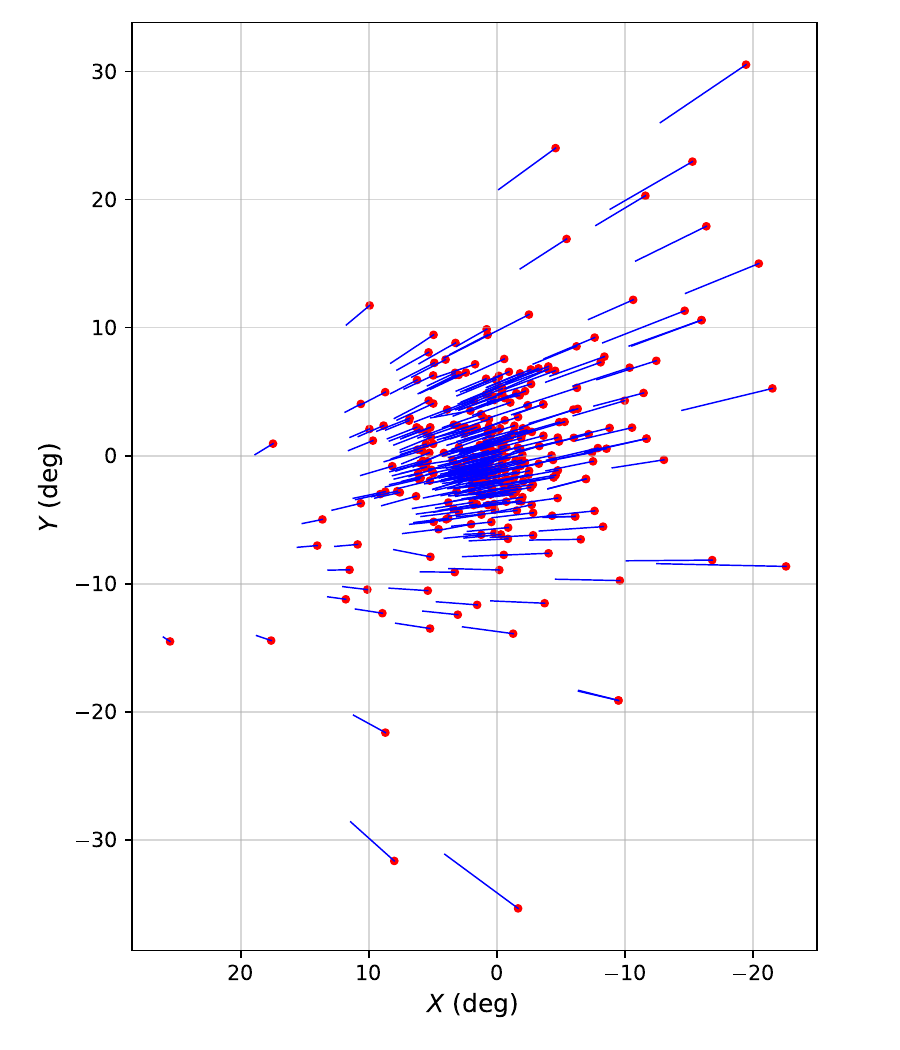}
  \end{tabular}
  \end{center}

\vskip -15pt  \figcaption{\emph{Top:} Distribution of survey targets on the plane of the
    sky. Objects considered members or possible members of the Hyades
    are marked with red circles.
    \emph{Bottom:} Proper motion vectors for members or possible members,
    showing the change in position over 100,000~yr. Stars are displayed
    in equatorial $XY$ coordinates on the tangential plane centered on
    the cluster. Red dots mark the initial positions.\label{fig:skysample}}

\end{figure}

As a result of the way the sample was assembled, the time coverage of
our spectroscopic observations is somewhat inhomogeneous, but we
estimate that our target list now includes the vast majority of true
members of the cluster down to $V \approx 14.5$, along with many others
in the general Hyades vicinity that have turned out to be unrelated
field stars, as we explain below. The sky positions of our stars are
shown at the top of Figure~\ref{fig:skysample}, in which objects that our
analysis has shown are true members or possible members are marked
with red circles. The bottom shows their change in position over 100,000~yr,
with the proper motion vectors pointing toward the convergent point
as a result of the common space motion.

Table~\ref{tab:sample} lists all targets with their SIMBAD names,
coordinates in ascending R.A.\ order, and other information gathered
from the third data release (DR3) of the Gaia catalog
\citep{Gaia:2023a}. A running number (\#) is added to facilitate referencing
across other tables in the paper. Membership information is also provided (see
below).

\setlength{\tabcolsep}{2pt}
\begin{deluxetable*}{llccccccccc}
\tablewidth{0pc}
\tablecaption{Sample of Targets in the Hyades Region \label{tab:sample}}
\tablehead{
\colhead{\#} &
\colhead{Name} &
\colhead{R.A.} &
\colhead{Dec.} &
\colhead{Gaia~DR3 ID} &
\colhead{$\pi_{\rm DR3}$} &
\colhead{$\mu_{\alpha}\cos\delta$} &
\colhead{$\mu_{\delta}$} &
\colhead{$G$} &
\colhead{$G_{\rm BP}\!-\!G_{\rm RP}$} &
\colhead{Memb.}
\\
\colhead{} &
\colhead{} &
\colhead{(deg)} &
\colhead{(deg)} &
\colhead{} &
\colhead{(mas)} &
\colhead{(mas yr$^{-1}$)} &
\colhead{(mas yr$^{-1}$)} &
\colhead{(mag)} &
\colhead{(mag)} &
\colhead{}
}
\startdata
1  &  vB 157     &  33.94275  &  25.78265  &   106051542927008128  &  $24.352 \pm 0.033$  &    $176.818 \pm 0.034$  &   $-65.175 \pm 0.032$\phs  &  5.68  &  0.59  &  NM \\ 
2  &  HD 14127   &  34.33297  &  \phn5.36082  &  2519811309169118848  &  $14.819 \pm 0.022$  &    $145.479 \pm 0.025$  &    $17.050 \pm 0.018$      &  8.42  &  0.71  &  NM \\ 
3  &  HD 15094   &  36.54724  &  11.17938  &    24656689987510912  &  $22.338 \pm 0.024$  &    $109.003 \pm 0.024$  & \phn$2.607 \pm 0.020$      &  9.05  &  1.04  &  NM \\ 
4  &  BD+14 400  &  36.64095  &  15.37348  &    76117304500935936  &  $16.880 \pm 0.181$  &    $181.504 \pm 0.196$  &    $13.112 \pm 0.186$      &  9.28  &  0.98  &  NM \\ 
5  &  HD 15128A  &  36.75458  &  31.28798  &   132522762106356224  &  $12.024 \pm 0.026$  & \phn$88.185 \pm 0.034$  &   $-30.672 \pm 0.024$\phs  &  7.77  &  0.67  &  NM 
\enddata
\tablecomments{For the target names, we have generally favored the SIMBAD
  designations most commonly used in the Hyades cluster, when
  available. For performing SIMBAD searches, those names should be
  prefixed with the cluster designation ``Cl Melotte 25'' for vB
  names, or ``Cl* Melotte 25'' otherwise. 
  Other non-cluster names can be used as listed.
  Coordinates (ICRS, epoch 2016.0), source IDs,
  parallaxes, proper motion components, $G$-band magnitudes, and
  $G_{\rm RP}\!-\!G_{\rm BP}$ colors are from the Gaia~DR3 catalog.
  The last column reports our
  assessment of membership in the cluster as `M' (member), `M?'
  (possible member), or 'NM' (non-member).
  An additional column in the electronic version of this table
  lists $\Delta\pi$, which is an additive adjustment to the Gaia~DR3
  parallaxes to correct for a systematic zero-point offset, as advocated by
  \cite{Lindegren:2021}.
  (This table is available in
  its entirety in machine-readable form.)}
\end{deluxetable*}
\setlength{\tabcolsep}{6pt}

\section{Observations}
\label{sec:observations}

The first observation for this project was obtained on the night of
1979 November~5. Over the course of the more than 45~yr since, we have
collected a total of 11987 usable spectra using four different
instrument/telescope combinations at the Center for Astrophysics.
Most (10156 observations) were gathered with three nearly identical
copies of the Digital Speedometer \citep{Latham:1992}. These were
echelle spectrographs delivering a resolving power of $R \approx
35,000$, coupled with intensified photon-counting Reticon detectors
that recorded a single order about 45~\AA\ wide, centered on the
\ion{Mg}{1}\,b triplet near 5187~\AA.  These instruments were attached
to the (now closed) 1.5m Wyeth reflector at the Oak Ridge Observatory
(in the town of Harvard, MA, USA), the 1.5m Tillinghast reflector at
the Fred L.\ Whipple Observatory (Mount Hopkins, AZ, USA), and the
4.5m-equivalent Multiple Mirror Telescope, also on Mount Hopkins,
before its conversion to a monolithic 6.5m mirror in 1998.
Signal-to-noise ratios varied greatly depending on brightness and sky
conditions. The reduction of these spectra used procedures described
by \cite{Latham:1985, Latham:1992}.

The stability of the velocity zero-point for the Digital Speedometers
was monitored by gathering exposures of the sky at morning and evening
twilight.  Based on those observations, small velocity corrections
typically under 2~\kms\ were applied to the raw velocities from run to
run, in order to place them all on the same native CfA system. Several
dozen minor planets in the Solar System were also observed regularly
with these instruments. Examination of those observations, reduced in
the same way as the science exposures, has determined that the native
CfA system is slightly offset from the IAU system by
0.14~\kms\ in the negative direction
\citep{Stefanik:1999}.\footnote{Note that the offset
was inadvertently given with the wrong sign in the original publication.}
In order to remove this shift, we
applied a correction of +0.14~\kms\ to all our raw velocities with the
Digital Speedometers described in the next section.

The Digital Speedometers were used until March of 2009, when the last
of those instruments was retired. Starting in September of that year,
we began using the Tillinghast Reflector Echelle Spectrograph
\citep[TRES;][]{Furesz:2008, Szentgyorgyi:2007} on the 1.5m telescope
in Arizona, which is a modern, bench-mounted, fiber-fed instrument
with a CCD detector. It delivers a resolving power of $R \approx
44,000$ in 51 orders, covering the wavelength range 3800--9100~\AA. We
collected a total of 1831 spectra with this instrument, through the end
of the 2025/2026 observing season. The reductions were performed
with a dedicated pipeline \citep[see][]{Buchhave:2012}. The velocity
zero-point of TRES was monitored with observations of IAU standard
stars each run, and observations of minor planets were again used to
transfer the raw velocities to the IAU system, as done with the
Digital Speedometers.

\begin{figure}
  \epsscale{1.17}
  \hspace*{-1mm}\includegraphics[width=0.48\textwidth]{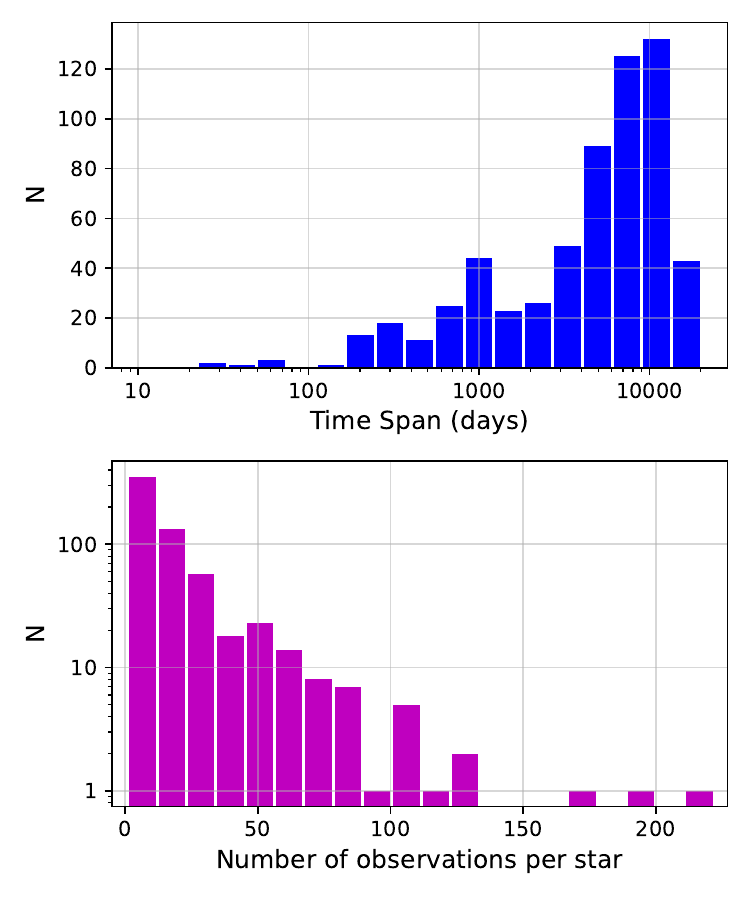}
  \figcaption{Histograms of the time span of our observations, and of
    the number of observations per target.\label{fig:stats}}
\end{figure}

In Table~\ref{tab:RVstats} we list for each object the time span of
the observations and the number of Digital Speedometer and TRES
spectra we gathered, along with other information discussed later.
The distributions of the time span and number of observations per
object are shown in Figure~\ref{fig:stats}.

\setlength{\tabcolsep}{3pt}
\begin{deluxetable*}{llccccccccccccl}
\tablewidth{0pc}
\tablecaption{Statistics of Spectroscopic Observations For Our Targets \label{tab:RVstats}}
\tablehead{
\colhead{\#} &
\colhead{Name} &
\colhead{Time span} &
\colhead{$N_{\rm DS}$} &
\colhead{$N_{\rm TRES}$} &
\colhead{$N_{\rm tot}$} &
\colhead{Mean RV} &
\colhead{$e/i$} &
\colhead{$P(\chi^2)$} &
\colhead{RV$_{\rm kin}$} &
\colhead{Memb.} &
\colhead{$T_{\rm eff}$} &
\colhead{$v \sin i$} &
\colhead{$M_1$} &
\colhead{Binarity}
\\
\colhead{} &
\colhead{} &
\colhead{(day)} &
\colhead{} &
\colhead{} &
\colhead{} &
\colhead{(\kms)} &
\colhead{} &
\colhead{} &
\colhead{(\kms)} &
\colhead{} &
\colhead{(K)} &
\colhead{(\kms)} &
\colhead{($M_{\odot}$)} &
\colhead{}
}
\startdata
1 &  vB 157     &  \phn850.8  &  7  &  0  &  7  &  $25.623 \pm 0.338$  &  0.931  &  0.4117  &  20.858  &  NM  &  6750 & 10 & \nodata &          \\ 
2 &  HD 14127   &  \phn843.9  &  6  &  0  &  6  &  $27.116 \pm 0.492$  &  0.956  &  0.4051  &  21.194  &  NM  &  6250 &  6 & \nodata &          \\ 
3 &  HD 15094   &  2807.1     & 10  &  0  & 10  &  $19.428 \pm 0.473$  &  0.881  &  0.5867  &  22.992  &  NM  &  5250 &  4 & \nodata &          \\ 
4 &  BD+14 400  &  \phn279.3  &  0  &  3  &  3  &  $16.071 \pm 0.145$  &  1.084  &  0.2982  &  23.075  &  NM  &  5250 &  4 & \nodata & DUP,RUWE \\ 
5 &  HD 15128A  &  3902.3     &  5  &  0  &  5  &  $23.376 \pm 1.182$  &  0.788  &  0.5653  &  22.095  &  NM  &  6750 & 50 & \nodata & DUP      
\enddata

\tablecomments{Columns $N_{\rm DS}$ and $N_{\rm TRES}$ contain the
  number of Digital Speedometer and TRES observations, and $N_{\rm tot}$
  is their sum. The following column is the weighted mean RV, followed by
  the $e/i$ statistic and $\chi^2$ probability described in
  Section~\ref{sec:variability}. The next two columns contain the
  expected (astrometric, or kinematic) RV of each target, and their membership
  assessment (Section~\ref{sec:membership}). We then list
  the $T_{\rm eff}$ and $v \sin i$ adopted for the templates of
  the single-lined objects, where `Barnard' means we used an observed
  spectrum of Barnard's star (GJ\,699) as the template.
  The templates for the double- or triple-lined objects are
  given in Table~\ref{tab:elemSB2b} below, or specified in Appendix~A or B.
  The next column lists approximate values of the stellar mass
  for members and possible members of the cluster with sufficient
  information to make those determinations. The masses were derived from Gaia
  photometry and a model isochrone for the Hyades, or rely on dynamical
  information, where available. For targets that
  are members of multiple systems, the value corresponds to the primary
  component (see Section~\ref{sec:qdistrib}).
  The final column gives
  information on binarity discussed later (Section~\ref{sec:binarity}).  
  (This table is available in its entirety in machine-readable form.)}

\end{deluxetable*}
\setlength{\tabcolsep}{6pt}

\section{Radial Velocities}
\label{sec:rvs}

\begin{figure}
  \epsscale{1.17}
  \hspace*{-1mm}\includegraphics[width=0.48\textwidth]{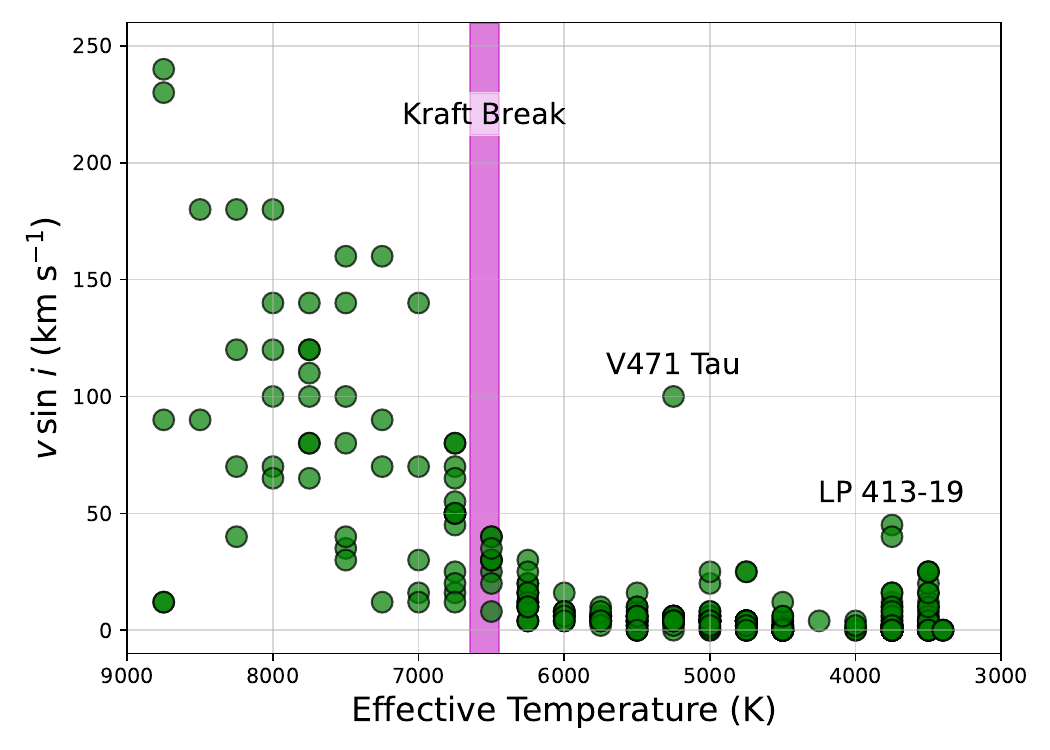}

  \figcaption{Distribution of the temperatures and $v \sin i$ values
  adopted for the templates of the Hyades members in our survey.
  The shaded area \citep[as per][]{Beyer:2024} marks the `Kraft Break'
  \citep{Kraft:1967}, where stars transition from slow to rapid rotation
  as their outer convective envelopes disappear, and magnetic breaking
  becomes ineffective. The two objects that stand out on the cool
  side of the break are spin-orbit synchronized stars belonging to
  the shortest period binaries in our sample: V471~Tau (the K dwarf in
  a system containing a white dwarf; $P = 0.52$\,d) and LP\,413-19
  (both components shown; $P = 0.48$\,d).
  \label{fig:vsiniteff}}

\end{figure}

For the single-lined objects in our sample, radial velocities and
corresponding uncertainties were obtained by cross-correlation using
the IRAF task {\tt XCSAO} \citep{Kurtz:1998}. The templates were
chosen from a large library of synthetic spectra based on model
atmospheres by R.\ L.\ Kurucz, and a line list that was fine-tuned by
Jon Morse to improve the match to real stars
\cite[see][]{Nordstrom:1994, Latham:2002}. The templates cover the
region between 5050\,\AA\ and 5350\,\AA, and are available over wide
ranges in effective temperature ($T_{\rm eff}$), projected rotational
velocity ($v \sin i$)\footnote{We use $v \sin i$ here as shorthand,
  although this is strictly the total line broadening, including
  contributions from other effects such as macroturbulence.  Our
  templates were calculated with a fixed macroturbulent velocity of
  $\zeta_{\rm RT} = 1~\kms$, representative of main-sequence stars.},
surface gravity ($\log g$), and metallicity ([m/H]).  As all but four
of our targets are on the main sequence, we adopted $\log g = 4.5$,
which is sufficiently close for our purposes. For the four giants
(vB\,28, vB\,41, vB\,70, and vB\,71), we used $\log g = 2.5$.  The metallicity
was held fixed at the solar value, as the composition of the Hyades is
only slightly higher \citep[${\rm [Fe/H]} =
  +0.18$;][]{Dutra-Ferreira:2016}. The optimal template for each star
was determined by running grids of correlations in temperature and $v
\sin i$, and selecting the one giving the highest correlation value
averaged over all exposures \citep[see][]{Torres:2002}. 
The temperature and $v \sin i$ values adopted in each case are included
in Table~\ref{tab:RVstats}, and their distribution is shown in
Figure~\ref{fig:vsiniteff}.
As the
Digital Speedometer spectra are limited to a narrow wavelength range
around the \ion{Mg}{1}\,b triplet, the velocities from the TRES
spectra were calculated using the 100~\AA\ order centered on the same
region, for consistency, and also because in our experience this is
the part of the spectrum that captures most of the velocity
information.

For objects in our sample showing two or three sets of lines, RVs were
derived using the two-dimensional cross-correlation technique {\tt TODCOR}
\citep{Zucker:1994} and its extension to three dimensions
\citep[{\tt TRICOR};][]{Zucker:1995}. The optimal template parameters were
determined in a similar way as above, and are listed below in Table~\ref{tab:elemSB2b}
for the double-lined targets. Those for triple-lined objects are given
in Appendix~A or B.
In the case of the Digital
Speedometers, previous work has shown that the small wavelength
coverage can sometimes introduce systematic errors in the velocities
from composite spectra,
caused by lines of the binary components shifting in and out of the
spectral order in opposite directions as a function of orbital phase.
These spurious velocity shifts are typically less than 5~\kms, and
were corrected based on numerical simulations as described by
\cite{Latham:1996} and \cite{Torres:1997d}. For TRES, with spectral
orders that are more than twice as wide, these effects are negligibly
small.

Finally, we note that for a few of our coolest targets we found that
better results were obtained using an observed spectrum of Barnard's
star (GJ~699, \ion{M4}{5}) as the template.

\begin{figure}
  \epsscale{1.17}
  \hspace*{-1mm}\includegraphics[width=0.48\textwidth]{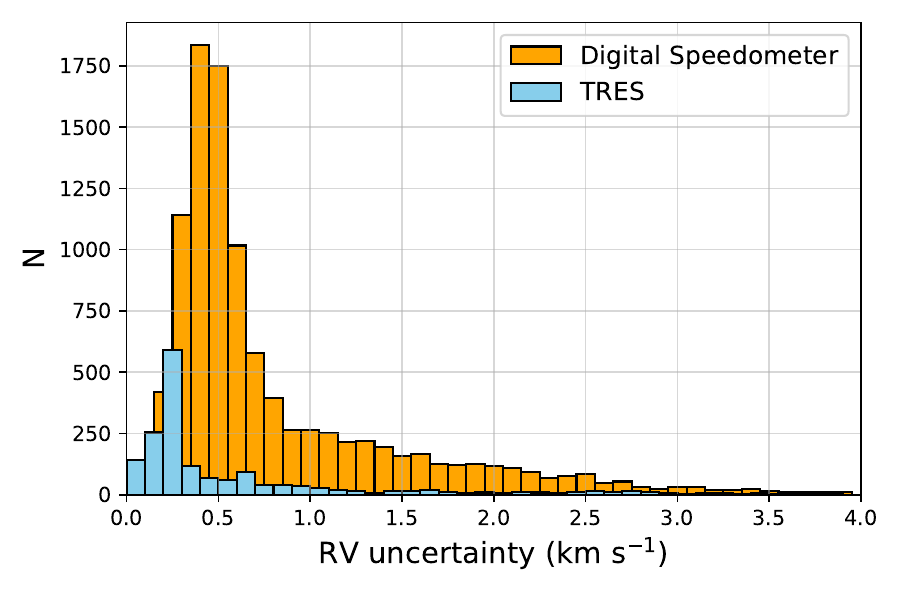}

  \figcaption{Histogram of the individual radial-velocity
    uncertainties for our single-lined objects,
    from the Digital Speedometers and the TRES
    instrument.\label{fig:RVerrors}}

\end{figure}

The individual radial velocities and associated uncertainties for the
single-lined objects, as returned by {\tt XCSAO}, are listed in
Table~\ref{tab:RVsb1}, and those for the double-lined objects are
found in Table~\ref{tab:RVsb2}. Typical uncertainties for sharp-lined
stars when using the Digital Speedometers are just under 0.5~\kms, and
those for TRES are about half as large.  Histograms of the RV
uncertainties for the Digital Speedometers and TRES are shown in
Figure~\ref{fig:RVerrors}.

\setlength{\tabcolsep}{10pt}
\begin{deluxetable}{llccc}
\tablewidth{0pc}
\tablecaption{Radial Velocity Measurements for Single-Lined Targets \label{tab:RVsb1}}
\tablehead{
\colhead{\#} &
\colhead{Name} &
\colhead{BJD} &
\colhead{RV} &
\colhead{Inst.}
\\
\colhead{} &
\colhead{} &
\colhead{(2,400,000+)} &
\colhead{(\kms)} &
\colhead{}
}
\startdata
1 & vB 157  &  48903.6887  &  $25.57 \pm 0.25$  & 1 \\
1 & vB 157  &  48946.5639  &  $25.36 \pm 0.33$  & 1 \\
1 & vB 157  &  49047.4652  &  $26.19 \pm 0.34$  & 1 \\
1 & vB 157  &  49226.8129  &  $25.96 \pm 0.33$  & 1 \\
1 & vB 157  &  49401.4669  &  $25.15 \pm 0.51$  & 1 
\enddata
\tablecomments{Radial velocity measurements and uncertainties. The
  last column indicates the instrument used (1 = Digital Speedometers;
  2 = TRES). The velocities for two single-lined targets from our survey
  published previously are not listed here, and may be found in the
  original publication: vB\,71 and vB\,72 \citep{Torres:1997c}. 
  (This table is available in its entirety in
  machine-readable form.)}
\end{deluxetable}
\setlength{\tabcolsep}{6pt}

\setlength{\tabcolsep}{2.8pt}
\begin{deluxetable}{llccc@{}c}
\tablewidth{0pc}
\tablecaption{Radial Velocity Measurements for Double-Lined Targets \label{tab:RVsb2}}
\tablehead{
\colhead{\#} &
\colhead{Name} &
\colhead{BJD} &
\colhead{RV$_1$} &
\colhead{RV$_2$} &
\colhead{Inst.}
\\
\colhead{} &
\colhead{} &
\colhead{(2,400,000+)} &
\colhead{(\kms)} &
\colhead{(\kms)} &
\colhead{}
}
\startdata
48 & LP~413-18  &   56224.7984 &  $66.78 \pm 0.20$ & $-13.95 \pm 1.74$\phs  &  2 \\
48 & LP~413-18  &   56262.9048 &  $35.45 \pm 0.28$ &  $35.45 \pm 2.47$      &  2 \\
48 & LP~413-18  &   56282.6123 &  $21.34 \pm 0.25$ &  $53.65 \pm 2.22$      &  2 \\
48 & LP~413-18  &   56309.7136 &  $45.29 \pm 0.24$ &  $20.20 \pm 2.09$      &  2 \\
48 & LP~413-18  &   56346.6366 &  $25.60 \pm 0.19$ &  $50.01 \pm 1.67$      &  2 
\enddata
\tablecomments{Radial velocity measurements and uncertainties for the
  primary and secondary components. The last column indicates the instrument used
  (1 = Digital Speedometers; 2 = TRES). 
  The velocities for double- or triple-lined targets from our survey
  published previously are
  not listed here, and are to be found in the original publications:
  vB\,120 \citep{Torres:2024b}; 
  vB\,162, vB\,23, vB\,34, vB\,182, vB\,117, vB\,119 \citep{Torres:2024a};
  PELS\,20 \citep{Torres:2024c}; 
  vB\,24 \citep{Torres:1997a}; 
  vB\,57 \citep{Torres:1997b}; 
  vB\,75 \citep{Torres:2019b}; 
  HAN\,346 \citep{Benedict:2021}.
  The RVs for four others (vB\,53, vB\,95, vB\,124, vB\,169) will be published
  in separate studies of those systems.
  (This table is available in its entirety in machine-readable form.)}
\end{deluxetable}
\setlength{\tabcolsep}{6pt}

\section{Orbital Solutions}
\label{sec:orbits}

For objects in our sample with enough observations and clear
evidence of velocity variations, we performed weighted least-squares
solutions to derive the elements of the single-lined or double-lined
orbits (SB1, SB2). The weights were taken to be inversely proportional
to the square of the RV uncertainties, $\sigma_{\rm RV}$. As these
errors are not always accurate, we adjusted them during the solutions
by solving for a multiplicative scale factor $F$ applied to $\sigma_{\rm
  RV}$ for each star, intended to achieve reduced $\chi^2$ values near
unity.  These factors were solved for simultaneously with the standard
orbital elements, which are $P$, $\gamma$, $K_1$, $e$, $\omega_1$, and $T_0$
for SB1s, and the additional parameter $K_2$ for SB2s.
Separate scale factors $F_1$ and $F_2$ were used for the primary and
secondary in the SB2s.

For many of our targets, orbits have been published previously, mainly
by Griffin and collaborators \citep{Griffin:1978, Griffin:1982,
  Griffin:1985a, Griffin:1988, Griffin:2001, Griffin:2012,
  Griffin:2013, Griffin:2016}, but also by other authors. We take the
opportunity here to update those orbits by adding our own
observations, which in a number of systems significantly improve the
phase coverage, and therefore the accuracy and precision of the
elements.  For several of these binaries, more than one orbit has been
published for the same object.  In other cases, our orbital solutions
based on our own measurements are new. When combining our velocities
with others that may have a different zero-point, we solved for one
additional parameter for each external set of velocities, representing
a possible systematic offset relative to the CfA system. We report
those offsets only when they are statistically significant. Similarly,
we have occasionally found and report significant systematic zero-point
differences between the primary and secondary velocities in some of
the SB2s, which are most likely caused by template mismatch.

The orbital elements we obtain for the SB1s and SB2s are listed in
Tables~\ref{tab:elemSB1} and \ref{tab:elemSB2a}.
Plots
of the orbits along with the observations are shown in Appendix~C. For
the double-lined binaries, other parameters inferred from the orbital
elements are given separately in Table~\ref{tab:elemSB2b}, together
with the secondary-to-primary spectroscopic flux ratios
$\ell_2/\ell_1$ at the mean wavelength of our observations,
$\sim$5187~\AA. The sources of external velocity measurements used
in some of our orbital solutions are listed in Table~\ref{tab:sources}.
In three cases (vB\,30, HD\,40512, vB\,102) the velocity residuals
from initial fits showed
obvious long-term drifts caused by outer companions. For those systems,
the final fits included the slope of the trend as an additional adjustable parameter,
which we solved together with the other orbital elements. Those results are
given in the notes to Table~\ref{tab:elemSB1}.

\setlength{\tabcolsep}{3pt}
\begin{deluxetable*}{rlcccccccccccc}
\tablecaption{Orbital Elements for the Single-lined Binaries in the Sample.\label{tab:elemSB1}}
\tablehead{
\colhead{\#} &
\colhead{Name} &
\colhead{$P$} &
\colhead{$\gamma$} &
\colhead{$K_1$} &
\colhead{$e$} &
\colhead{$\omega_1$} &
\colhead{$T_0$ (BJD)} &
\colhead{$f(M)$} &
\colhead{$M_2 \sin i$} &
\colhead{$a_1 \sin i$} &
\colhead{$\sigma$ (\kms)} &
\colhead{$F$} &
\colhead{Source}
\\
\colhead{} &
\colhead{} &
\colhead{(day)} &
\colhead{(\kms)} &
\colhead{(\kms)} &
\colhead{} &
\colhead{(deg)} &
\colhead{(2,400,000+)} &
\colhead{($M_{\sun}$)} &
\colhead{($M_{\sun}$)} &
\colhead{($10^6$ km)} &
\colhead{$N_{\rm obs}$} &
\colhead{Memb.} &
\colhead{}
}
\startdata
%
 7  &  G 36-30   &  1228.6        &  31.721     &  6.518  &  0.4874  &  118.1        &  46635.6         &  0.02347  &  0.2863  &  96.15     &  0.342  &  1.164  &  4,14 \\  
 7  &  G 36-30   &  \phm{222}1.0  &  \phn0.092  &  0.068  &  0.0086  &  \phn\phn1.2  &  \phm{2222}2.8   &  0.00063  &  0.0026  &  \phn0.86  &    84   &   NM    &       \\ [1.5ex]

14  &  HD 17922  &  318.879       &  23.301     &  7.528  &  0.3361  &  107.7        &  52427.53        &  0.01178  &  0.2275  &  31.09     &  0.290  &  1.113  &  8    \\  
14  &  HD 17922  &  \phn\phn0.059 &  \phn0.075  &  0.048  &  0.0063  &  \phn\phn1.3  &  \phm{2222}0.97  &  0.00023  &  0.0015  &  \phn0.20  &    92   &   M     &       \\ [1.5ex]

25  &  vB 158    &  20843         &  45.203     &  3.417  &  0.645   &  192.6        &  46095           &  0.0385   &  0.3377  &  749       &  0.271  &  1.019  &  5    \\  
25  &  vB 158    &  \phn\phn835   &  \phn0.054  &  0.092  &  0.016   &  \phn\phn2.4  &  \phm{222}55     &  0.0027   &  0.0080  &  \phn28    &    76   &   NM    &       
\enddata

\tablecomments{Uncertainties for the orbital elements and derived
  quantities are given in the second line for each system. The symbol
  $T_0$ represents a reference time of periastron passage for
  eccentric orbits, or a time of maximum primary velocity for circular
  orbits.  $M_2 \sin i$ is the coefficient of the minimum secondary
  mass multiplying the factor $(M_1+M_2)^{2/3}$. The first line of the
  penultimate column ($F$) represents the scale factor applied to the
  internal RV uncertainties to reach a reduced $\chi^2$ of
  unity. Our membership assessment is listed immediately below $F$.
  The last column lists the external RV sources used in
  these solutions; the corresponding bibliographic references are listed in Table~\ref{tab:sources}.
  Additional columns reporting the zero-point offsets of
  each external source relative to the CfA velocity system are
  included in the electronic version of this table. The offsets
  are in the same order as the source codes. For the three systems showing
  long-term RV drifts superimposed on the orbital motion, we obtained the following slopes:
  $d\gamma/dt = -0.147 \pm 0.020~\kms\ {\rm yr}^{-1}$ (vB\,30),
  $d\gamma/dt = -0.0394 \pm 0.0082~\kms\ {\rm yr}^{-1}$ (HD\,40512),
  and $d\gamma/dt = -0.0716 \pm 0.0038~\kms\ {\rm yr}^{-1}$ (vB\,102).
  (This table is
  available in its entirety in machine-readable form.)}

\end{deluxetable*}
\setlength{\tabcolsep}{6pt}  

\setlength{\tabcolsep}{4pt}
\begin{deluxetable*}{rlccccccccccc}
\tablecaption{Orbital Elements for the Double-lined Binaries in the Sample.\label{tab:elemSB2a}}
\tablehead{
\colhead{\#} &
\colhead{Name} &
\colhead{$P$} &
\colhead{$\gamma$} &
\colhead{$K_1$} &
\colhead{$K_2$} &
\colhead{$e$} &
\colhead{$\omega_1$} &
\colhead{$T_0$ (BJD)} &
\colhead{$\sigma_1$ (\kms)} &
\colhead{$N_1$} &
\colhead{$F_1$} &
\colhead{Source}
\\
\colhead{} &
\colhead{} &
\colhead{(day)} &
\colhead{(\kms)} &
\colhead{(\kms)} &
\colhead{(\kms)} &
\colhead{} &
\colhead{(deg)} &
\colhead{(2,400,000+)} &
\colhead{$\sigma_2$ (\kms)} &
\colhead{$N_2$} &
\colhead{$F_2$} &
\colhead{Memb.}
}
\startdata
%
48  &  LP 413-18  &  16.2493     &  34.72     &  28.9        &  43.1         &  0.501   &  284.0        &  56385.84           &  0.161  &    8    &  0.745  &  \nodata  \\          
48  &  LP 413-18  &  \phn0.0054  &  \phn0.19  &  \phn1.1     &  \phn1.9      &  0.024   &  \phn\phn3.2  &  \phm{2222}0.11     &  1.420  &    8    &  0.745  &  M        \\ [1.5ex]

49  &  LP 413-19  &  0.47613265  &  35.16     & 119.6        &  125.9        &  0.0     &  \nodata      &  59590.21672        &  3.192  &    16   &  0.914  &  \nodata  \\          
49  &  LP 413-19  &  0.00000018  &  \phn0.69  & \phn\phn1.5  &  \phn\phn1.9  &  \nodata &  \nodata      &  \phm{2222}0.00056  &  4.005  &    16   &  0.917  &  M?       \\ [1.5ex]

132 &  HG 7-97    &  4.52818     &  54.63     & 12.38        &  12.95        &  0.128   &  261          &  51041.90           &  1.388  &    14   &  0.842  &  \nodata  \\          
132 &  HG 7-97    &  0.00044     &  \phn0.31  & \phn0.62     &  \phn0.65     &  0.044   &  \phn16       &  \phm{2222}0.19     &  1.471  &    14   &  0.890  &  NM       
\enddata

\tablecomments{Uncertainties for the orbital elements
  are given in the second line for each system. The symbol
  $T_0$ represents a reference time of periastron passage for
  eccentric orbits, or a time of maximum primary velocity for circular
  orbits. $F_1$ and $F_2$ represent multiplicative scaling factors
  applied to the internal RV errors for the primary and secondary in
  order to achieve reduced $\chi^2$ values near unity. The last column
  lists the external RV sources used in these solutions (first line), if any,
  and our membership assessment below.
  Bibliographic references for the external sources are given separately in
  Table~\ref{tab:sources}.
  (This table is available in its
  entirety in machine-readable form.)}
\end{deluxetable*}
\setlength{\tabcolsep}{6pt}  

\setlength{\tabcolsep}{6pt} 
\begin{deluxetable*}{rlccccccccc}
\tablecaption{Additional Properties for the Double-lined Binaries in the Sample.\label{tab:elemSB2b}}
\tablehead{
\colhead{\#} &
\colhead{Name} &
\colhead{$M_1 \sin^3 i$} &
\colhead{$M_2 \sin^3 i$} &
\colhead{$a_1 \sin i$} &
\colhead{$a_2 \sin i$} &
\colhead{$a_{\rm tot}$} &
\colhead{$q$} &
\colhead{$\ell_2/\ell_1$} &
\colhead{$T_{\rm eff}$} &
\colhead{$v \sin i$}
\\
\colhead{} &
\colhead{} &
\colhead{($M_{\sun}$)} &
\colhead{($M_{\sun}$)} &
\colhead{($10^6$ km)} &
\colhead{($10^6$ km)} &
\colhead{($R_{\sun}$)} &
\colhead{} &
\colhead{} &
\colhead{(K)} &
\colhead{(\kms)}
}
\startdata
%
48  &  LP 413-18  &  0.245    &  0.164    &  5.60    &  8.34   &  20.04     &  0.671  &  0.192 & 3750 & 0 \\
48  &  LP 413-18  &  0.020    &  0.012    &  0.13    &  0.27   &  \phn0.52  &  0.017  &  0.018 & 3750 & 0 \\ [1.5ex]

49  &  LP 413-19  &  0.374    &  0.356    &  0.7830  &  0.824  &  2.310     &  0.950  &  0.84  & 3750 & 45 \\
49  &  LP 413-19  &  0.012    &  0.010    &  0.0098  &  0.012  &  0.022     &  0.019  &  0.03  & 3750 & 40 \\ [1.5ex]

132 &  HG 7-97    &  0.00380  &  0.00364  &  0.765   &  0.800  &  2.248     &  0.956  &  0.986 & 4250 & 2 \\
132 &  HG 7-97    &  0.00045  &  0.00042  &  0.037   &  0.039  &  0.084     &  0.059  &  0.085 & 4250 & 1 
\enddata

\tablecomments{Uncertainties for the derived properties are given in the
second line for each system. The mass ratio is defined as $q \equiv M_2/M_1$.
The last three columns list the measured flux ratio at $\sim$5187~\AA\
($\ell_2/\ell_1$), and
the $T_{\rm eff}$ and $v \sin i$ adopted for
the primary and secondary templates (first and second line for each pair),
where `Barnard' means we used an observed
spectrum of Barnard's star (GJ\,699) as the template.
Five additional columns in the electronic version of this table include
quantities $\Delta$RV$_{\rm 1,2,CfA}$, $\Delta$RV$_{\rm 1,2,Other1}$,
and $\Delta$RV$_{\rm 1,2,Other2}$, which
represent systematic primary/secondary offsets for the CfA measurements and
for those from up to two external sources (the latter in the same order
as they appear in Table~\ref{tab:elemSB2a}). A further two columns,
$\Delta$RV$_{\rm Other1}$, $\Delta$RV$_{\rm Other2}$, represent
the zero-point offsets of up to two
external sources (also in the order in which they appear in
Table~\ref{tab:elemSB2a}) relative to the CfA system. Offsets are
only reported when at least one of them is statistically significant,
or in the case of $\Delta$RV$_{\rm Other}$,
when one of the external sources is from Griffin; otherwise they
were set to zero in our solutions.
 (This table is available in its entirety in machine-readable
  form.)}

\end{deluxetable*}
\setlength{\tabcolsep}{6pt}  

\setlength{\tabcolsep}{6pt}
\begin{deluxetable}{@{\hskip -0.5ex}r@{\hskip 1.5ex}lr@{\hskip 1.5ex}l}
\tablewidth{0pc}
\tablecaption{External Radial Velocity Sources.\label{tab:sources}}
\tablehead{
\multicolumn{2}{c}{Source} &
\multicolumn{2}{c}{Source}
}
\startdata
3   &  \cite{Griffin:1978}    &  16  &  \cite{Tomkin:2002}     \\
4   &  \cite{Griffin:1985a}    &  17  &  \cite{Griffin:2016}    \\			  	 
5   &  \cite{Griffin:2012}    &  18  &  \cite{Paulson:2004}    \\			  	 
6   &  \cite{Griffin:1982}    &  19  &  \cite{Mermilliod:2009} \\			 
7   &  \cite{Griffin:2001}    &  20  &  \cite{Mermilliod:2007} \\			 
8   &  \cite{Griffin:2013}    &  21  &  \cite{Tokovinin:1990}  \\			 
9   &  \cite{Tomkin:2003}     &  22  &  \cite{Vaccaro:2015}    \\
10  &  \cite{Tomkin:2005}     &  23  &  \cite{Tal-Or:2019}     \\				 
11  &  \cite{Imbert:2006}     &  24  &  \cite{Conti:1969}      \\			 
12  &  \cite{Balona:1987}     &  25  &  \cite{Diaz:2012}       \\			 
13  &  \cite{Fekel:2005}      &  26  &  Elodie public archive\tablenotemark{a} \\
14  &  \cite{Halbwachs:2018}  &  27  &  Sophie public archive\tablenotemark{b} \\
15  &  \cite{Halbwachs:2020}  &      &                        
\enddata
\tablenotetext{a}{\url{http://atlas.obs.fr/elodie}}
\tablenotetext{b}{\url{http://atlas.obs.fr/sophie}}

\tablecomments{These are the external RV sources used in some of our spectroscopic
orbital solutions (see Tables~\ref{tab:elemSB1} and
\ref{tab:elemSB2a}). Sources \#1 and \#2, omitted here, correspond to our own
Digital Speedometer and TRES measurements, respectively.}

\end{deluxetable}
\setlength{\tabcolsep}{6pt}

An additional set of RVs that we considered using is that of \cite{Bender:2008}.
These authors performed high-resolution near-infrared spectroscopy on more than
two dozen targets selected from our survey, which were known only as single-lined
spectroscopic binaries at the time. They detected the faint secondaries in 25 systems, and measured between 1 and 5 velocities in each case. Unfortunately, however, we have
found those measurements difficult to use, mainly because of
concerns about the velocity zero-point, the fact that only the RVs for the secondaries were reported, the significant uncertainties in some cases, and the
limited phase coverage in others.
Furthermore, in about a third
of the systems our own observations also reveal the presence of the secondaries,
allowing us to measure their velocities and achieve much more complete phase
coverage. For these reasons, we have chosen not to incorporate these near-infrared
measurements into our orbital solutions.

About a dozen of our objects are hierarchical
triple or quadruple systems in which the
inner and outer orbits can be determined. In some cases, available
astrometry for the outer orbits allows the dynamical masses to be
derived. Other targets are binaries with incomplete phase coverage,
for which astrometric information can be folded in to complement the
spectroscopy, and enable the orbit to be solved. These more
complicated systems are dealt with separately in Appendix~A.
Table~\ref{tab:triples} lists all systems in our survey with multiplicity
higher than two, for which the inner and outer orbits are known
either from this work or the work of others.

\setlength{\tabcolsep}{6pt}
\begin{deluxetable*}{clcccl}
\tablewidth{0pc}
\tablecaption{Triple and Quadruple Systems in the Survey with Known Inner and Outer Orbits. \label{tab:triples}}
\tablehead{
\colhead{\#} &
\colhead{Name} &
\colhead{$P_{\rm inner}$ (d)} &
\colhead{$P_{\rm outer}$} &
\colhead{Memb.} &
\colhead{Notes and sources}
}
\startdata
101 & LP 301-69   &    5.418682    &    4810 d  & NM  &   Triple, Appendix~A \\
162 & PELS 20       &    2.39436657  &   43.13 yr &  M  &   Triple, \cite{Torres:2024c} \\ 
198 & vB 22      &    5.60921410  &  3077.8 d  &  M  &   Triple, Appendix~A \\
250 & HD 27691      &    4.00018569  &   246.9 yr &  M  &   Triple, outer orbit by \cite{Josties:2021} \\
288 & HAN 346\,BC   &    0.7492425   &   988.0 d  &  M  &   Quadruple, \cite{Benedict:2021} \\ 
357 & vB 75      &   21.254396    &  40.752 yr &  M  &   Triple, \cite{Torres:2019b} \\ 
440,439 & HD 29205\,A,B   &   10.691399792 &     965 yr & NM  &   Triple, premature outer orbit by \cite{Izmailov:2019} \\ 
443 & vB 102     &  734.09        &   124.9 yr &  M  &   Triple, outer orbit by \cite{Tokovinin:2021c} \\ 
450 & vB 304       &   60.80936     &    8310 d  &  M  &   Triple, Appendix~A \\
466 & vB 185       &  276.7678      &    79.2 yr &  M  &   Triple, Appendix~A \\
526 & vB 124A   &  143.586       &   95.10 yr &  M  &   Quadruple, \cite{Tomkin:2007} \\
526 & vB 124B   &  496.7         &   95.10 yr &  M  &   Quadruple, \cite{Tomkin:2007} \\
585 & HD 34031 &   78.5400      &   848.9 d  & NM  &   Triple, Appendix~A \\
622 & vB 169A   &    4.4475858   &  6809.9 d  & NM  &   Quadruple, \cite{Fekel:2002} \\
622 & vB 169B   &    4.7835361   &  6809.9 d  & NM  &   Quadruple, \cite{Fekel:2002} 
\enddata
\tablecomments{The inner and outer periods are given for each subsystem.}
\end{deluxetable*}
\setlength{\tabcolsep}{6pt}

Several other objects in our sample have astrometric orbits with
orbital periods of decades or centuries, to which our velocities can
contribute little. We list them
in Table~\ref{tab:wdsorbits}, with their periods, eccentricities, and
membership information. For completeness, the table also includes
all other targets in the survey with astrometric orbits of any period, including ones
for which the same orbit has been determined spectroscopically.

\setlength{\tabcolsep}{6pt}
\begin{deluxetable*}{rlccllcl}
\tablewidth{0pc}
\tablecaption{Visual Binary Systems in the Survey with Known Astrometric Orbits.\label{tab:wdsorbits}}
\tablehead{
\colhead{\#} &
\colhead{Name} &
\colhead{WDS name} &
\colhead{$P$ (yr)} &
\colhead{$a^{\prime\prime}$ ($\arcsec$)} &
\colhead{$e$} &
\colhead{Memb.} &
\colhead{Notes}
}
\startdata
 5,6 &  HD 15128 A,B & 02270+3117 &  1041   &  2.47  &  0.86  &  NM  &  Premature orbit; constant RVs for A and B \\
 7   &  G 36-30      & 02433+1926 &  1214 d  &  0.03124 & 0.5210 & NM &  SB1 with the same period \\
 40  &  vB 3         & 03305+2006 &  31.46  &  2.94  &  0.395 &  NM  &  Premature orbit \\
 50  &  vB 5         & 03376+2121 &  110.0  &  0.848 &  0.434 &  M   &  Premature orbit; partial RV coverage \\
 107 &  vB 9         & 04007+2023 &  16.728 &  0.113 &  0.287 &  M   &  SB1 with the same period 
\enddata
\tablecomments{System properties (orbital periods, semimajor axes, and
eccentricities) are taken from the Sixth Catalog
of Orbits of Visual Binary Stars
(\url{https://www.astro.gsu.edu/wds/orb6/orb6frames.html}),
consulted on 2026 January 30. 
Bibliographic
references may be found therein, or where indicated in the notes.
Orbital periods are given in years, with
a few exceptions given in days. (This table is available in its entirety
in machine-readable form.)}
\end{deluxetable*}
\setlength{\tabcolsep}{6pt}

Finally, in a few of our targets with double-lined spectra, we
are able to measure the velocities of both components using {\tt TODCOR},
but we do not have enough of them to derive an orbital solution. In
such cases, it is still possible to infer an accurate center-of-mass
velocity $\gamma$, as well as the mass ratio $q \equiv M_2/M_1$, by
employing a simple method described by \cite{Wilson:1941}, although
the idea goes much further back \citep[e.g.,][]{Zurhellen:1907, Goos:1908}.
It involves fitting a straight line through pairs of primary and secondary
velocities, which then allows one to solve for $\gamma$ and $q$ from the
slope and intercept.
Knowledge of $\gamma$, even if the full orbit is not determined,
is very useful because it can be sufficient in some cases to immediately
rule out membership in the cluster.
Because both components have their own RV uncertainties, which can be very
different in some systems,
here we have implemented the procedure using orthogonal distance regression.
The targets for which we used it are noted in Appendix~B, and all turn out to
be non-members.

\subsection{The Griffin velocity zero-point}
\label{sec:griffinoffset}

Throughout his extensive, decades-long program of spectroscopic
observations in the Hyades region, R.\ Griffin maintained a consistent
zero-point for his velocities that he reported was 0.8~\kms\ more
positive than the IAU system \citep[see, e.g.,][]{Griffin:1988, Griffin:2012}. We
have observations for 47 of the SB1 systems he published over the
years, and 15 of his SB2s.  This allows us to compare his offset
estimate with the shifts we have determined independently for each of
these systems. We obtained a weighted average offset for the 62
binaries of ${\rm Griffin}-{\rm CfA} = +0.778 \pm 0.037~\kms$, which
is in excellent agreement with his assessment, on the assumption that
our velocities are on the IAU system.

\subsection{Comparison with Gaia~DR3 orbits}
\label{sec:gaiaorbits}

The Gaia~DR3 catalog contains orbital or trend solutions for some
800,000 binary or multiple systems of the astrometric,
spectroscopic, or eclipsing kind, as well as combinations in some cases
\citep[e.g.,][]{Gaia:2023b}.  By necessity, these `non-single star'
solutions were carried out in a fully automated manner, and this
effort has been largely successful. It is a remarkable achievement,
given the complexity of the problem and limitations from either the
sparseness, irregular distribution, or quality of the measurements, the broad range
of stellar characteristics, or even confusion from higher-order
multiplicity in many instances \citep[see, e.g.,][]{Babusiaux:2023,
  Gosset:2025}.

Gaia has reported full spectroscopic orbital solutions for 34 of our
targets, derived from the mission's RV measurements, or in some cases
a combination of RVs and astrometric measurements. We consider 20 of
these binaries to be true members of the Hyades (see below,
Section~\ref{sec:membership}).  An additional 19 targets have
astrometric-only solutions from Gaia, of which 14 belong to the
cluster. Four of those
are among the 34 that also have independent
spectroscopic orbits, although the periods from the two kinds of
solutions are not always the same. In one case (vB\,67B, a non-member), the
spectroscopic period from Gaia is 5.8\,d and the astrometric one is
1112\,d, implying it may be a hierarchical triple system. Our own four
RV measurements for vB\,67B show evidence of changes over the long term,
but curiously, not over the short term, which should presumably have
larger amplitude.

\begin{figure}
  \epsscale{1.17}
    \hspace*{-1mm}\includegraphics[width=0.48\textwidth]{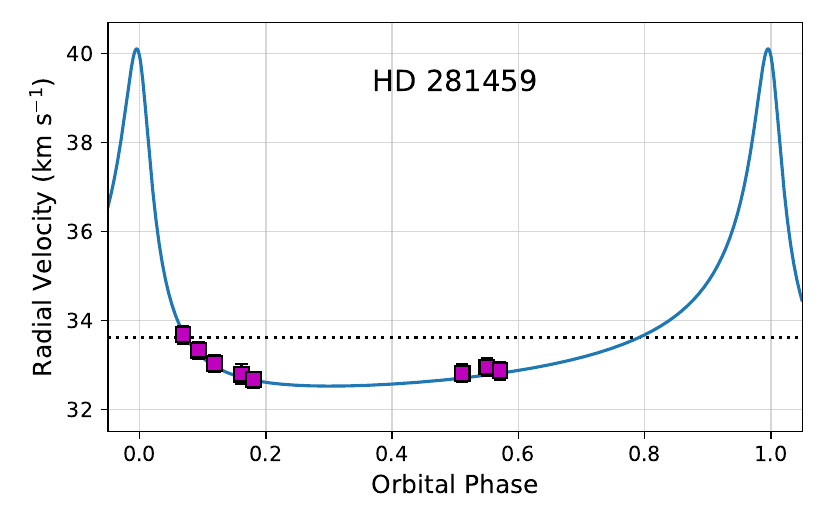}

\figcaption{Our RV measurements for HD~281459, shown against the
  spectroscopic orbit from the Gaia~DR3 non-single star solution. RV
  uncertainties are typically smaller than the symbol
  size. \label{fig:HD281459}}

\end{figure}

A comparison of our own orbital solutions against those from Gaia
reveals that in eight cases
the periods and other elements from Gaia
are incorrect, or deviate significantly from ours. All are members of
the cluster. On the other hand, 11 of our targets have new orbital
solutions from Gaia that our survey has missed,
due to insufficient
observations or limited time or phase coverage. Only three of those
are regarded as members of the Hyades.
Two have astrometric-only
orbits (HAN\,291, vB\,123), and the other, HD\,281459, has a spectroscopic
orbit. Our eight RV measurements for this last one are in excellent
agreement with the Gaia solution (see Figure~\ref{fig:HD281459}),
which has a period of 721\,d and is quite eccentric ($e = 0.75$).

\section{RV variability}
\label{sec:variability}

To assess whether the velocity of a star in our sample is variable, we
have chosen to rely on the $e/i$ (external/internal) statistic, which
is defined for a given object as the ratio between the standard
deviation of the measured RVs and the measurement precision. It has
commonly been used to identify variables in a number of other
spectroscopic surveys \citep[e.g.,][]{Mermilliod:1992, Geller:2015,
  Torres:2021}. We calculated it here as described by
\cite{Hole:2009}, using the individual measurement precision for each
observation. Some authors have occasionally adopted the $\chi^2$
probability as a measure of variability. We have found, however, that
it is much more sensitive to outliers than $e/i$, and we therefore
prefer the latter as a more conservative metric.

\begin{figure}
  \epsscale{1.17}
  \hspace*{-1mm}\includegraphics[width=0.48\textwidth]{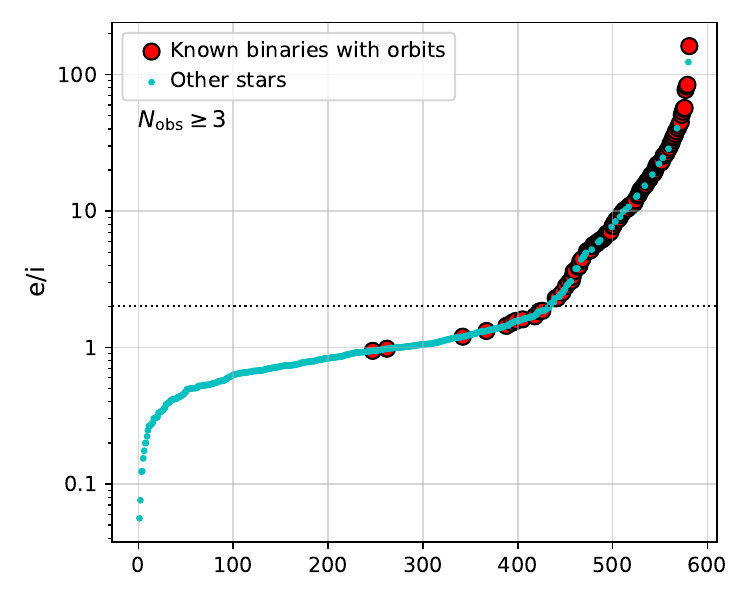}

\figcaption{Values of $e/i$ sorted in increasing order for the targets
  in our sample. Known binaries are marked as labeled, and the dotted
  line at $e/i = 2$ represents the adopted threshold for this work,
  above which stars are considered to be binaries.\label{fig:ei}}

\end{figure}

Figure~\ref{fig:ei} shows the $e/i$ values, sorted in increasing
order, for all stars in our sample with three or more observations.
Objects with known spectroscopic orbits are indicated with larger red
dots. The distribution displays a noticeable change in slope
(or ``knee'') at a value near $e/i = 2$. Most of the objects with higher
values than this are in fact known spectroscopic binaries with orbits,
as reported in the previous section.  This appears, therefore, as a
natural threshold for variability, and we adopt it for our work. It
is marked by the dotted line in the figure. There are, however, about
a dozen objects that have $e/i$ below this threshold, and that are
also known spectroscopic binaries.  Not surprisingly, they tend to be
systems with long orbital periods and small velocity semiamplitudes.
It is quite possible there are other such low-amplitude binaries in
our sample with $e/i < 2$, but they would be spectroscopically
undetectable given our measurement precision. The figure also shows
there are some three dozen objects not previously identified as
binaries that have $e/i > 2$. In what follows we consider them to be velocity
variables.

The $e/i$ values for all our targets are included in
Table~\ref{tab:RVstats}. We additionally include the
probability associated with the $\chi^2$ value for each star, $P(\chi^2)$,
for readers who would prefer to use that statistic as a measure of
variability.

\vskip 30pt
\section{Membership}
\label{sec:membership}

In order to identify which of our targets belong to the Hyades, we
relied primarily on their proper motions and parallaxes from Gaia~DR3,
their $G$ magnitudes and $G_{\rm BP}-G_{\rm RP}$ colors, and their RVs
as determined here. For the latter, we adopted either the weighted
mean for each object (Table~\ref{tab:RVstats}), or the center-of-mass
velocity, for systems with known orbits. The parallaxes were corrected
for the known systematic zero-point offset following
\cite{Lindegren:2021}.

Three conditions were required for membership. The first was that the
location in the absolute magnitude vs.\ color diagram must be
consistent with what is expected for known members. We ignored
extinction, which is negligibly small for the Hyades, and allowed for
objects to be slightly above the main-sequence, as would be expected
if they are binaries or higher order multiples, 
or slightly below, to account for errors in the magnitudes or the parallax.
The second condition for membership
was that the radial velocity should be close to the value expected
from the assumption of a common space velocity for all cluster stars.
Such predicted RVs \citep[or ``astrometric'' RVs; see,
  e.g.,][]{Dravins:1999} depend on the sky position of the star and
the known components of the space velocity vector of the cluster, along
with the coordinates of the convergent point, all of which we adopted
from \cite{Gaia:2018}, with their formal errors. The difference
$\Delta$RV between the spectroscopic and astrometric velocities was
considered to satisfy the criterion for membership when it was less
than three times its formal uncertainty. The uncertainty included not
only the errors of both velocities, but also a term added in
quadrature to account for the internal dispersion of the cluster in
the radial direction \citep[$\sigma_{\rm cl,RV} =
  0.40~\kms$;][]{Gaia:2018}.\footnote{As will be shown later in
  Section~\ref{sec:dispersion}, our own estimate of $\sigma_{\rm cl,RV}$
  is somewhat smaller, especially in the central regions of the
  cluster, although we prefer to retain the larger value
  in this section to be conservative.}
The third condition for membership, also
under the assumption of a common space motion, was that the component
of the proper motion perpendicular to the direction to the convergent
point, $\mu_{\perp}$, must be small. As with the velocities, we
adopted a 3$\sigma$ threshold, where the uncertainty for each star $i$ was constructed
from the quadrature sum of the errors coming from the proper motion
components and a term for the dispersion in the plane of the sky.
This 2D dispersion term, converted to angular measure (mas yr$^{-1}$),
was taken to be $\pi_i \sqrt{2}\sigma_{\rm cl,RV}/4.74047$,
assuming that
the velocity distribution in the cluster is isotropic.  As the proper
motions depend on distance, and the Hyades has a significant depth
along the line of sight, we normalized $\mu_{\perp}$ by multiplying it
by the ratio $\pi_{\rm cl}/\pi_i$ of the parallaxes, where the mean
cluster parallax was taken also from \cite{Gaia:2018}.

Application of the criteria outlined above is complicated by the
presence of binaries, whether recognized or not.  Even if binarity is
recognized (e.g., as indicated by a large $e/i$ value, or a visible
trend in the velocities), the mean RV we compute for any given object
will generally deviate from the expected value unless the
center-of-mass velocity of the system is known from a full orbital
solution, and unless there are no additional companions.  Furthermore,
even if the spectroscopic orbit is known, binarity can also change the measured
proper motion components, and therefore $\mu_{\perp}$, particularly if
the orbital period is long. This is because in most cases the proper
motions from Gaia~DR3 did not account for orbital motion when it was not
possible to solve for an astrometric orbit simultaneously with the
rest of the astrometric parameters. As a result, objects in our sample with known
spectroscopic orbits may well satisfy the $\Delta$RV condition, and
even have a consistent location in the color-magnitude diagram, but
may not meet our $\mu_{\perp}$ requirement, making it appear that they
are non-members.

To aid in accounting for these complications, we considered additional
information that might hint at, or in some cases confirm, the binary
nature of an object. This included the Renormalized Unit Weight Error
(RUWE) from Gaia \citep{Lindegren:2018}, which is a measure of the
quality of the astrometric solution.  RUWE values in excess of 1.4 can
often be a sign of unmodeled binary motion. For each object, we also
checked the {\tt ipd\_frac\_multi\_peak} parameter from the Gaia~DR3
catalog, which indicates the fraction of the time the object appeared
to have more than one peak in scans oriented in different directions.
We also consulted the Washington Double Star catalog
\citep[WDS;][]{Worley:1997, Mason:2001} for information on close
companions\footnote{Consulted on 2024 August 12.}, as well as other
sources available to us \citep{Morzinski:2011}.
In a number of cases, visual orbits have already been
derived, confirming the stars' physical association.  Yet another
valuable resource is the catalog of astrometric accelerations of
\cite{Brandt:2021}, which provided homogenized proper motions from
Hipparcos and Gaia at mean catalog epochs separated by almost 25~yr.
Any significant difference ($> 3\sigma$) in either right ascension or
declination was considered here to be a strong indication that the
object is a binary, and therefore that our $\mu_{\perp}$ estimate
might be affected. We note, however, that this information is only
available for the relatively bright objects accessible to Hipparcos,
which constitute less than half of our target list.  Finally, in
assessing the robustness of the mean RVs, we considered the number and
distribution of measurements by examining plots of the velocity
history for each object for subtle trends that may have gone
unnoticed.

In view of the potential for biases from target multiplicity, it was
deemed more practical to carry out the membership assessment manually,
object by object. For those displaying any of the conditions described
above, we chose to relax our general $3\sigma$ criteria on $\Delta$RV
and $\mu_{\perp}$ to some degree, to account for their binary nature.
While this does bring an element of subjectivity to the procedure,
we believe it also reduces the risk of excluding objects that might
reasonably still be bona fide members of the cluster, especially ones
with spectroscopic orbits that are the focus of the rest of this
paper.

Not all of our 625 targets have all the information we needed in order to decide
on their membership status. There are 21 that are missing either the
Gaia parallaxes or proper motions, or one or more of the Gaia
magnitudes. Several of these are close companions to other targets. In
these cases, we have drawn the missing information from the Gaia~DR2
or Hipparcos catalogs \citep{Gaia:2018, vanLeeuwen:2007}, when
possible, or done our best to make a judgment based on what is
available.

\begin{figure}

\begin{center}
  \begin{tabular}{c}
  \hspace*{-5mm}\includegraphics[width=0.50\textwidth]{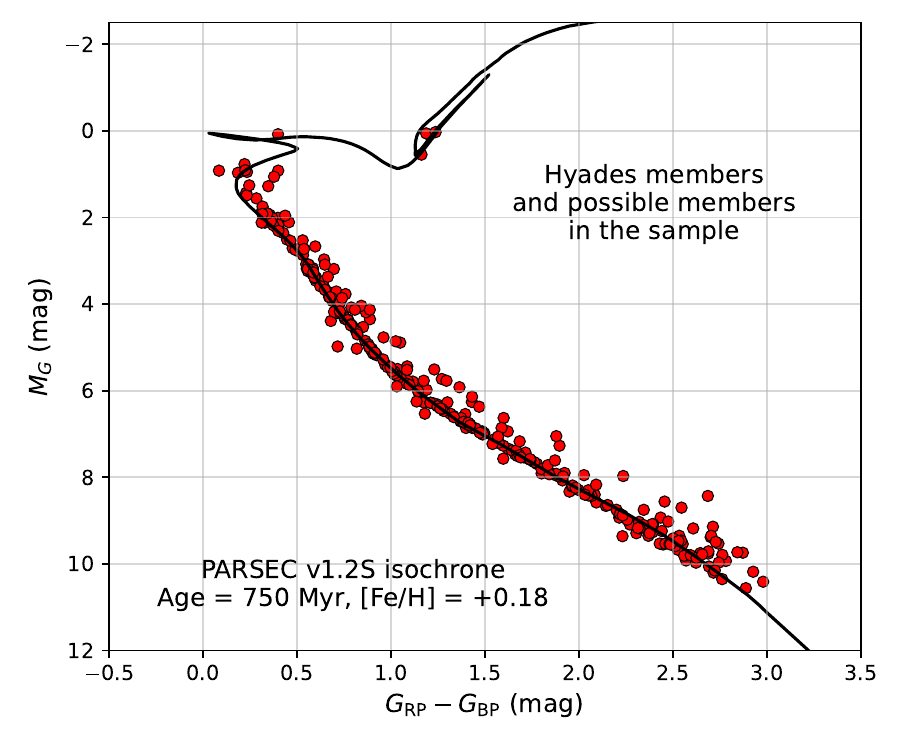} \\ [-2ex]
  \hspace*{-5mm}\includegraphics[width=0.48\textwidth]{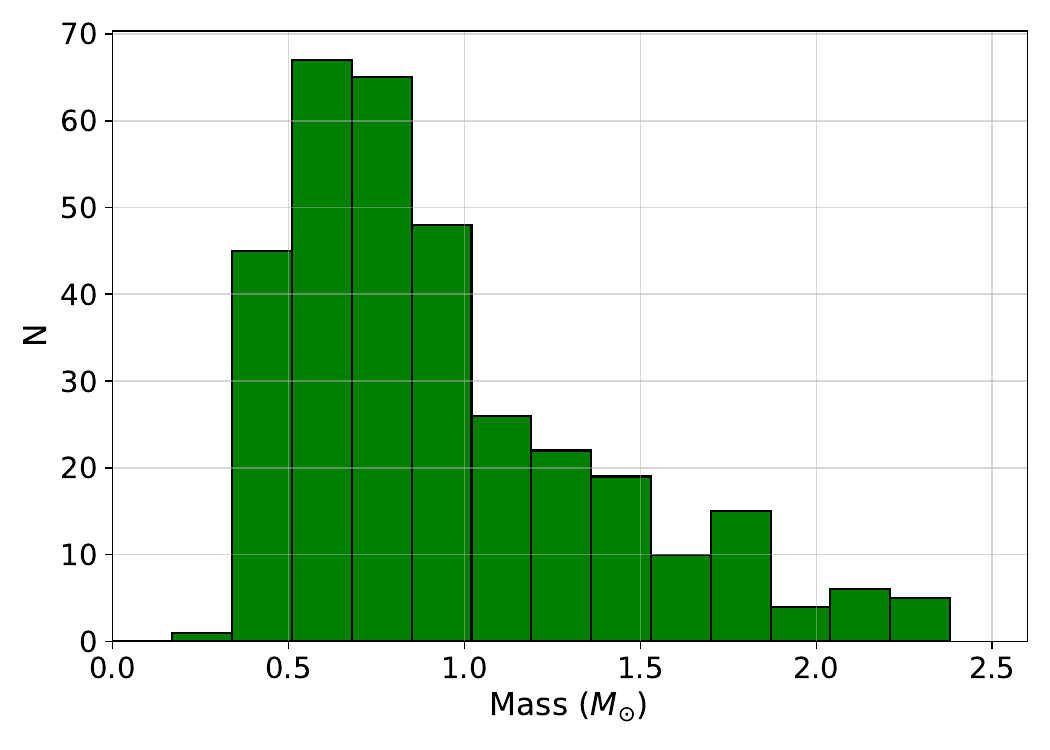}
  \end{tabular}
  \end{center}

\vskip -15pt \figcaption{\emph{Top:} Color-magnitude diagram of the Hyades from Gaia photometry,
for all members (`M') and possible members (`M?') in our sample.
Also shown is the model isochrone for the cluster we use throughout this
paper, including color corrections as prescribed by \cite{Wang:2025}.
\emph{Bottom:} Distribution of stellar masses for objects in the top panel
with sufficient information to make an estimate. For binary and multiple
systems, the value shown is for the primary component (see
Table~\ref{tab:RVstats}).
\label{fig:cmd}}

\end{figure}

We find that a total of 282 stars are clearly non-members, and we flag
them as ``NM'' in Table~\ref{tab:sample} and Table~\ref{tab:RVstats}.
Borderline cases, of which there are 16, are noted as ``M?''. The
remaining 327 objects are considered confirmed members of the
Hyades. 
Figure~\ref{fig:cmd} (top) shows the Gaia color-magnitude diagram of
the members and possible members from our survey.
The bottom panel displays a histogram of the stellar masses,
estimated mostly from Gaia photometry (Table~\ref{tab:RVstats});
they range from about 0.4 to 2.4\,$M_{\sun}$.
An independently constructed membership list of 920 objects based on Gaia
DR3 was compiled by \cite{Gaia:2021} using different criteria, and
offers a chance for an intercomparison. It includes 267 of the
targets in our sample, with most of the others being below our magnitude limit.
All but one of those 267 objects are confirmed members based on our own criteria.
The exception (vB\,184) is in the "M?" category, and shows 
a long-term trend in the RVs and a very large RUWE value, both of which
signal it has a companion.

On the other hand, there are 76 targets in
our sample that we consider to be members or possible members, and that
are missing from the Gaia membership list.
The brightest one, vB\,71 ($G = 3.58$), is $\theta^1$\,Tau, a well-known
binary member of the cluster.
All except for 8 of the others also show indications of one or more companions,
from spectroscopic and/or astrometric information. We suspect this may have caused
them to be missed in the Gaia list. We discuss these signs of binarity
in our sample next.

\section{Binarity}
\label{sec:binarity}

Statistically significant changes in the radial velocity of a star, as
described in Section~\ref{sec:variability}, can be evidence of its
binary nature, although this generally only works for
periods up to a few decades. Here we have taken advantage of other
sources of information that increase our completeness, and that also
extend the census of multiple systems in the Hyades into the realm of
the astrometric binaries.  In particular, the Gaia~DR3 catalog
\citep{Gaia:2023a} contains a wealth of information in both regimes.
This comes in the form of various flags relevant to binarity, as well
as the actual detection of companions separate from the target,
sometimes with their own measured parallaxes, proper motions, and
photometry.  This can allow for the confirmation of the physical
association. However, Gaia is not complete at the smallest
separations, and for DR3, close pairs are increasingly missing for
separations below about 1\farcs5 \citep[see][]{Fabricius:2021}, or are
increasingly lacking measurements of their parallaxes and proper
motions.  Nevertheless, as many of our targets are fairly bright, they
are accessible to the visual binary observers, who have been recording
the presence of many companions over the last two centuries, in some
cases down to separations as small as 0\farcs1. These measurements,
also used earlier in Section~\ref{sec:membership}, are collected in the
WDS catalog, which includes lunar occultation
observations that can reach even smaller separations.  Other indirect
evidence of binarity comes from changes in the magnitude or direction
of the proper motion at the mean epochs of the Hipparcos and Gaia surveys,
as reported in the catalog of astrometric accelerations by
\cite{Brandt:2021}.

We have summarized the available information on binarity within our
sample with a series of codes in Table~\ref{tab:RVstats}, which we
defined as follows. Targets that are either single-, double-, or
triple-lined, and for which we have determined spectroscopic orbits
(Tables~\ref{tab:elemSB1}--\ref{tab:elemSB2b}, and Appendix~A) are
indicated with ``SB1'', ``SB2'', or ``ST3'', respectively. 
For triple systems that are only single- or double-lined, we used
``ST1'' and ``ST2''. Targets discussed in Appendix~A are distinguished with a
``+'' sign at the end of the code, and those we have published previously
have an added ``p'' at the end. For stars
with no spectroscopic orbit but with $e/i > 2$, we use the code
``VAR''.  The few cases in which long-term trends in the RVs were
detected visually, even if $e/i < 2$, are flagged as``LT''. Stars with
significant changes in their proper motions ($> 3\sigma$ in either
component) between the Hipparcos and Gaia~DR3 catalog epochs, as
listed by \cite{Brandt:2021}, are given the code ``ACC'' (astrometric
acceleration). If the Gaia~DR3 catalog reported a RUWE value in excess
of 1.4, we take that as an indication of unmodeled binary motion in
the satellite astrometry, and assign the code ``RUWE''. For other
stars, more direct evidence of duplicity may come from the presence of
more than one peak in the Gaia scans that were recorded along different
directions (parameter {\tt ipd\_frac\_multi\_peak}); those situations
are indicated with ``DUP''.

Of the many stars with visual companions listed in the WDS, a few have
more than one at different separations. Here we have considered only
companions under 5\arcsec, provided that Gaia observed them
separately, that the physical association to our target is confirmed by the
similarity of the parallaxes and proper motions, and that they were
not observed spectroscopically by us as a separate target in our
sample.  These cases are indicated with ``VB'' in
Table~\ref{tab:RVstats}.  If the Gaia mission did not observe these companions
separately, or if it did, but no parallax or proper motion are
available, then we only accepted them if the separations are less than
1\farcs5, our adopted completeness threshold for Gaia.\footnote{Given
the relatively large proper motion of the Hyades, in some
cases it is possible to confirm the physical association of a companion
lacking a parallax or proper motion by verifying that it has a small motion relative to
the target, as measured at multiple epochs by visual observers.
We have not pursued this possibility here.}
The same holds when
two different visual companions are listed in the WDS (``VT''),
indicating a triple system. A good number of these visual companions
are listed in the WDS as having astrometric orbital solutions in the
literature, which confirms their association. We indicate them with
``VBO'' or ``VTO'', and in the case of the interesting hierarchical
quadruple systems vB\,124 and vB\,169, as ``VQO'' (both of the inner
orbits as well as the outer orbit are known).
In many cases the astrometric companions are the same as the
spectrosopic ones, i.e., the same orbits have been determined by both
techniques. For such instances, the code indicating availability of an
astrometric orbit is written in lower case letters.

The Gaia team has reported ``non-single star'' solutions of several
kinds for about three dozen of our targets \citep[see][]{Holl:2023,
  Halbwachs:2023}, as indicated earlier.  This includes full
spectroscopic orbits that we missed due to insufficient RV
observations (flagged here as ``GaiaSBO''), astrometric orbits in
which Gaia was able to follow and model the center-of-light of the object
(``GaiaVBO''), and objects showing obvious curvature in their paths on
the plane of the sky, sometimes supported by a similar trend in the
RVs from the satellite. These latter cases are similar in nature to
those listed by \cite{Brandt:2021}, and we assign them the
code``GaiaACC''. Finally, the Gaia catalog contains entries for
additional visual companions that are not listed in the WDS. If they
are within 5\arcsec\ and are confirmed to be associated based on their
parallaxes and proper motions, we accepted them and used the flag
``GaiaVB''. If they cannot be confirmed in that way due to a lack of information, we accepted them as true companions only if they are within 1\farcs5 of the primary, which
makes it more likely that they are related.

A total of
347 of our targets have one or more indicators of binarity, although
many do not belong to the cluster. For objects we consider members or
possible members of the Hyades, the percentage of stars with some
evidence of multiplicity is quite high: 58\% (198 out of 343 objects).

\section{Completeness}
\label{sec:completeness}

Because the RV sampling for our targets is neither uniform nor complete, and
there are limits to our velocity precision that degrades for rapid
rotators, many spectroscopic binaries will have been missed. This
is especially true for ones with long orbital periods, small
secondary masses or mass ratios, and/or high eccentricities.  To
determine our sensitivity limits as a function of each of these parameters, we
carried out simulations closely following the procedures of
\cite{Torres:2021}.  The reader is referred to that work for the full
details \citep[see also][]{Geller:2015}.  Briefly, we considered the
232 targets in our sample that are members of the Hyades, have three
or more observations, and no indication of variability. We excluded
objects for which spectroscopic or astrometric orbits have been published
by Gaia or by others, even if we ourselves detected no variability. For each star,
we generated 10,000 synthetic orbits with random orbital elements
taken from suitable distributions for field stars, as described in the
above references. For synthetic binaries with periods shorter than
the tidal circularization period ($P_{\rm circ} = 5.9$\,d; see Section~\ref{sec:elogp}),
we assumed circular orbits. Approximate primary masses sufficient for
our purposes for each
synthetic binary were obtained from a PARSEC~v1.2S model isochrone for
the Hyades from \cite{Chen:2014}, using photometry for our targets as
listed in the Gaia~DR3 catalog. We accounted for the color corrections
to the models advocated by \cite{Wang:2025}.
For each orbit, we predicted the RVs
at the actual times of observation for each star, and added Gaussian
noise corresponding to the actual RV error at each epoch.  We then
used the $e/i$ criterion described earlier to determine whether the
binary would be detected. The fraction of detected synthetic binaries
represents our completeness.

\begin{figure*}
  \epsscale{1.17}
  \hspace*{-1mm}\includegraphics[width=\textwidth]{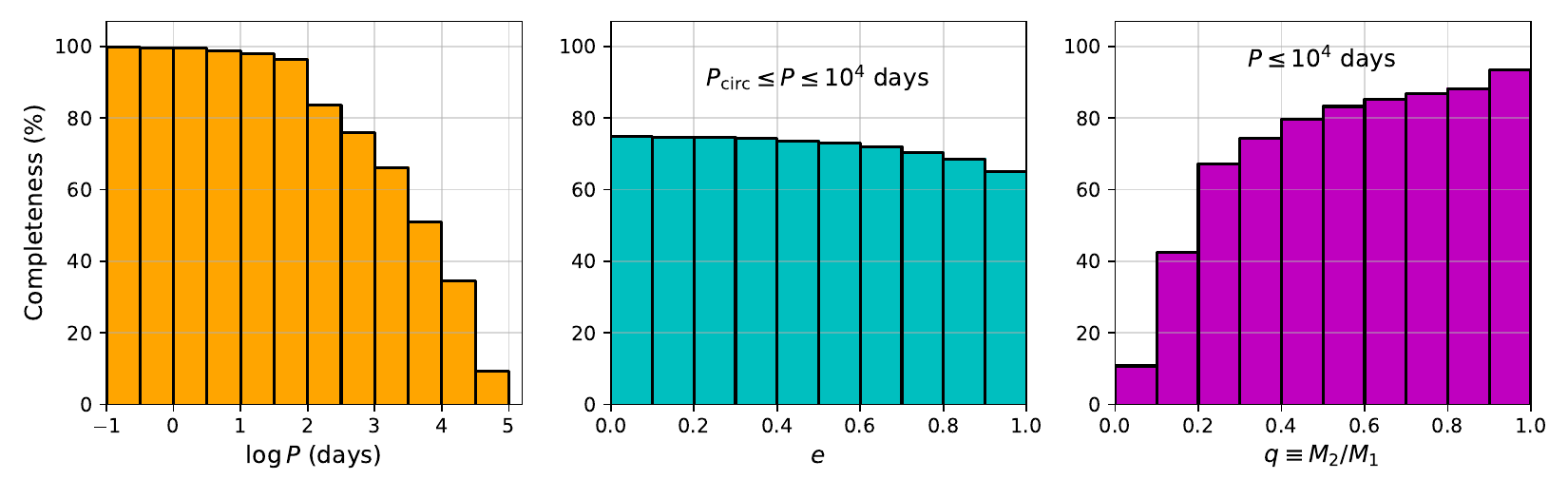}

  \figcaption{Results of our Monte Carlo simulations illustrating the
    sensitivity of our survey to spectroscopic binaries in the Hyades. The panels
    show our detection completeness as a function of orbital period,
    eccentricity (for $P_{\rm circ} \leq P \leq 10^4$\,d), and mass
    ratio ($P \leq 10^4$\,d).  Binaries with periods shorter than
    the circularization period $P_{\rm circ}$, taken here to be 5.9\,d
    (see Section~\ref{sec:edistrib}), are excluded from the
    calculation of our sensitivity as a function of eccentricity
    because their orbits have typically been made circular by
    tidal forces. \label{fig:completeness}}

\end{figure*}

The results indicate that our survey is highly sensitive to binaries
with periods under 10\,d (99.2\% completeness), still very sensitive to
those with periods up to 100\,d (97.8\%), and that we should detect 88\% of
systems with periods up to 1000\,d. Completeness begins to fall off for
longer periods, such that we are able to detect only about half of the
systems with $1000\,{\rm d} < P < 10^4\,{\rm d}$, and very few beyond
that. Our overall sensitivity up to periods of $10^4$\,d is 75\%.
This is very similar to the completeness level found by others in spectroscopic
surveys of about the same duration and velocity precision as ours
\citep[e.g., in the M67 cluster;][]{Geller:2021}.
A histogram of our completeness as a function of orbital period
is shown in the left panel of Figure~\ref{fig:completeness}.

While this paper does report a handful of spectroscopic orbits for
cluster members with periods even longer than $10^4$\,d ($\sim$27~yr),
in some cases they incorporate astrometric observations to constrain or determine
the period (vB\,185, $P \approx 28,\!600$\,d; vB\,80, $P \approx 63,\!000$\,d),
and in others they rely also
on earlier RV measurements by R.\ Griffin (vB\,176, vB\,50, $P \approx
16,\!600$\,d and 47,900\,d), or the orbits were reported originally by
other authors \citep[vB\,124, $P \approx 34,\!800$\,d;][]{Tomkin:2007}.
Some of these, and others (vB\,75, PELS\,20, $P \approx 14,\!900$\,d and
15,800\,d), are hierarchical triple or quadruple systems in which the
inner orbits have also been determined.  Because the completeness of
our survey is far lower at these longer periods, we have chosen to
restrict our attention to binaries with $P < 10^4$\,d.

Our completeness as a function of orbital eccentricity out to binary
periods of $10^4$\,d decreases only slightly toward higher
eccentricities.  On the other hand, our sensitivity to binaries of
different mass ratios is high for equal mass binaries, but not
surprisingly, much poorer for small mass ratios, especially below $q =
0.2$ (see Figure~\ref{fig:completeness}).

\section{The Multiplicity Fraction in the Hyades}
\label{sec:frequency}

Because of its proximity, the discovery of binaries in the Hyades is
likely to be substantially more complete than in other more distant clusters. This
is especially true for the astrometric technique, thanks not only to
the Gaia mission but also to the efforts of the visual binary
observers since the 19th century, and to the observers of the precious but relatively rare
lunar occultation events. Here we define the multiplicity fraction as
the fraction of objects in our sample that are in multiple systems,
including hierarchical triples and quadruples, provided that at least
one of their periods is shorter than $10^4$\,d.
We will also refer to it simply as the binary frequency.

\begin{figure}
  \epsscale{1.17}
  \hspace*{-1mm}\includegraphics[width=0.48\textwidth]{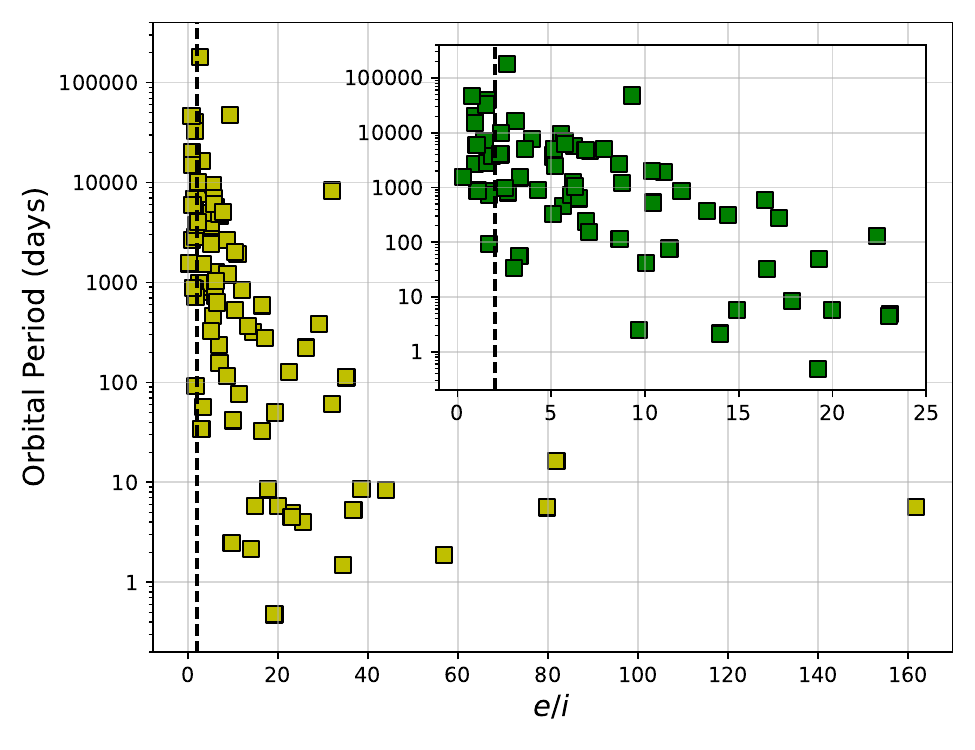}

  \figcaption{Orbital period as a function of the $e/i$ statistic for members or
  possible members of the Hyades with known astrometric or
  spectroscopic orbits. The inset zooms in on the inner portion of the
  diagram ($e/i < 25$), and shows that most targets with $e/i > 2$
  (marked with a dashed line) tend to have periods under $10^4$\,d,
  with very few exceptions. We may expect, then, that other objects
  with $e/i > 2$, which we consider to be binaries even though the
  orbits are unknown, will also typically have
  periods under $10^4$\,d, and we include them in the calculation of the
  multiplicity fraction.
  \label{fig:period_ei}}

\end{figure}

Of the 343 targets in our sample that we consider to be members or
possible members of the Hyades,
90 have known spectroscopic orbits with periods shorter than our cutoff:
53 are SB1s, 28 are SB2s, 7 are triples, and 2 are
quadruples.\footnote{Two
of the SB1s (vB\,30 and HD\,40512) show long-term velocity trends superimposed on
the orbital motion, indicating they are actually triple systems. However,
this reclassification does not change the total number of systems, so our
frequency calculations are unaffected. The third such case with a RV drift (vB\,102)
has a known outer orbit.}
A further 5 objects
have variable RVs ($e/i > 2$) and are considered to be binaries
as well. However, examination of the RV histories of these 5 stars shows that two of them
(vB\,328, vB\,184) have slow upward or downward drifts over at least $10^4$\,d,
indicating that their periods must be significantly longer. We therefore
excluded them. For the other 3 (vB\,290, vB\,141, REID\,88B) we assumed
here that their periods are shorter, based on the fact that the vast
majority of binaries with known orbits and $e/i > 2$ have periods
under $10^4$\,d (see Figure~\ref{fig:period_ei}).

The total number of spectroscopic systems is then $53 + 28 + 7 + 2 + 3 = 93$,
which gives a raw multiplicity of $93/343 = 27$\%. After correction for
our incompleteness up to $10^4$\,d (Section~\ref{sec:completeness}), we obtained
$36 \pm 4$\%. This represents the spectroscopic binary fraction in the Hyades
for stars of all spectral types (AFGKM). However, binary frequency estimates for
other open clusters and field populations have typically focused on binaries
with ``solar-type'' primaries (approximately FGK). If we restrict our estimate 
to just those spectral types for a more fair comparison
(corresponding approximately to temperatures
of 4000--7000\,K, or masses of $\sim$0.7--1.5\,$M_{\sun}$), the
result is only a little different. We count 65 spectroscopic systems in our sample
among the 217 FGK targets, for a raw frequency of $65/217 = 30$\%.
After correction for incompleteness, this becomes $40 \pm 5$\%, which is slightly
higher than that found in other open clusters, and significantly higher
than the halo and solar-neighborhood (field) samples, although the former
has not been corrected for incompleteness. Table~\ref{tab:binfreq}
summarizes these determinations, expanded from a similar table by \citealt{Torres:2021}.
As suggested above, the higher binary fraction in the Hyades may be caused
in part by the more complete inventory of multiple systems in this cluster.

Aside from the spectroscopic systems, 3 of our targets were not detected as
RV variables, but do have astrometric orbits with periods shorter than our
cutoff (vB\,122, HAN\,291, vB\,123). The first is from the WDS,
and the other two are from Gaia.
Many other objects in our sample are known to have close visual companions,
although in most cases the information is limited to the angular separations
and photometric properties. Nevertheless, these systems 
are of interest for computing the total (spectroscopic + astrometric)
multiplicity fraction in the Hyades.
At the mean distance to the cluster, Kepler's third law and an assumed
typical binary mass of $\sim$1\,$M_{\sun}$ indicates that a period
of $10^4$\,d corresponds to a semimajor axis of about 0\farcs2, which we may
assume is also a typical angular separation for such systems. We therefore considered
targets with companions up to this separation, unless they are already in
one of the other categories mentioned above. A total of 12 objects meet this
condition.
We obtained a corrected total multiplicity fraction of $41 \pm 4$\%
for stars of all spectral types, and $44 \pm 5$\% for FGK stars. These
should be considered lower limits, as the completeness for the astrometric
systems is unknown.

\setlength{\tabcolsep}{6pt}
\begin{deluxetable}{lccc}
\tablecaption{Spectroscopic Binary Frequency in the Hyades Compared with Other Populations.\label{tab:binfreq}}
\tablehead{
\colhead{Population} &
\colhead{Age (Gyr)} &
\colhead{Frequency (\%)} &
\colhead{Source}
}
\startdata
$\alpha$ Per & 0.09    &  $32 \pm 9$    &     1        \\ 
Blanco~1  &  0.1       &  $20 \pm 6$    &     2        \\
Pleiades  &  0.125     &  $25 \pm 3$    &     3        \\
M35       &  0.15      &  $24 \pm 3$    &     4        \\
Coma Berenices & 0.45  &  $22 \pm 8$    &     5        \\ 
Praesepe  &  0.7       &  $30 \pm 6$    &     6        \\ 
Hyades    &  0.7       &  $40 \pm 5$    &  This paper  \\
NGC~7789  &  1.6       &  $31 \pm 4$    &     7        \\
NGC~2506  &  2.0       &  $35 \pm 5$    &     8        \\
NGC~6819  &  2.5       &  $22 \pm 3$    &     9        \\
M67       &  4         &  $34 \pm 3$    &    10        \\
NGC~188   &  7         &  $33 \pm 4$    &    11        \\
        Field     &  \nodata  &  \phm{\tablenotemark{a}}$14 \pm 2$\tablenotemark{a}    &    12        \\
Halo      &  $\sim$10  &  $15 \pm 2$    &    13        
\enddata

\tablecomments{Binary frequencies for mostly solar-type stars in different populations,
generally up to orbital periods of $10^4$\,days, and generally corrected for incompleteness.
Sources are: 
(1) \cite{Mermilliod:2008a};
(2) \cite{Mermilliod:2008b};
(3) \cite{Torres:2021};
(4) \cite{Leiner:2015};
(5) \cite{Mermilliod:2008c};
(6) \cite{Mermilliod:1999};
(7) \cite{Nine:2020};
(8) \cite{Linck:2024};
(9) \cite{Milliman:2014};
(10) \cite{Geller:2021};
(11) \cite{Narayan:2026};
(12) \cite{Raghavan:2010};
(13) \cite{Latham:2002}.
The estimates for $\alpha$~Per, Blanco~1, Coma Berenices, Praesepe,
and the halo sample have not been corrected for incompleteness.}
\tablenotetext{a}{We inferred this value from the work of
\cite{Raghavan:2010}, by scaling their 44\% total binary frequency for
all periods (including triples and quaduples) by the fractional
area of their log-normal period distribution up to periods of $10^4$~d,
for consistency with the typical cutoff period for the other samples.}

\end{deluxetable}
\setlength{\tabcolsep}{6pt}  

\section{Distribution of binary properties}
\label{sec:binaryproperties}

In the following subsections, we analyze the distributions of binary orbital periods,
eccentricities, and mass ratios for confirmed cluster members and possible
members in our sample, as well as the eccentricity-period relation.
The total number of orbits with $P \leq 10^4$\,d is 97,
counting also 4 outer orbits of systems with multiplicity higher than two.
This includes one outer orbit from the WDS (vB\,122), as well as three
spectroscopic or astrometric orbits from Gaia (HD\,281459, HAN\,291, vB\,123).
Out of these 97 orbits, we have ignored the one for the white
dwarf binary V471\,Tau, in which mass transfer has likely altered the orbital
elements.

\subsection{The Period Distribution}
\label{sec:pdistrib}

\begin{figure}
  \epsscale{1.17}
  \hspace*{-1mm}\includegraphics[width=0.48\textwidth]{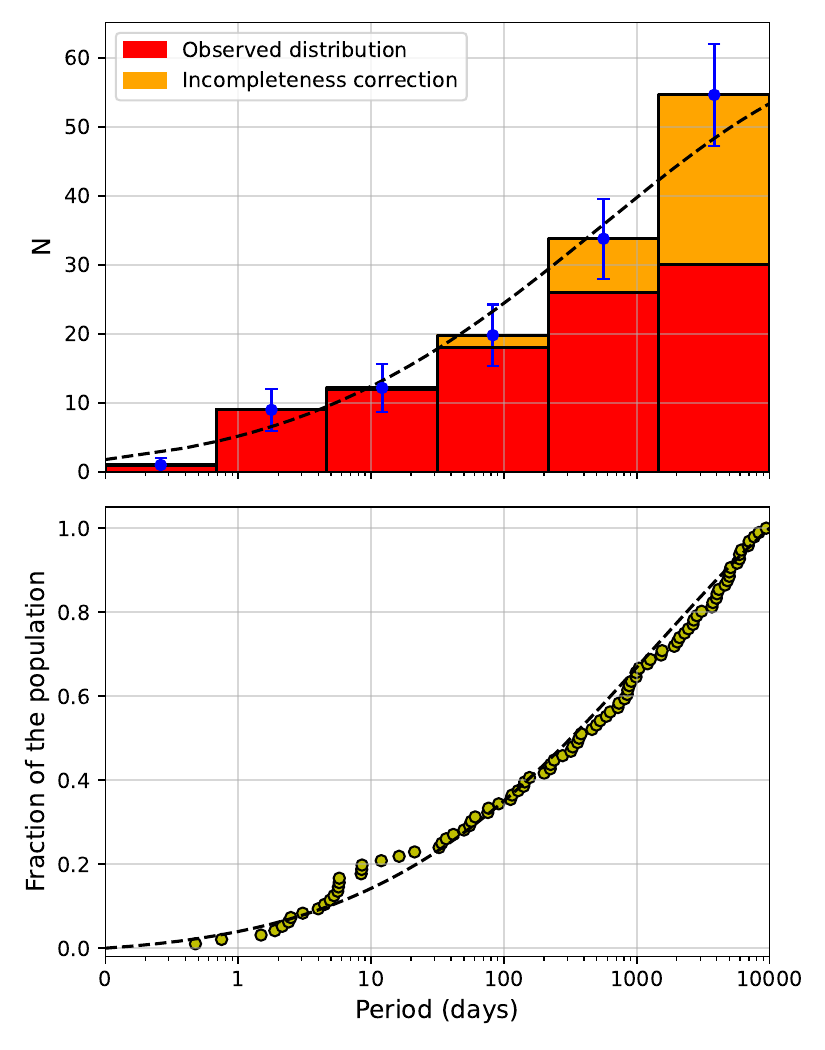}

  \figcaption{\emph{Top:} Distribution of orbital periods for the 96
    binary members of the Hyades cluster with $P \leq 10^4$\,d.
    The optimal number of bins was determined
    following the Freedman-Diaconis rule.
    Error bars on the corrected distribution
    are from counting statistics, and the dashed line is the
    log-normal distribution for solar-type binaries in the field, from
    the work of \cite{Raghavan:2010}.
    \emph{Bottom:} Cumulative distribution function of observed periods,\
    compared against the same field distribution as above (dashed line; see the text).\label{fig:pdistrib}}

\end{figure}

Figure~\ref{fig:pdistrib} presents the histogram of the 96 Hyades
binaries with orbital periods up to $10^4$\,d, in which we
include triple and quadruple systems as well, and we count the period
of each subsystem as if it were from an independent binary. We assume here,
and in the following section, that
internal dynamics in these higher-multiplicity systems have not 
modified the orbital elements significantly over the age of the Hyades.
The bottom part of each bin corresponds to the
binaries actually detected, and the top part adds the incompleteness
correction from Figure~\ref{fig:completeness}, applied individually to each
binary period. The dashed line
represents the log-normal distribution for solar-type binaries in the field
from \cite{Raghavan:2010}, with a mean of $\log P = 5.03$ and a
standard deviation of $\sigma_{\log P} = 2.28$. We have normalized the
field distribution to the same number of detected binaries up to
$10^4$\,d, with the added incompleteness corrections.

In the bottom panel of Figure~\ref{fig:pdistrib},
we show the cumulative distribution of the observed periods, together with the
log-normal field distribution (dashed line) from \cite{Raghavan:2010},
modified so that it
reflects the same level of incompleteness as the real distribution.
This latter curve was obtained by multiplying the Gaussian field distribution
by our completeness function from Figure~\ref{fig:completeness}, and integrating.
The Anderson-Darling test indicates that the shapes of
the Hyades and field distributions are
statistically indistinguishable ($p~{\rm value} = 0.62$).

In an earlier study, \cite{Duquennoy:1991} presented a figure with the
binary period distribution in the Hyades based on 45 systems taken
from the work of \cite{Griffin:1985a}.  They noted an apparent excess of short-period binaries
between 1 and 10\,d compared to the distribution of solar-type binaries
in the field, as it was then known, and speculated this might be an evolutionary effect.
They suggested that tidal forces may have shortened the periods of
some systems, causing a pile-up at the present time, with the
possibility that some of these binaries may eventually coalesce.
Alternatively, a
similar outcome could result from the mechanism invoked to produce
W~UMa systems (angular momentum loss in spin-orbit coupled binaries).
However, it does not appear that the period distribution of
\cite{Duquennoy:1991} was corrected for incompleteness, so it is
unclear how significant that feature was.
Nevertheless, with a sample that is now more than twice the
size of the one they had at the time, and proper corrections for
incompleteness, the cumulative distribution function of
Figure~\ref{fig:pdistrib} still shows a small kink at that location,
although as mentioned earlier, it does not appear to be
statistically significant.

\subsection{The Eccentricity Distribution}
\label{sec:edistrib}

\begin{figure}
  \epsscale{1.17}
  \hspace*{-1mm}\includegraphics[width=0.48\textwidth]{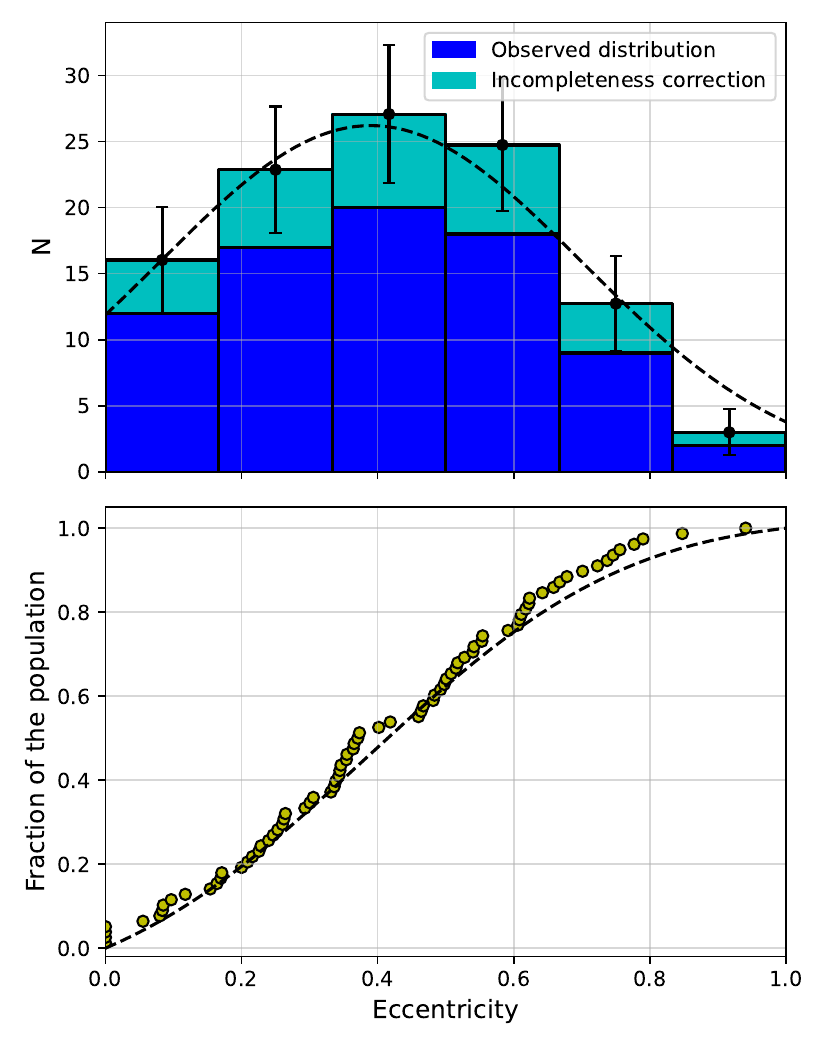}

  \figcaption{\emph{Top:} Distribution of orbital eccentricities for the 78
  binary members of the Hyades cluster with periods
  $P_{\rm circ} \leq P \leq 10^4$\,d, including triples and quadruples,
  and excluding the two giants with known spectroscopic orbits.
    Error bars on the corrected distribution
    are from counting statistics, and the dashed line is the
    Gaussian model for solar-type binaries in the field, from
    the work of \cite{Geller:2012}. \emph{Bottom:} Cumulative
    distribution function (see the text).\label{fig:edistrib}}

\end{figure}

Figure~\ref{fig:edistrib} presents the distribution of eccentricities
for the 78 binaries with periods between the tidal circularization period
($P_{\rm circ} = 5.9$\,d; Section~\ref{sec:elogp}) and $10^4$ days,
out of the 96 systems considered
in the previous subsection. Here we have also excluded the two giants in the
Hyades with known orbits (vB\,41, vB\,71), as stronger tidal forces due
to their much more extended structures may have affected their
eccentricities differently than the dwarfs.

The top portion of each bin in the histogram represents the
correction for incompleteness, derived in the same way as explained
for the periods. We have indicated with a dashed line the eccentricity
distribution for field binaries, which \cite{Geller:2012} have
shown is well approximated by a Gaussian with a mean of 0.39 and a
standard deviation of 0.31. This is the same distribution adopted for
our incompleteness corrections in Section~\ref{sec:completeness}.
The cumulative distribution function of eccentricities is shown in the bottom panel,
where the dashed line represents the field distribution with the
same incompleteness as the binaries in our sample.
There is very close agreement between this model and the observed distribution:
the Anderson-Darling test returns a $p$ value of 0.68.

\subsection{The Mass Ratio Distribution}
\label{sec:qdistrib}

The mass ratios measured directly for the SB2s in our sample
$(q \equiv M_2/M_1 = K_1/K_2)$ were supplemented
here with mass information from the SB1s, which comes only in the form
of the spectroscopic mass function,
$f(M) = M_2^3 \sin^3 i/ (M_1+M_2)^2 = (P/2\pi G) (1-e^2)^{3/2} K_1^3$.
Inferring a mass ratio from $f(M)$ requires knowledge of the primary
mass and the inclination angle.
While statistical inversion methods may be applied to convert the distribution of
$f(M)$ into one of $q$ if rough estimates of $M_1$ are available
\citep[see, e.g.,][and references therein]{Lucy:1979, Mazeh:1992,
Shahaf:2017}, here we have chosen to proceed differently and
follow the methodology of \cite{Torres:2021}.
This uses the fact that in a cluster such
as the Hyades, with a known age and metallicity, it is possible to obtain
sufficiently accurate values for
both $M_1$ and $M_2$, and therefore $q$, by using a model isochrone to
deconvolve the brightness
measurements for the combined light of each binary,
constrained by the
measured values of $f(M)$, or equivalently, by the minimum secondary mass,
$M_2 \sin i = (P/2\pi G)^{2/3} \sqrt{1-e^2} K_1 (M_1+M_2)^{2/3}$. 
The model isochrone we used is from the PARSEC~v1.2S series \citep{Chen:2014},
with color corrections as determined by \cite{Wang:2025}.
Accurate parallaxes are also required, and the photometry must be free
from contamination by other nearby sources.

In addition to the Gaia~DR3 photometry ($G$, $G_{\rm BP}$, $G_{\rm RP}$)
that is available for all our SB1s, we also used the 2MASS $JHK_{\rm S}$
magnitudes \citep{Cutri:2003}, as these near-infrared measurements are more
sensitive to the low-mass companions that many of these binaries have. 
Complete details on the methodology to infer the mass ratios and their
uncertainties may be found in the paper by \cite{Torres:2021}.
The determination of $q$ for each SB1 was repeated 100 times, propagating
the uncertainties in the photometry and in $M_2 \sin i$ in a Monte Carlo
fashion.

Initial determinations revealed poor photometric
fits for some of the binaries, which we traced to inaccurate Gaia parallaxes,
most likely a result of biases from the very binary nature of the objects. We
found much improved fits by replacing the trigonometric parallaxes for all
objects with their ``kinematic" parallaxes (moving cluster method), derived
using the Gaia proper motions and the assumption that all Hyades members
share a common motion in space.
The space motion vector and convergent point coordinates were adopted
also from the Gaia mission itself \citep{Gaia:2018}.
In a few cases for which Gaia~DR3 did not report a
proper motion, we used highly precise values from \cite{Altmann:2017}, which are based on
positions from an earlier data release by the Gaia mission (DR1) combined
with positions from ground-based catalogs going back decades. This tends
to average out the orbital motion in most cases, yielding more accurate
proper motions.

For our study of the mass ratio distribution we have included triple
and quadruple systems in which
the mass ratios for the outer subsystems are also known. In these cases, we
have added up the masses of the inner components and considered them
as a single object for the purpose
of computing the mass ratio involving the outer component. This assumes that
these mass ratios in the outer orbits of triple and quadruple systems are
drawn from the same population as those of binary systems.

Among the 96 member systems in our sample (V471\,Tau excluded)
with known inner or outer orbits
and periods under $10^4$\,d, 54 are SB1s, 39 are SB2s (including triples
and quadruples; see above), and another 3 have only
astrometric orbits from Gaia or the WDS (HAN\,291, vB\,123, vB\,122).
These last three provide
no useful dynamical mass information for our
purposes,\footnote{The orbits for HAN\,291 and vB\,123, reported by
Gaia, only pertain to the motion of the photocenter, and vB\,122 has no
parallax or proper motion in the Gaia DR3 catalog.} and were excluded.
Two of the SB1s are in triple systems with known outer orbits
(vB\,22, vB\,102; see Table~\ref{tab:triples}). For the first, in which
the inner pair is an eclipsing binary with known absolute masses, astrometric
observations have allowed us to determine the mass of the tertiary, and
therefore the mass ratio in the outer orbit (see Appendix~A).
For the other, the detectable wobble of the inner pair
on the plane of the sky allowed \cite{Tokovinin:2021c} to derive the mass
ratio for the inner binary. We have added these two systems to the
39 with directly measured values of $q$.

On the other hand, not all 52 of the remaining SB1s are suitable for the purpose of
this section. We removed 6 of these objects for a variety of reasons.
Three of the ones with long periods (vB\,91, vB\,271, HAN\,432) have components
of similar brightness ($\Delta m \lesssim 1$~mag),
and suffer from severe blending of the spectral lines
such that the velocities could only be measured from the blend, as if it
were a single object. In these cases, the mass functions are unreliable because the velocity semiamplitudes will always be underestimated. Another object
(vB\,190) lacks a parallax and a proper motion from Gaia. Our attempt to use
proper motions from an alternate source to infer its kinematic parallax resulted
in a large mass ratio ($\sim$0.8), inconsistent with the fact that the companion
is 5.5~mag fainter than the primary.
Two additional systems (vB\,9, vB\,41) were removed for giving very poor fits to
the photometry.
The remaining $52 - 6 = 46$ SB1s were analyzed together with the 41 SB2s.

\begin{figure}
  \epsscale{1.17}
  \hspace*{-1mm}\includegraphics[width=0.48\textwidth]{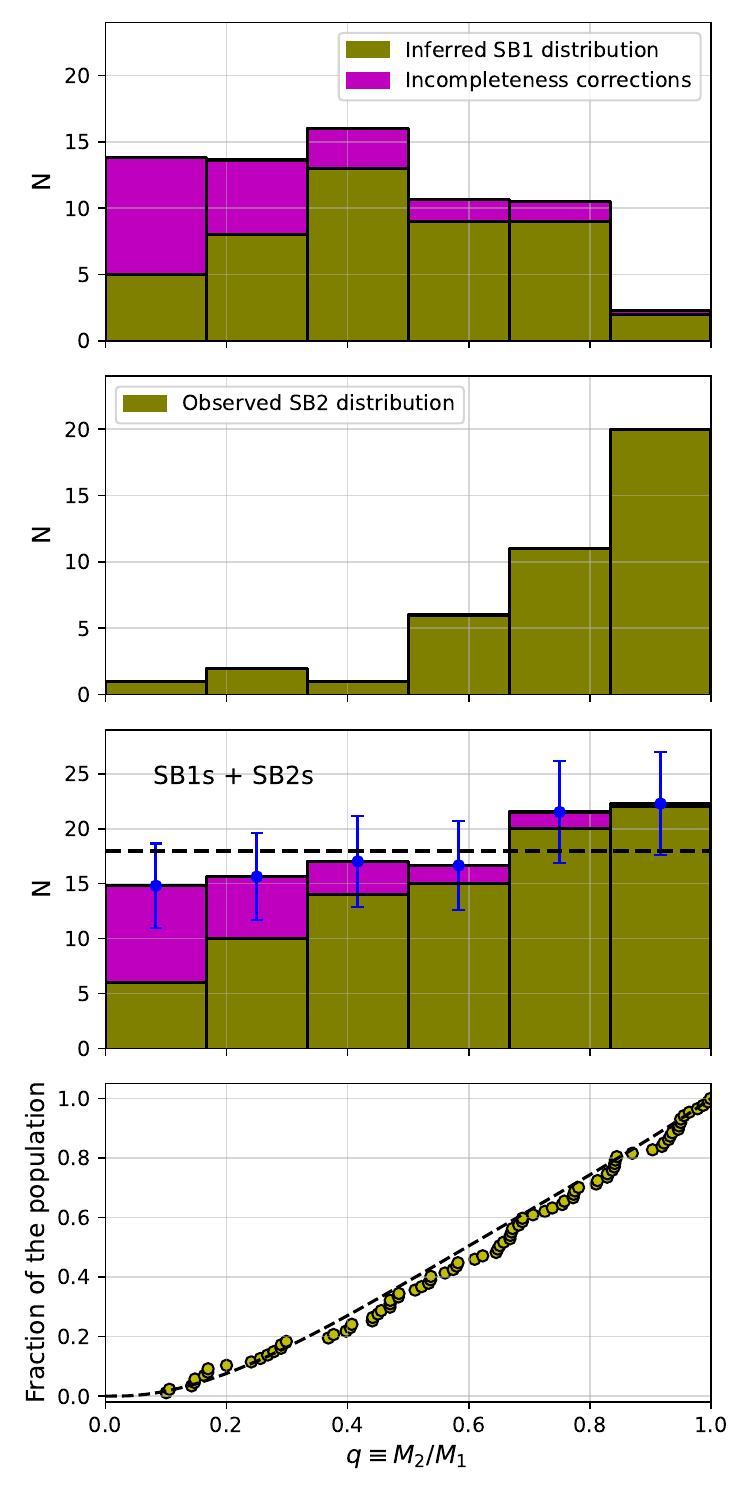}

  \figcaption{Distribution of mass ratios in the Hyades cluster,
  from 87 binary and multiple systems. From top to bottom, the
  panels show the inferred distribution for single-lined systems,
  the observed distribution for double-lined systems, the sum,
  and finally the cumulative distribution (see the text).
  Incompleteness corrections are indicated in the first and third panels.
    Error bars on the corrected distribution for SB1s+SB2s
    are from counting statistics, and the dashed line (shown also
    in the bottom panel) corresponds
    to a flat distribution. \label{fig:qdistrib}}

\end{figure}

Our procedure for inferring mass ratios for the SB1s generates a distribution
of 100 estimates of $q$ for each binary. The median values for each system are
plotted in histogram form in the top panel of Figure~\ref{fig:qdistrib}
(bottom section of each bin). Incompleteness corrections were computed for
each of those 100 estimates,
and the median values of those corrections for each binary are shown in the top portion of
each bin. As expected, the distribution with incompleteness corrections included
rises toward small mass ratios,
as such systems are increasingly harder to detect spectroscopically.

The second panel of Figure~\ref{fig:qdistrib} displays the distribution
of mass ratios for the SB2s. In this case, it is unclear whether incompleteness
corrections should be applied because these systems are often recognized as double-lined
early-on in the observations, and are typically then followed up more frequently
until a publishable orbital solution has been obtained. We show the distribution
without any adjustments. The preference for equal-mass systems reflects the fact
that components of similar mass also have similar brightness, making the secondaries
easier to detect and measure.

The third panel adds together the distributions for the SB1s and SB2s. There
is perhaps a slight rise toward equal-mass systems, although the
Anderson-Darling test indicates no significant
difference compared to a flat distribution ($p$ value = 0.48), which we show with a
dashed line in the figure. The bottom panel presents the cumulative probability
distribution of observed mass ratios, with the flat model adjusted for incompleteness.

Returning to the three systems with blended lines that were excluded
above, the relatively bright secondaries in all three cases imply that the mass ratios
must be quite large. As an experiment to see how their exclusion might
influence the results, we added them to
the sample with randomly chosen $q$ values larger than 0.9, which places
them in the rightmost bin of the SB1+SB2 histogram. We then repeated the
Anderson-Darling test to compare against a flat distribution.
Although adding these three binaries tends to reinforce the visual impression
of an increase in the frequency toward large mass ratios, the $p$ value
we obtain is still large (0.19), indicating no meaningful distinction
between the two distributions.

Finally, for the benefit of the reader, we collected in Table~\ref{tab:statistics}
the properties of the binary and multiple systems used for the study of
the distributions of mass ratios, periods, and eccentricities in this
section and in the previous two. In addition to $P$, $e$, $q$, and $M_1$
for each system, we list the correction factors
applied to account for incompleteness. The table has 96 entries, which
is the number of systems used for the period distribution. Not all were
used for the eccentricity and mass ratio statistics, as explained earlier.

\setlength{\tabcolsep}{6pt}
\begin{deluxetable*}{rlccccccc}
\tablewidth{0pc}
\tablecaption{Material Used for Binary Statistics.\label{tab:statistics}}
\tablehead{
\colhead{\#} &
\colhead{Name} &
\colhead{$P$ (d)} &
\colhead{$e$} &
\colhead{$q$} &
\colhead{$M_1$ ($M_{\sun}$)} &
\colhead{$C(P)$} &
\colhead{$C(e)$} &
\colhead{$C(q)$}
}
\startdata
 14 &  HD 17922   &  \phn318.8792    &  0.3361 &  0.441 &  1.186 &  1.216 &  1.349 &  1.224 \\
 48 &  LP 413-18  &  \phn\phn16.2493 &  0.5009 &  0.671 &  0.540 &  1.029 &  1.362 &  1.181 \\
 49 &  LP 413-19  &  \phm{222}0.4761 &  0.0000 &  0.950 &  0.449 &  1.000 & \nodata &  1.059 \\
 68 &  LP 357-4   &      1523.8262   &  0.4924 &  0.829 &  0.411 &  1.479 &  1.361 &  1.152 \\
103 &  vB 8       &  \phn855.8604    &  0.3733 &  0.377 &  1.429 &  1.358 &  1.351 &  1.283 
\enddata
\tablecomments{Primary masses ($M_1$) were estimated using photometry
from Gaia, and the isochrone for the Hyades employed in this work. The
mass ratios were estimated as described in this section.
For the triple systems (vB\,22, HAN\,346, vB\,304), which have two entries each,
$M_1$ in the second entry is the total mass of the ``primary'' in the outer orbit.
For the hierarchical quadruple system vB\,124, which also has two entries,
the mass ratios pertain to each of the inner binaries.
The last three columns contain the correction factors applied to account
for incompleteness in the distributions of $P$, $e$, and $q$. Empty
fields in the $q$ column indicate the system was not used to generate
the corresponding distribution, for reasons explained in the text. In
those cases, the $C(q)$ column is also empty. Systems with no $C(e)$ value
are either giants (vB\,41, vB\,71) or ones with $P < P_{\rm circ}$,
which were excluded from the eccentricity distribution.
(This table is available in its entirety in machine-readable form.)}
\end{deluxetable*}
\setlength{\tabcolsep}{6pt}

\vskip 30pt
\subsection{Tidal Circularization: the $e$--$\log P$ Diagram}
\label{sec:elogp}

The empirical property that the shortest-period spectroscopic binaries
tend to have circular orbits was recognized by observers more than a
century ago \citep[see, e.g.,][]{Campbell:1910, Ludendorff:1910,
Schlesinger:1910}, even if the physical reason appeared somewhat mysterious
at the time. It is now well understood that this is caused by tidal
forces, with their strong dependence on the orbital separation
(inverse cube relation), and additional dependence on the stellar radii
and masses. However, there is still considerable debate about
the efficiency of the different tidal mechanisms that have been proposed.
Tidal forces result in the dissipation of orbital
energy through friction, with gradual reduction of the eccentricity
and period over time. The
effect is easily seen in the familiar eccentricity-log period \mbox{($e$--$\log P$)}
diagram, in which most binaries tend to be circular up to a certain
period, beyond which there is a rise toward higher eccentricities with
a broad distribution of values. This transition period has been
postulated to depend on stellar age \citep[e.g.,][]{Mathieu:1988,
Latham:1992, Mathieu:2004, Mazeh:2008}, and these works and others published since have
presented evidence to that effect from studies of several coeval populations
of solar-type binaries. Those data sets have included pre-main-sequence (PMS) objects,
several open clusters, field binaries in the solar neighborhood, and a
sample of older halo binaries \citep{Meibom:2005, Meibom:2006,
Geller:2012, Milliman:2014, Leiner:2015, Nine:2020, Geller:2021}.

\cite{Meibom:2005} introduced a robust way of measuring this transition
or ``circularization'' period, $P_{\rm circ}$, and used it to make
determinations for several of the samples of coeval binaries mentioned above
(see their
Figure~9, which has been updated more recently by \citealt{Nine:2020}
and by \citealt{Zanazzi:2022}).  One of the \cite{Meibom:2005} data sets
combined 21 systems in the Hyades with a slightly larger number of
binaries in Praesepe, which is of similar age. 
They discussed all of these determinations in the context of the two
leading tidal mechanisms: the equilibrium tide, operating in
stars with convective envelopes \citep{Zahn:1966, Zahn:1977,
Zahn:1989}, and the dynamical tide, acting in hot stars with radiative
envelopes
\citep{Zahn:1975, Zahn:1977, Goodman:1998, Terquem:1998, North:2003}.
They found that the measurements are not consistent with the
predictions by \cite{ZahnBouchet:1989}, who proposed that essentially
all of the circularization should occur during the pre-main-sequence
phase, when stars have larger sizes and deeper convective envelopes.
For binaries with
typical
initial eccentricities near 0.3, \cite{ZahnBouchet:1989} expected
the ``cutoff'' period separating circular from eccentric binaries
to be between about 7.2 and 8.5\,days,
depending somewhat on the masses and initial conditions. However,
while the younger samples
from \cite{Meibom:2005} (PMS binaries, the Pleiades, M35)
were consistent with those expectations, the
older ones (M67, NGC\,188, field binaries, and halo binaries) had even
longer $P_{\rm circ}$ values, suggesting that tidal forces continue to
act during the main-sequence phase, reducing the eccentricities even
further.  Notably, the ${\rm Hyades}+{\rm Praesepe}$ sample was an
outlier, with a $P_{\rm circ}$ value they estimated to be $3.2 \pm 1.2$\,d,
or less than half of that expected. 

An early version of the $e$--$\log P$ diagram for the Hyades alone,
based on 46 systems (including triples), was presented by \cite{Stefanik:1992}, who
remarked that the transition period was somewhat uncertain, and may be
between 5.75 and 8.50\,d, based on the binaries with the first eccentric
and last circular orbits.
An update was shown
by \cite{Griffin:2012} (see their Figure~69), although without any
discussion other than the mention of a few objects of interest at the
short period end. The 64 systems they displayed also included the outer
orbits of several triples known at the time.  The present work offers
an opportunity to revisit the determination of $P_{\rm circ}$ with a
larger and cleaner sample, now including much improved knowledge about
higher-order multiplicity. Our sample
features several new binaries in the
critical period range ($<10$\,days): LP\,413-19 (0.47\,d), vB\,34 (3.06\,d),
HD\,283810 (4.92\,d), and HD\,36525 (5.68\,d). Our goal in this section is
to include, to the extent possible, only unevolved main-sequence systems
that have not suffered
changes in their eccentricity due to internal dynamics.
Therefore, in addition to excluding the triples and quadruples with known orbits,
and other systems mentioned at the beginning of Section~\ref{sec:binaryproperties},
we have removed the two giants that have spectroscopic orbits (vB~41
and vB~71).

\begin{figure}
\epsscale{1.17}
  \hspace*{-1mm}\includegraphics[width=0.48\textwidth]{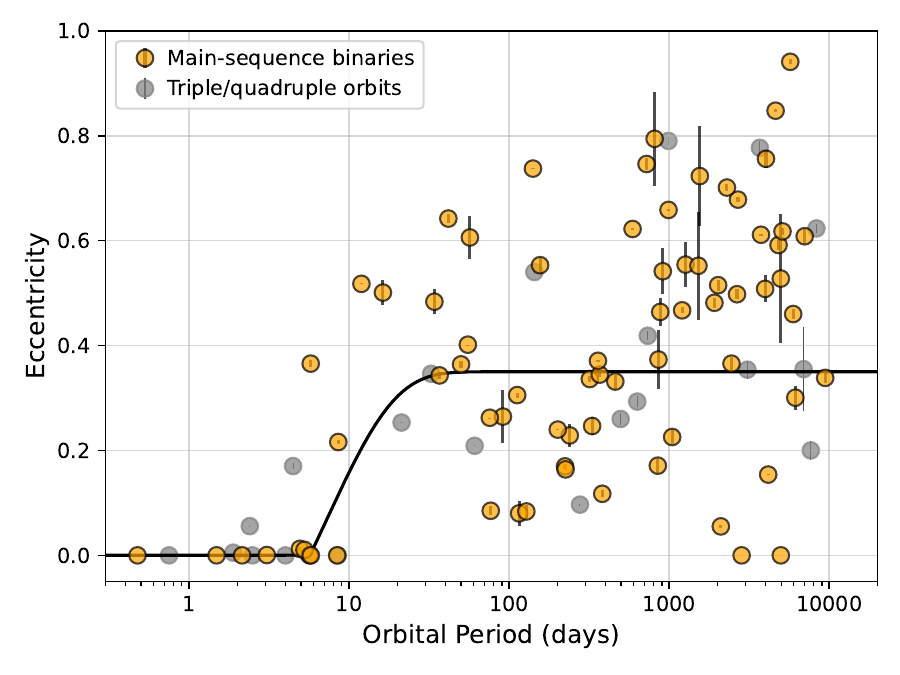}

\figcaption{Eccentricity vs.\ period diagram for our sample of 72
binary members in the Hyades. The curve represents our fit to the circularization
function of \cite{Meibom:2005}, giving $P_{\rm circ} = 5.9$\,d.  The gray symbols represent
22 additional inner and outer orbits of triple and quadruple systems,
which were not used to determine the circularization
 period. \label{fig:elogp}}
\end{figure}

The data for the remaining 72 binary members with periods up to
$10^4$\,d were used to determine $P_{\rm circ}$ using the formalism
of \cite{Meibom:2005}, for consistency with the earlier studies of
other coeval populations. We obtained $P_{\rm circ} = 5.9 \pm
1.1$\,d, where the uncertainty is statistical only.\footnote{The true uncertainty in the circularization period is difficult
to quantify, as it is likely dominated by the finite size of our
sample, rather than errors in the eccentricity measurements.
Furthermore, the value of $P_{\rm circ}$ can be sensitive to the
properties of just one or two systems at critical locations in the
$e$--$\log P$ diagram (see below). 
\label{foot:elogp}} This is longer than the value of 3.2\,d
reported by \cite{Meibom:2005}, but is still well below the predictions
by \cite{ZahnBouchet:1989}.
The $e$--$\log P$ diagram for the Hyades is shown
in Figure~\ref{fig:elogp}.

The spectral types for our Hyades binaries range between A and M,
based on our spectroscopic temperatures, whereas the previously
reported coeval samples used to estimate $P_{\rm circ}$ have been
described loosely as consisting of binaries with solar type
primaries (FGK), although some of the samples extend beyond that range.
As before, this corresponds approximately to $T_{\rm eff} = 4000$--7000\,K,
or masses of $\sim$0.7--1.5\,$M_{\odot}$.
On the other hand, the tidal theory of \cite{ZahnBouchet:1989}
focused on masses in the range 0.5--1.25~$M_{\sun}$.
We find that if we restrict the temperatures in our sample to
correspond to the range 0.7--1.5\,$M_{\sun}$ (49 binaries)
or to 0.5--1.25\,$M_{\sun}$ ($\sim$3500--6300\,K, 55 binaries),
the result for $P_{\rm circ}$ in both cases is only slightly shorter:
$4.9 \pm  1.3$\,d.

Recent studies of tidal circularization in binaries have cast doubt on
the long-held belief
that the circularization period depends
primarily on the age of the population being considered.  This 
perception that age is the dominant factor has been driven largely by
the findings of \cite{Meibom:2005}, and others, that binaries in older
samples such as M67 or the halo are circularized out to much longer
periods than younger ones.
In a recent investigation based on several hundred eclipsing binaries from the
Kepler and TESS missions, \cite{Zanazzi:2022} agreed with earlier authors
that there is indeed a statistically significant
difference between the eccentricity distributions of young and old
binaries \cite[see also][]{Penev:2022, Narayan:2026}.
A different 
 interpretation was put forward
by \cite{Bashi:2023} and \cite{IJspeert:2024}, who analyzed even
larger samples of binaries from Gaia and from TESS, respectively. They
concluded that $P_{\rm circ}$ depends primarily on the effective
temperature, rather than on age, although the two properties are
of course statistically correlated. Their hotter binaries displayed shorter
circularization periods than the cooler ones.

As a test, we divided our Hyades sample in half by
temperature, and determined the circularization period in each subset.
We obtained $P_{\rm circ}$ values of $5.9 \pm 1.5$\,d for the hotter half
and $6.0 \pm 1.5$\,d for the cooler half.
Within the uncertainties there is no difference, although
admittedly our samples here are very small.

It is worth reiterating
that inferring a circularization period from
relatively small samples of just a few dozen binaries, as has been the
case for most previous studies in clusters, is fraught with uncertainty
stemming in part from
the irregular distribution of orbital periods, over which
there is no control, and possibly also the methodology used.
Limitations from the finite sample sizes have previously
been pointed out, e.g., by \cite{Zanazzi:2022}. On the other hand, the
one parameter we can control perfectly in the case of the Hyades is
the age. Field samples, such as those analyzed in
the larger studies above, are less homogeneous as they must rely
instead on age estimates that are
necessarily less precise, and are subject to various observational uncertainties,
including a lack of information on multiplicity, or imperfect
knowledge of the stellar parameters or of interstellar reddening.

To further illustrate the dependence of $P_{\rm circ}$ 
on the distribution of orbital periods, and its sensitivity to the
location of just one or two systems in the $e$--$\log P$ diagram, we
point out that our result above (5.9\,d) changes significantly if we remove
two binaries with relatively short periods that stand out for their large eccentricities:
vB\,121 ($P = 5.75$\,d, $e = 0.37$) and vB\,62 ($P = 8.55$, $e = 0.22$).
Without these two systems, we obtain $P_{\rm circ} = 8.6$\,d, which would
then become consistent with the predictions of \cite{ZahnBouchet:1989}.
The primary of vB\,121 is a late F star, for which we find no
evidence of additional components that might be perturbing the eccentricity,
in either the WDS or Gaia~DR3 catalogs,
or from our own spectroscopy.
vB\,62 has the same spectral type, and in this case there is a distant
and faint tertiary companion ($\rho \approx 11\arcsec$, $\Delta G \approx 8$) that
appears to be physically associated, according to information in the
Gaia\,DR3 catalog. It is therefore possible, in principle, that the inner
eccentricity is being pumped up by the third star \citep[e.g.,][]{Mazeh:1990}.
However, we estimate the modulation timescale for this to be on the order
of $2 \times 10^{10}$~yr \citep[see, e.g.,][]{Rappaport:2013}, and this is
long enough that the effect on the eccentricity of vB\,62 since birth is
expected to have been minimal. 

Thus, there is currently no reason to exclude these two systems from our
estimate of $P_{\rm circ}$, and it is possible that both started out
with an exceptionally large eccentricity, and have not yet had time
to circularize.

\vskip 30pt
\section{The RV Dispersion in the Hyades}
\label{sec:dispersion}

In Section~\ref{sec:membership} we used an estimate from Gaia of the internal
velocity dispersion of the Hyades in the radial direction,
to assist in establishing membership. Here we employ our own RV measurements
to obtain an independent estimate of $\sigma_{\rm cl,RV}$, which we
will show is somewhat smaller.
We began by removing from the mean RV of each star the large
geometric effect coming from the range in viewing perspectives from one side of
the cluster to the other, which can change the velocities by about
$\pm$10~\kms\ relative to a location at the cluster center.
As in Section~\ref{sec:membership}, we relied
for this on the convergent point coordinates and the cluster space
velocity vector from the Gaia mission. We refer to the differences
between the measured mean RVs and these calculated (``astrometric'')
RVs as $\Delta$RV.

\begin{figure}
  \epsscale{1.17}
  \hspace*{-1mm}\includegraphics[width=0.48\textwidth]{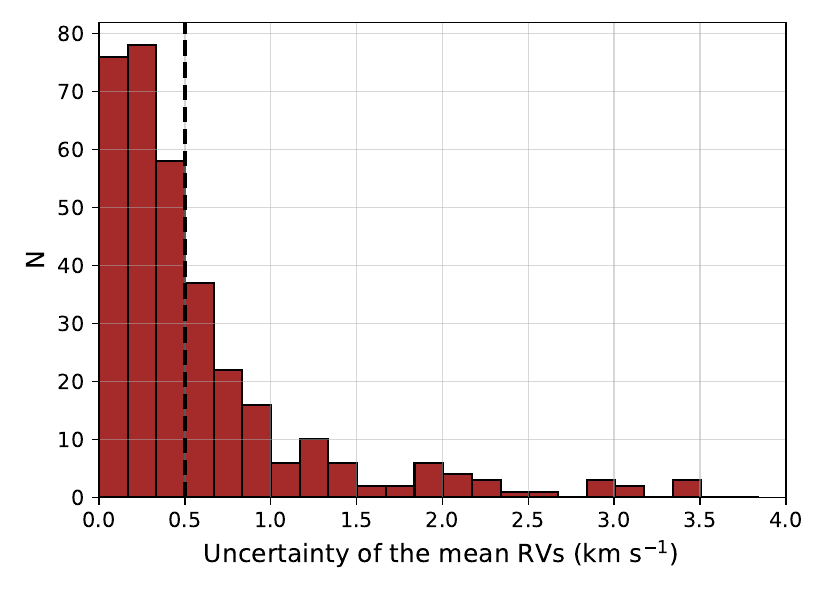}

  \figcaption{Distribution of the uncertainties of the mean
  RVs for the cluster members in our sample with
  $N_{\rm obs} \ge 3$. Our determination of the cluster velocity
  dispersion is based on stars with errors under 0.5~\kms
  (dashed line). \label{fig:meanrverrors}}

\end{figure}

Hotter and rapidly rotating stars, as well as our fainter targets,
tend to have much larger RV uncertainties that would unduly inflate our estimate of
the velocity dispersion. Therefore, we chose to restrict
the sample to stars (or multiple systems) with the most precise mean velocities.
Figure~\ref{fig:meanrverrors} displays the histogram of the uncertainties
of the mean velocities for the 339 member stars with at least three RV
measurements. We adopted a cutoff of 0.5~\kms, marked in the figure
with a dashed line.
In the following we will focus our attention on the main-sequence FGK stars
in our survey ($T_{\rm eff}$ approximately 4000--7000~K), which
dominate the population.

The standard deviation of the 184 $\Delta$RV values for the FGK stars
with the best mean RVs is
$0.79 \pm 0.04~\kms$. However, this overestimates the velocity dispersion
for several reasons. Unrecognized binaries, as well as any unrecognized
non-members that remain in the sample, will tend to increase the dispersion.
Even when their presence is known, visual companions, with their
orbital periods much longer than the duration of our survey, can affect the RV
of a target at a low level, shifting it away from the value expected if
it were a single star. Because the Hyades cluster is so well studied, a sizable
fraction of our targets have such close visual companions listed either
in the WDS or Gaia catalogs (see Table~\ref{tab:RVstats}), and these are
likely to corrupt our estimate of $\sigma_{\rm cl,RV}$. We identified and
removed 41 cases, after which the standard
deviation was reduced to $0.53 \pm 0.03~\kms$.
Furthermore, out of concern that there may be a few non-members among the stars we
classified as possible members, we chose to
remove 8 of those objects with less secure membership credentials.
With this, the dispersion became $0.44 \pm 0.03~\kms$.
Measurement errors in the velocities will also inflate the dispersion.
We corrected for this effect following \cite{McNamara:1986}, and obtained
a further reduced dispersion of $0.34 \pm 0.04~\kms$ from 135 stars.

It is important to note that while the astrometric RVs represent
the true line-of-sight motion of the center of mass of a star, this
is generally not the case for our measured RVs,
As explained in Section~\ref{sec:rvs}, our velocity zero-point was established
using asteroid observations, which represent reflected sunlight. Because of
this, all our measured RVs have the gravitational redshift (GR) and convective
blueshift (CB) of the Sun removed. Velocities we measure for a solar-type star should
accurately be on the IAU system, but those for any other type of star will
not. There will be a small offset equal to the difference between the
star's GR and CB relative to those of the Sun, and
this will lead to an overestimate of the internal dispersion.

\begin{figure}
  \epsscale{1.17}
  \hspace*{-1mm}\includegraphics[width=0.48\textwidth]{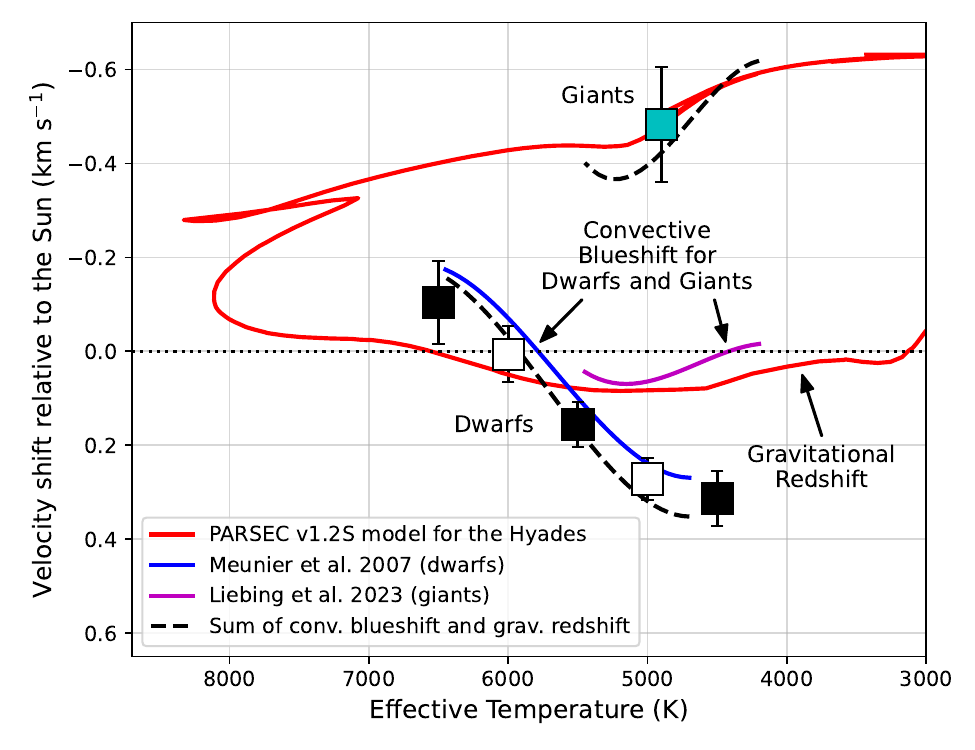}

  \figcaption{Magnitude of the GR and CB affecting the RVs of stars
  in the Hyades, relative to those effects in the Sun. The red
  curve for the GR is based on the same PARSEC~v1.2S model isochrone
  used earlier in this work. The temperature dependence of the CB
  relative to the Sun is from empirical determinations by other
  authors, and is represented by a blue curve for FGK dwarfs
  \citep{Meunier:2017} and a magenta curve for giants \citep{Liebing:2023}.
  The dashed lines correspond to the combination of the two effects,
  separately for dwarfs and giants.
  Filled black squares mark the average $\Delta$RV (measured RVs minus
  astrometric RVs) for stars in three separate 1000~K temperature bins
  (see the text). Open squares are shown for non-independent 1000~K
  bins overlapping with the previous ones. \label{fig:gravred}}

\end{figure}

Figure~\ref{fig:gravred} shows the expected difference
between the GR of a star in the Hyades and the value
for the Sun (0.63~\kms). The curve is based on masses
and radii read from the same
PARSEC~v1.2S isochrone used previously. For FGK stars on the main sequence
($T_{\rm eff} \approx 4000$--7000~K),
the offset can be up to about 0.1~\kms\ in the positive direction, and less
in the negative direction. It
becomes larger in the negative direction as stars approach the turnoff
and proceed to the giant branch. CBs have a stronger
dependence on temperature. We show that relation in the same figure,
in which the curve was derived from Figure~3 of the work by \cite{Meunier:2017},
based on measurements for F7--K4 dwarfs, after subtracting the value for the
Sun ($-0.34~\kms$).
The dashed line in Figure~\ref{fig:gravred} adds the two contributions together.

Examination of our $\Delta$RV values reveals a clear dependence on
temperature, in the same direction as predicted above. Separating our
measurements into three temperature bins containing the majority
of our targets (4000--5000~K, 5000--6000~K,
6000--7000~K), we find average $\Delta$RV values of $+0.313 \pm 0.058~\kms$,
$+0.144 \pm 0.050~\kms$, and $-0.104 \pm 0.088~\kms$, respectively. 
These averages are represented in Figure~\ref{fig:gravred} with filled
squares, and are seen to be fairly consistent with expected trend shown
by the dashed line (GR + CB).
Intermediate temperature bins overlapping by half with the
previous three (4500--5500~K, 5500--6500~K; open squares) suggest that
the average $\Delta$RV varies in a continuous fashion.

The smaller magnitude of the CB expected for K giants,
as determined by \cite{Liebing:2023}, is shown by the magenta line in
Figure~\ref{fig:gravred}. The dashed line at the top of the figure
combines this effect with the GR for giants relative to the Sun, over
the same temperature range.
The average of our measured $\Delta$RV values for the four giants in
the Hyades is indicated with the cyan square, and again matches the
prediction.

We used a simple linear fit to the observed trend in $\Delta$RV to
subtract out the GR and CB effects from the individual values for the
FGK dwarfs in our sample, and recalculated
the internal RV dispersion of the cluster. Our estimate then becomes
$\sigma_{\rm cl,RV} = 0.31 \pm 0.04~\kms$, which is
somewhat smaller than the result from the Gaia mission used in
Section~\ref{sec:membership} (0.40~\kms).

Numerous other empirical estimates of the velocity dispersion of the Hyades
in one dimension have been reported, based either on RVs, on proper
motions, or a combination of both. They have generally assumed that
the dispersion is isotropic, at least in the central part of the
cluster, which most authors consider to be the case.
These determinations have typically ranged from
about 0.3 to 0.7~\kms\ \citep{Schwan:1990, Perryman:1998, Lindegren:2000,
Madsen:2003, Roser:2011, Reino:2018, Leao:2019, Oh:2020, Evans:2022}. Smaller
estimates around 0.2~\kms\ have been obtained through modeling
of the light or mass density distribution, by
adopting a Plummer potential or similar, along with some estimate
for the total mass and for the tidal or core radius of the cluster
\citep[e.g.,][]{Gunn:1988, Perryman:1998, Oh:2020}. In the next section
we will show that our own estimate of the overall cluster dispersion,
$\sigma_{\rm cl,RV}$, may still be overestimated, and that it appears to be significantly
smaller in the inner regions of the Hyades.

In other clusters, the observed internal velocity dispersion has been used to
determine the total mass through the virial theorem, on the assumption
of dynamical equilibrium \citep[e.g.,][]{Geller:2015, Torres:2021}.
If applied to the Hyades, with values for the half-mass radius of
4.1~pc \citep{Roser:2011} or 5.75~pc \citep{Evans:2022}, the result
is $M_{\rm tot} \sim$ 700--1000~$M_{\sun}$, about a factor of 2 larger than
most other estimates. The reason is that the Hyades cluster
is not in virial equilibrium, as has been noted by many authors,
but rather, is in a state of dissolution
and near the end of its life. It has long been known to display mass segregation
\citep[e.g.,][]{Reid:1992, Perryman:1998, Roser:2011, Evans:2022}, and
has already lost a significant fraction of its lighter stars through
evaporation or tidal stripping in the Galactic field.
By some estimates, as much as 60\% of the original mass of the cluster
has already been stripped away \citep[e.g.,][]{Roser:2011, Lodieu:2019,
Oh:2020}. Many
of the escaped stars have ended up in the prominent tidal tails of
the Hyades, which extend for several hundred parsecs on either side in
the direction of Galactic rotation \citep{Roser:2011, Jerabkova:2021}.

\section{Cluster Rotation or Shear -- Impact on $\sigma_{\rm cl,RV}$}
\label{sec:rotation}

In all likelihood, the natal molecular cloud from which the Hyades
emerged had some angular momentum to begin with, and it would not
be surprising to see that property reflected in the motion of stars in the
present-day cluster.
The possibility of rotation in the Hyades has been examined for
at least six decades \citep{Wayman:1965}, although the results have
been mixed. \cite{Gunn:1988} used
proper motions and RVs, but were unable to confirm rotation at
a significant level. Negative results were also obtained by
others \citep{Perryman:1998, Lindegren:2000, Vereshchagin:2013,
Reino:2018}. On the other hand, \cite{Leao:2019} presented what they
interpreted as a significant signature of rotation by examining
the $\Delta$RV differences between the RVs from high-resolution
ground-based spectroscopy and the corresponding astrometric
velocities based on Gaia\,DR2, for a sample of 56 stars. They
reported a statistically significant correlation between $\Delta$RV
and the right ascension of their stars (their Figure~3), with a slope of
$0.0340 \pm 0.0032~\kms$ per degree. Other possible explanations
for this trend, such as an inaccurate space velocity vector,
or cluster expansion, were tested and ruled out.

\begin{figure}
  \epsscale{1.17}
  \hspace*{-1mm}\includegraphics[width=0.48\textwidth]{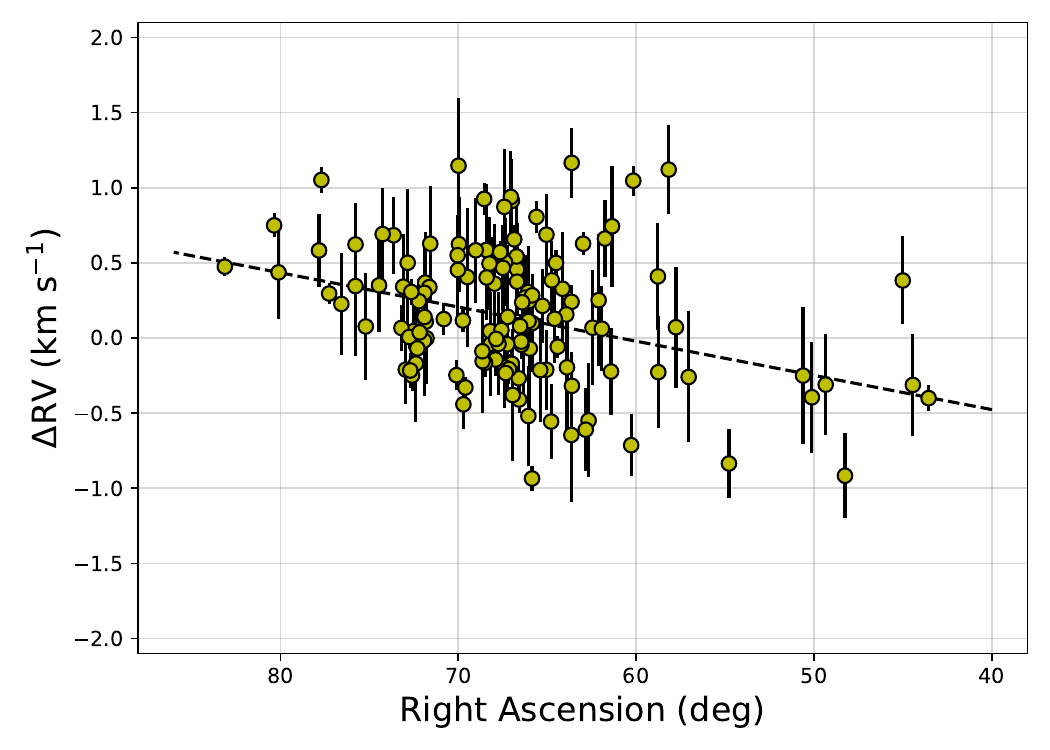}

  \figcaption{$\Delta$RV values (difference between the measured
  spectroscopic RVs and the astrometric RVs, with GR and CB removed)
  shown as a function of right ascension for 135 stars in the Hyades.
  The dashed line is a least-squares fit, giving a slope of $0.0228
  \pm 0.0017~\kms$ per degree. The Pearson correlation coefficient
  is 0.35 ($p~{\rm value} = 3 \times 10^{-5}$). For the interpretation
  of this trend, see the text. \label{fig:rotation}}
\end{figure}

Inspired by this result, we examined our own $\Delta$RV values
from the previous section, which, as explained above,
include an additional correction
compared to those of \cite{Leao:2019} to remove the effects of
GR and CB. We have also gone to greater lengths to exclude triple
systems with unknown spectroscopic orbits, which may bias the RVs.
Figure~\ref{fig:rotation} shows the run of $\Delta$RV versus
right ascension, revealing a similar trend as found by \cite{Leao:2019},
only with a smaller but still statistically significant
slope of $0.0228 \pm 0.0017~\kms$ per degree.

A more sophisticated analysis of the Gaia\,DR2 proper motions and
RVs was carried out by \cite{Oh:2020}. Their model
allowed for rotation, shear, expansion or contraction, as well as
for an anisotropic velocity dispersion. Their results revealed only mild
anisotropy in the velocity field except in the tails, but more
importantly, no significant rotation. Instead, they measured significant
azimuthal and radial shear in the Galactic frame, and indicated that
any rotation of the magnitude claimed by \cite{Leao:2019} would have
been easily detected in their analysis. In view of this,
they expressed the opinion that the \cite{Leao:2019} result is
a consequence of shear rather than rotation.

More recently, \cite{Hao:2024} carried out an independent analysis
based on data from Gaia\,DR3, and reported a statistically
significant measure of rotation of $0.09 \pm 0.03~\kms$ within the
tidal radius of the Hyades, which they assumed to be 9~pc.
\cite{Jadhav:2024}, however, found no rotation, despite also
using Gaia\,DR3 data. Finally, in another Gaia\,DR3 study, 
\cite{Olivares:2025} concluded that the cluster does display
a modest degree of rotation with a rate of $0.0321 \pm 0.0110~\kms$~pc$^{-1}$.

Whether the trend in Figure~\ref{fig:rotation} is caused by
rotation or by shear, it has a direct impact on the estimate of the internal velocity
dispersion $\sigma_{\rm cl,RV}$ that we reported previously. As stars
farther from the cluster center dominate the correlation, we separated the sample
into two groups by linear distance from the center and recomputed the dispersion in
each half. In the inner region within about 5.5~pc (68 stars), where
$\Delta$RV shows no trend with distance, the dispersion corrected
for GR, CB, and observational errors is
$\sigma_{\rm cl,RV} = 0.21 \pm 0.05~\kms$. In the outer region
it is $0.38 \pm 0.05~\kms$, with about two thirds of those 67 stars lying
beyond the tidal radius of the cluster \citep[$\sim$10~pc;][]{Perryman:1998,
Roser:2011}, some possibly in the tails. Interestingly,
the dispersion for the subset of stars closest to the center matches
the smallest estimates inferred by others from steady-state dynamical
modeling using a Plummer model \citep[e.g.,][]{Gunn:1988, Perryman:1998,
Oh:2020}.

\section{Discussion and Conclusions}
\label{sec:conclusions}

This study marks the completion of the spectroscopic survey of the
Hyades region conducted at the CfA over 45+ years. During that time, we
collected $\sim$12,000 spectra for 625 objects covering an area on the sky
of nearly 4000 square degrees. Among the 343 cluster members
and possible members in the sample, there are 90 systems with known
spectroscopic orbits having periods shorter than $10^4$\,d.
About three dozen other systems
with known spectroscopic orbits are non-members, or are members with
even longer periods. Several additional objects in our
survey have astrometric orbits, generally also with longer periods.
The precision and accuracy of our RV measurements is high enough that
the subtle effects of gravitational redshift and convective blueshift are
clearly discernible in both the dwarfs and the giants.

As a result of this work, and that of others, we now have the most
complete inventory of spectroscopic binaries in the cluster.
The frequency of solar-type (FGK) spectroscopic binaries up to periods
of $10^4$\,d is determined to be $40 \pm 5$\%, after a correction for incompleteness.
This is only slightly higher than estimates for other open clusters,
and considerably higher than measured for field
samples (the solar neighborhood, and the halo), which collectively
span two orders of magnitude in age.

Taking into account our sensitivity, we find that the distributions of
both the orbital periods and the eccentricities for systems up to $10^4$\,d are
consistent with those of solar-type binaries in the field.
In particular, the shape of the eccentricity distribution in the Hyades is
quite similar to that in other open clusters of very different ages,
including the Pleiades, M67, and NGC~188
\citep{Torres:2021, Geller:2021}, although for M67 and NGC~188
the distributions are also formally consistent with being uniform.
The eccentricity distribution in the Hyades resembles
the Rayleigh distribution found by \cite{Wu:2025}
from a large sample of main-sequence binaries from Gaia\,DR3, even though
that sample is a mixture of objects of unknown ages, compositions, and
origins. As suggested by \cite{Wu:2025}, this type of distribution is
most likely primordial, and may reflect a universal process. As examples
of this process, they proposed excitation by the circumbinary disks, or 
ejection of brown dwarfs that may have been part of the binary systems
at birth.

The distribution of mass ratios provides useful constraints on theoretical
models of binary star formation. Unlike the periods and eccentricities of binaries,
the mass ratios are largely insensitive to dynamical evolution within a cluster.
However, a survey of the literature reveals rather different results from
one study to the next.
In the Hyades, we find a distribution that is essentially flat, with only a
hint of an upward trend with $q$. This is similar to what was found
in the Pleiades (also flat). On the other hand, the distributions in M67, and
especially NGC\,188, show a rise toward small mass ratios \citep{Geller:2021}.
Similar bottom-heavy shapes for field samples have been found by
\cite{Gullikson:2016}, \cite{Murphy:2018}, and others, with the distributions
peaking at $q \sim 0.2$--0.3.
On the other hand, an independent study of the Hyades by \cite{Albrow:2024},
based on the location of stars above the main-sequence ridge line in the
color-magnitude diagram of the cluster, found a distribution that rises toward
mass ratios of unity.
In the widely cited study by \cite{Raghavan:2010}, solar-type binaries in the field
show a fairly flat mass ratio distribution between 0.2 and 0.95, much like
the Hyades, but then display a significant enhancement for equal-mass (`twin') binaries.
\cite{Moe:2017} found a similar enhancement toward $q = 1$, with slight
variations depending on the orbital period. We do not see such an excess of
twin binaries in the Hyades.
A sharp excess of mass ratios near
unity was also found by \cite{ElBadry:2019} in a large-scale study based on main-sequence
binaries from Gaia\,DR2, although their results focused on much wider binaries than 
those considered here, with separations larger than 50~au
($P \sim 10^5$\,d) and up to 50,000~au.

We determine a tidal circularization period in the cluster of $5.9 \pm 1.1$\,d,
using the methodology of \cite{Meibom:2005}. This is significantly longer than
their estimate of 3.2\,d. Tidal theory has yet to provide a coherent explanation
for the results published previously from various coeval samples of binaries
in clusters and in the field,
which appear to suggest a dependence of $P_{\rm circ}$ on age. Part of the
difficulty may be related to the empirical determination of $P_{\rm circ}$ itself, which
is highly sensitive to the size of the sample and the distribution of orbital periods,
or even the methodology used. The evolution of the eccentricity in a binary depends not
only on the separation, but also on the primary mass as well as the 
nature of the secondary component \citep[see, e.g.,][]{Mazeh:2008}. The
cluster samples referenced above are not completely homogeneous (and the
field samples are even less so), such that there is likely to be some added
scatter in the $P_{\rm circ}$ estimates due to these factors,
beyond their statistical uncertainties.

An additional drawback for the determination of $P_{\rm circ}$ is
the contamination of the binary sample by triple systems, as the tertiaries
can influence the eccentricities of the inner pairs. While we have made an 
effort here to weed out such cases, especially among the shorter-period
binaries that are the most critical, there will inevitably be triples
that have escaped our detection, even in a nearby and well-studied cluster like
the Hyades. It is difficult to quantify how this might impact $P_{\rm circ}$,
but the limitation must be kept in mind. To make things worse, binaries with
short periods almost always seem to have tertiary components.
\cite{Tokovinin:2006} reported that
the fraction of spectroscopic binaries with third components is as high
as 96\% for binaries with periods under 3\,d, and still 34\% for periods longer than 12\,d.
Similar results have been found by \cite{Laos:2020}. This appears to
be a rather fundamental difficulty, as removing all triples from the
sample, even if that were possible, would most likely preclude a determination
of the circularization period.

Open clusters are known to be fragile structures. Unless they are
very massive to begin with, most disintegrate
after a few hundred million years through interactions with the
Galactic environment \citep[e.g.,][]{Friel:1995}.
At its age of about 700~Myr, the Hyades cluster has completed
three revolutions around the Galactic center, and clearly shows
signs of stress. 
It has already lost a substantial fraction of its
members, and is gradually being torn apart in the Galactic field.
It displays leading and trailing tidal tails with a total
extent of almost 1~kpc \citep{Jerabkova:2021}.
The cluster is not in a state of dynamical equilibrium.
In the central regions within about 5.5~pc (roughly the half-mass radius),
we measure a line-of-sight velocity dispersion of $0.21 \pm 0.05~\kms$,
close to what is expected on theoretical grounds
\citep{Gunn:1988, Perryman:1998, Roser:2011}.
Farther out, the dispersion is larger: we obtain $0.38 \pm 0.05~\kms$ for the
outer half of our sample, as the kinematics of the cluster become
increasingly dominated by stars that are slowly drifting away.
Studies of the dynamical evolution
of the Hyades indicate that it is approaching the end of its life.
The most dire predictions \citep[e.g.,][]{Oh:2020} have it completely
dissolving into the Galactic field within the next 30~Myr.

The full-sky astrometric and spectroscopic survey performed by Gaia,
along with extensive and complementary RV monitoring from the ground, have
provided valuable knowledge about the structure, kinematics, and
dynamical evolution of the cluster, and about its composition, including
binarity. Future data releases from the Gaia mission promise to
increase that knowledge even further, adding to the importance of the Hyades
as a valuable benchmark for stellar astrophysics.

\section*{Appendix A: Orbital Solutions for Special Cases}
\label{sec:appendixB}

Here we describe our orbital solutions for more complex systems, including
ones for which we incorporated astrometric measurements to strengthen or
add value to the results from the RVs alone. The velocities for the single-
and double-lined objects discussed below are listed in Tables~\ref{tab:RVsb1}
and \ref{tab:RVsb2} of the main text.

\subsection*{(101) LP 301-69}  
\label{sec:L301-69}

\begin{figure}
  \epsscale{1.17}
  \hspace*{-1mm}\includegraphics[width=0.48\textwidth]{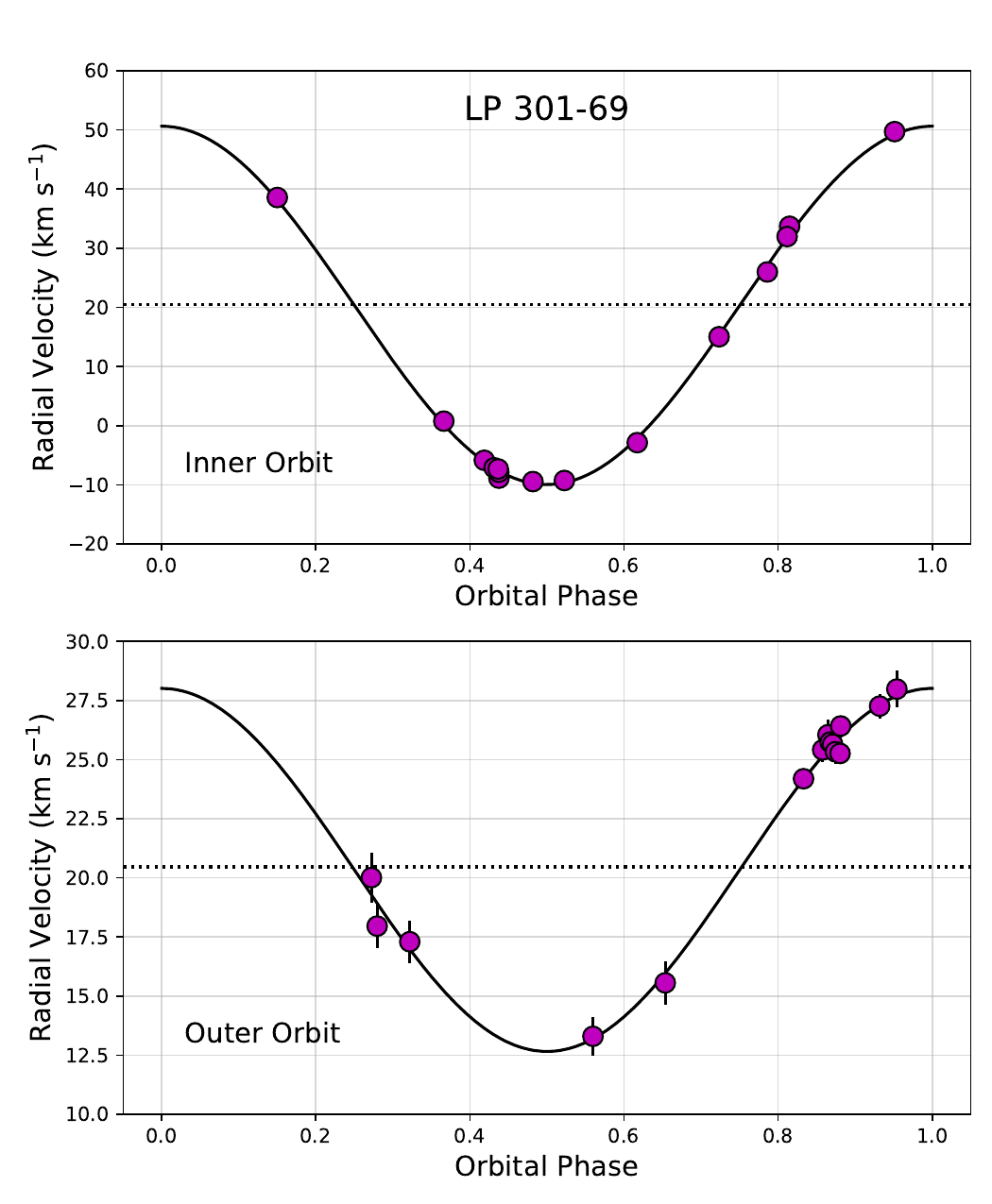}

  \figcaption{RV measurements for LP\,301-69, together with
  our models for the inner and outer orbits. In each case, the
  motion in the other orbit has been subtracted for display
  purposes. The dotted line represents the center-of-mass velocity.
  \label{fig:LP301-69rv}}

\end{figure}

The 15 RV measurements we have of this object, which are listed in
Table~\ref{tab:RVsb1} of the main text, clearly indicate variability
with a period of 5.4\,d. The very large scatter we initially found from
a single-lined orbital solution is due to a third star in the system
with a period of about 13~yr and a near-circular orbit. The inner
orbit appears to be circular as well. A simultaneous triple star solution
was carried out in a Markov chain Monte Carlo (MCMC) framework, 
with both eccentricities set to zero, to reduce the
number of free parameters. Light travel time corrections were applied
to the times of observation in the inner orbit. The results are
presented in Table~\ref{tab:L301-69}.  Figure~\ref{fig:LP301-69rv}
displays our model with the observations. The center-of-mass velocity
of LP\,301-69, as well as its Gaia parallax and proper motion, all rule
out association with the Hyades. The large Gaia RUWE value of 4.309 is
likely a reflection of the perturbation from the wide companion.

\setlength{\tabcolsep}{1pt}
\begin{deluxetable}{l@{\hskip -1ex}c@{\hskip -1ex}c}
\tablewidth{0pc}
\tablecaption{Orbital Parameters for LP 301-69\label{tab:L301-69}}
\tablehead{
\colhead{Parameter} &
\colhead{Value} &
\colhead{Prior}
}
\startdata
 $P_{\rm A}$ (day)                       & $5.418682 \pm 0.000024$            & [5, 6]         \\
 $T_{\rm max,A}$ (BJD)                   & $54154.9764 \pm 0.0095$\phm{2222}  & [54152, 54156] \\
 $K_{\rm Aa}$ (\kms)                     & $30.30 \pm 0.26$\phn               & [10, 50]       \\
 $P_{\rm AB}$ (day)                      & $4810 \pm 50$\phn\phn              & [2000, 6000]   \\
 $T_{\rm max,AB}$ (BJD)                  & $56539 \pm 47$\phn\phn\phn         & [53000, 57500] \\
 $K_{\rm A}$ (\kms)                      & $7.67 \pm 0.44$                    & [1, 20]        \\
 $\gamma$ (\kms)                         & $+20.33 \pm 0.29$\phn\phs          & [0, 50]        \\
 $f_{\rm CfA}$                           & $0.90 \pm 0.27$                    & [$-5$, 5]        \\[0.5ex]
\hline \\ [-1.5ex]
\multicolumn{3}{c}{Derived Properties} \\ [1ex]
\hline \\ [-1.5ex]
 $a_{\rm Aa} \sin i$ ($10^6$ km)         & $2.258 \pm 0.019$                  & \nodata        \\
 $M_{\rm Ab} \sin i /(M_{\rm Aa} + M_{\rm Ab})^{2/3}$ & $0.2499 \pm 0.0022$   & \nodata        \\
 $a_{\rm A} \sin i$ ($10^6$ km)          & $507 \pm 31$\phn                   & \nodata        \\
 $M_{\rm B} \sin i /(M_{\rm A} + M_{\rm B})^{2/3}$ & $0.607 \pm 0.035$        & \nodata        
\enddata

\tablecomments{The values listed correspond to the mode of the
  posterior distributions from our MCMC analysis, with uncertainties
  representing the 68.3\% credible intervals. Times of maximum
  primary velocity are referred to BJD~2,400,000. $f_{\rm CfA}$ 
  represents a multiplicative scale factor applied to the internal 
  RV errors.  Priors in square
  brackets in the last column are uniform over the ranges specified,
  except for the one for $f_{\rm CfA}$, which is log-uniform.}

\end{deluxetable}
\setlength{\tabcolsep}{6pt}

\vskip 30pt
\subsection*{(145) HD 26090}  
\label{sec:+28624}

This is a close visual binary
(WDS~04089+2911, $\rho \sim 0\farcs3$) with a long orbital period and
a fairly high eccentricity, in which the spectroscopic observations
are concentrated mostly near periastron. The companion was discovered
and first measured visually in 1891 \citep{Burnham:1894}. HD\,26090 is
double-lined. Additional RVs were
obtained by \cite{Griffin:2001}, and partially overlap in time with ours.
The phase coverage provided by the RVs is limited, and the orbital period
cannot be determined from the velocities alone. However, the astrometric
measurements made since Burnham's time complement the spectroscopy well,
and enable the period to be better constrained.
Several astrometric orbits have been reported in the literature,
the most recent of which
appears to be the one by \cite{Muller:1978}. It gives a period of
60~yr, and is considered preliminary. Griffin adopted that period, and
held it fixed to derive a preliminary spectroscopic orbital solution
from his RV measurements obtained between 1986 and 2000.

The spectroscopic observations at the CfA began with the Digital
Speedometers in 1983, and continued through 2006. Additional spectra
of HD~26090 were obtained between 2012 and 2024 with the TRES
instrument. The templates adopted for the {\tt TODCOR} analysis
had temperatures of 6000 and 5500~K, and $v \sin i$ values of
4~\kms\ for both components.
The flux ratio we determined from the higher-quality TRES
spectra is $\ell_2/\ell_1 = 0.419 \pm 0.028$, at the mean wavelength
of our observations. The combined velocity coverage of the CfA material
is a little over 40~yr, or about 2/3 of the binary orbit.

As in other cases discussed in this section, a simultaneous
astrometric-spectroscopic orbital solution is the preferred approach
to take advantage of the strengths of each type of observation.
Our joint MCMC solution
incorporates the Griffin RVs in addition to ours, as well as the
astrometric measurements. To account for a possible difference in
the zero-points between our RVs and Griffin's, we solved for an
offset $\Delta$RV$_{\rm Griffin}$ simultaneously with the other
elements, to place his measurements on the same system as ours.
We also solved for separate multiplicative scale factors for the
nominal RV errors of the primary and secondary from each data set.

The pair's relative positions have only been measured at times when the
angular separation was near maximum, while spectroscopy tightly
constrains the orbit at periastron. The early visual measurements are
less reliable, particularly some of the angular separations, and
several of those are best excluded. Other measurements prior to the
1950s have been assigned uncertainties of 5\arcdeg\ in position angle
and 0\farcs02 in separation. More recent ones are better. The results
are presented in Table~\ref{tab:+28624}, where the symbols have their
usual meaning.

\setlength{\tabcolsep}{6pt}
\begin{deluxetable}{lcc}
\tablewidth{0pc}
\tablecaption{Orbital Parameters for HD\,26090 \label{tab:+28624}}
\tablehead{
\colhead{Parameter} &
\colhead{Value} &
\colhead{Prior}
}
\startdata
 $P$ (days)                        & $23379 \pm 390$\phn\phn      & [5000, 40000]   \\
 $T_{\rm peri}$ (BJD$-$2,400,000)  & $49873.2 \pm 9.6$\phm{2222}  & [35000, 55000]  \\
 $a^{\prime\prime}$ (\arcsec)      & $0.2550 \pm 0.0049$          & [0.1, 2.0]      \\
 $\sqrt{e}\cos\omega_1$            & $-0.6163 \pm 0.0069$\phs     & [$-1$, 1]         \\
 $\sqrt{e}\sin\omega_1$            & $-0.6143 \pm 0.0078$\phs     & [$-1$, 1]         \\
 $\cos i$                          & $+0.121 \pm 0.027$\phs       & [$-1$, 1]         \\
 $\Omega$ (deg)                    & $169.2 \pm 1.5$\phn\phn      & [0, 360]        \\
 $K_1$ (\kms)                      & $6.953 \pm 0.058$            & [1, 20]         \\
 $K_2$ (\kms)                      & $7.86 \pm 0.11$              & [1, 20]         \\
 $\gamma$ (\kms)                   & $+38.001 \pm 0.033$\phn\phs  & [20, 60]        \\
 $\Delta$RV$_{\rm Griffin}$ (\kms)     & $-1.102 \pm 0.083$\phs       & [$-5$, 5]         \\[0.5ex]
\hline \\ [-1.5ex]
\multicolumn{3}{c}{Error Scaling Factors} \\ [1ex]
\hline \\ [-1.5ex]
 $f_{\rm CfA,1}$                   & $1.09 \pm 0.10$              & [$-5$, 5]         \\
 $f_{\rm CfA,2}$                   & $1.037 \pm 0.096$            & [$-5$, 5]         \\
 $f_{\rm Griffin,1}$               & $1.08 \pm 0.17$              & [$-5$, 5]         \\
 $f_{\rm Griffin,2}$               & $1.17 \pm 0.19$              & [$-5$, 5]         \\
 $f_{\theta}$                      & $2.60 \pm 0.30$              & [$-5$, 5]         \\
 $f_{\rho}$                        & $2.42 \pm 0.33$              & [$-5$, 5]         \\[0.5ex]
\hline \\ [-1.5ex]
\multicolumn{3}{c}{Derived Properties} \\ [1ex]
\hline \\ [-1.5ex]
 $i$ (deg)                         & $83.1 \pm 1.5$\phn           & \nodata         \\
 $e$                               & $0.7569 \pm 0.0049$          & \nodata         \\
 $\omega_1$ (deg)                  & $224.89 \pm 0.66$\phn\phn    & \nodata         \\
 $a$ (au)                          & $20.97 \pm 0.23$\phn         & \nodata         \\
 $M_1$ ($M_{\sun}$)                & $1.193 \pm 0.032$            & \nodata         \\
 $M_2$ ($M_{\sun}$)                & $1.055 \pm 0.022$            & \nodata         \\
 $M_{\rm tot}$ ($M_{\sun}$)        & $2.249 \pm 0.051$            & \nodata         \\
 $q \equiv M_2/M_1$                & $0.885 \pm 0.015$            & \nodata         \\
 $\pi_{\rm orb}$ (mas)             & $12.16 \pm 0.23$\phn         & \nodata         \\
 Distance (pc)                     & $82.2 \pm 1.6$\phn           & \nodata         
\enddata

\tablecomments{When expressed in Julian years, the period is $64.0 \pm 1.1$~yr,
  and the time of periastron passage is $1995.423 \pm 0.026$.  The
  values listed in the table correspond to the mode of the posterior
  distributions from our MCMC analysis, with uncertainties
  representing the 68.3\% credible intervals. $\Delta_{\rm Griffin}$
  is the offset to be applied to the Griffin RVs to place them on the
  same system as the CfA velocities. The $f$ symbols represent
  multiplicative scaling factors for the nominal errors of the radial
  velocity and astrometric measurements, solved for simultaneously
  with the other elements. Priors in square brackets are uniform over
  the ranges specified, except those for the error inflation factors
  $f$, which are log-uniform. }

\end{deluxetable}
\setlength{\tabcolsep}{6pt}

\begin{figure}
  \epsscale{1.17}
  \hspace*{-1mm}\includegraphics[width=0.48\textwidth]{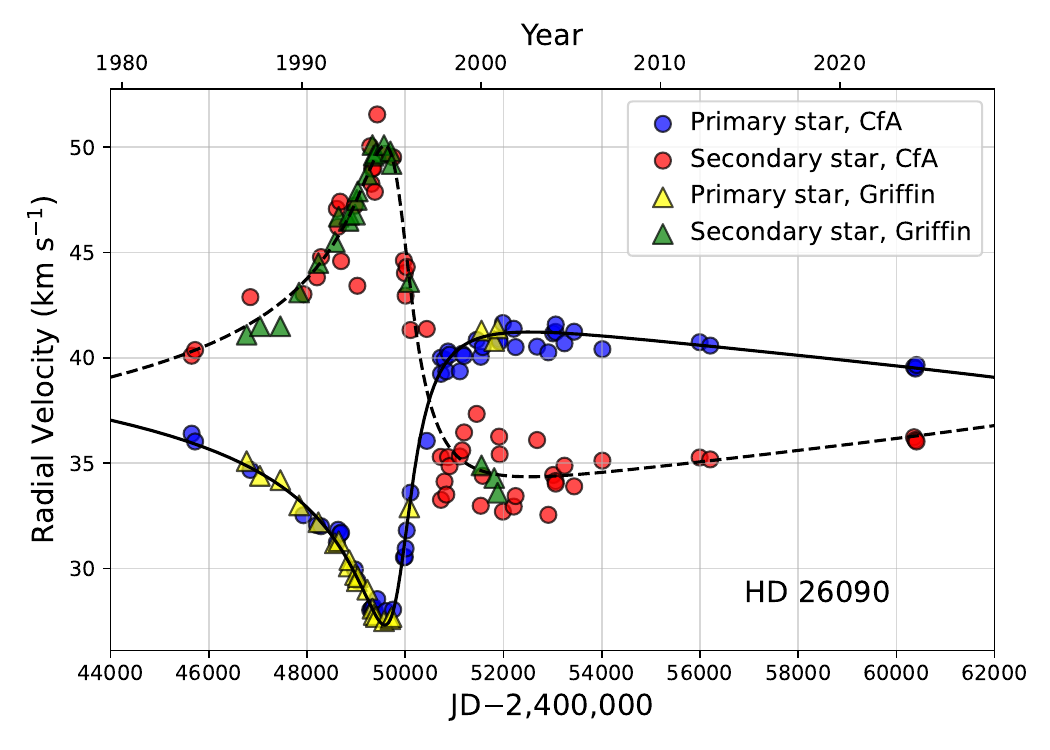}

  \figcaption{Radial velocity measurements of HD~\,6090 from this paper
    and from \cite{Griffin:2001}, along with our model for the
    spectroscopic orbit.  \label{fig:+28624rvs}}

\end{figure}

\begin{figure}
  \epsscale{1.17}
  \hspace*{-1mm}\includegraphics[width=0.48\textwidth]{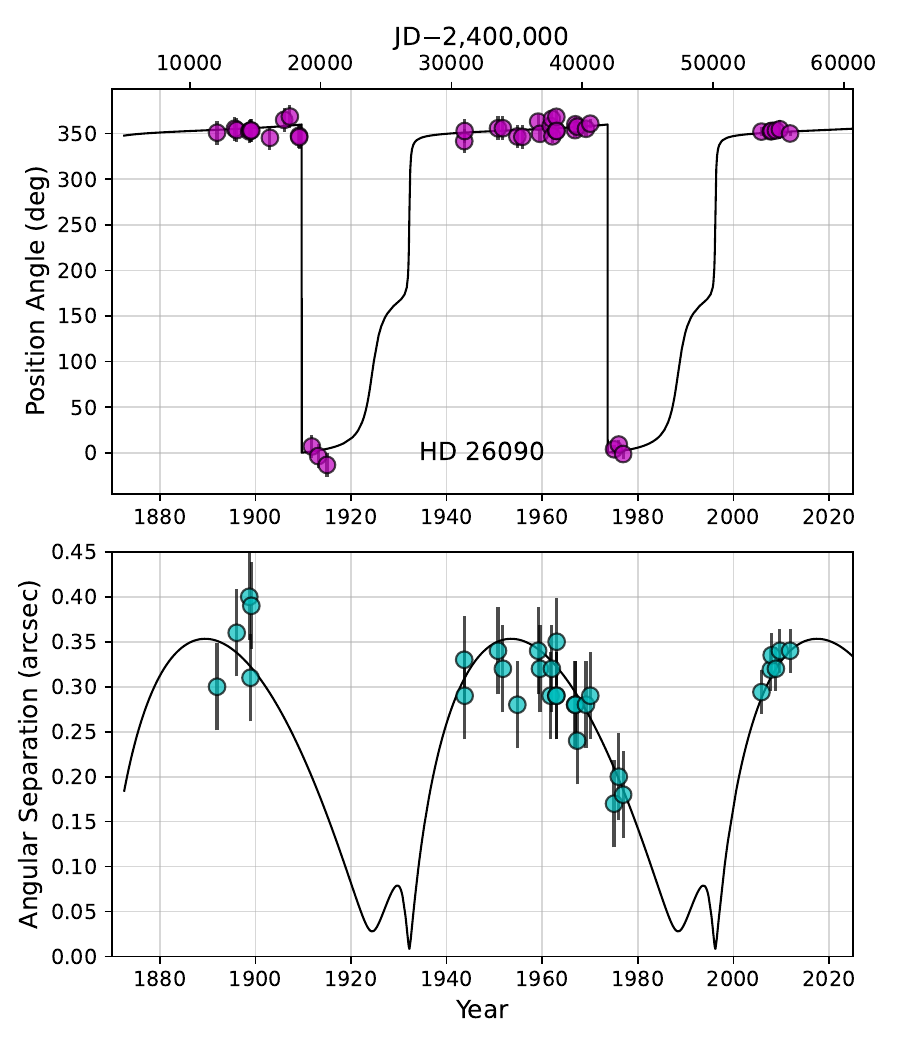}

  \figcaption{Measured position angles and angular separations for HD\,26090
    from the WDS, together with our orbit model. \label{fig:+28624pasep}}

\end{figure}

The RV measurements from the CfA and from Griffin are displayed in
Figure~\ref{fig:+28624rvs}, showing virtually complete coverage of the
1995 periastron passage. The astrometric measurements (position angles
and separations) are shown as a function of time in
Figure~\ref{fig:+28624pasep}, and are seen to cover two orbital
cycles.

In addition to the dynamical masses we obtained, which are formally good to better
than 3\%, we inferred an orbital parallax of $\pi_{\rm orb} = 12.16 \pm
0.23$~mas, implying a distance of more than 80~pc. The original Hipparcos
catalog \citep{ESA:1997} listed a trigonometric parallax of $10.68 \pm
1.43$~mas, which was later revised to $12.84 \pm 1.07$~mas
\citep{vanLeeuwen:2007}.  Interestingly, the Gaia~DR2 catalog
\citep{Gaia:2018} reported a much larger value of $24.03 \pm
0.50$~mas, although the most recent release \citep[DR3;][]{Gaia:2023a}
has it back down again at $13.95 \pm 0.11$~mas. None of these values
except our own took into account the binarity of the object. The large
implied distance, the 15\arcdeg\ projected separation from the center
of the Hyades, and the fact that the center-of-mass velocity of
+38.0~\kms\ deviates by almost 2.5~\kms\ from what is expected for a
star at this position in the cluster, led \cite{Griffin:2001} to argue
that HD\,26090 is not a member of the Hyades. We agree with that
assessment.

\subsection*{(198) vB 22}  
\label{sec:VB22}

This is a well-known member of the Hyades, also referred to as HD\,27130.  It
is a hierarchical triple system in which the inner, 5.6\,d binary was
found by \cite{McClure:1982} to be eclipsing, the brightest such
system in the cluster. This makes it special because it enables
precise model-independent masses and radii to be determined for the
components. This can provide useful constraints on models of stellar
evolution, particularly in a cluster like the Hyades with a known age
and chemical composition \citep[see, e.g.,][]{Lebreton:2001,
  Pinsonneault:2003, Torres:2019b, Torres:2024a}.

The long history of RV observations of vB\,22, spanning more than a
century, has been described in detail by \cite{Griffin:1985a} and by
\cite{Griffin:2012}. Double-lined spectroscopic orbits for vB\,22 were
published in both of those studies, and also previously by
\cite{McClure:1982} and \cite{Mayor:1987}, as well as more recently by
\cite{Brogaard:2021}.  Studies of the light curve have been carried
out by \citep{McClure:1982} and \cite{Schiller:1987}, with the latest
being the one by \cite{Brogaard:2021}.

The first evidence of a possible drift in the center-of-mass velocity
of the inner binary was reported by \cite{Griffin:1985a}, and although
they did not go so far as to claim the presence of a third star,
surely they would have had that in mind, and recommended further
monitoring of the system. Our own spectroscopic observations of vB\,22
at the CfA started in October of 1992, and reveal the lines of both
stars. 
By the next season, it was clear from the run of the residuals
from a double-lined orbit that the center-of-mass velocity was
drifting upward, and over the following two seasons the change
reversed course, leaving little doubt as to the cause. By 2002 the
residuals had returned to their maximum value from some 8~yr earlier.
An orbit for the tertiary was eventually published by
\cite{Griffin:2012}, with a period of 3079\,d (8.4~yr).  The third star
has remained undetected spectroscopically, and is believed to be
a faint M dwarf.

We continued to monitor vB\,22 until 2013, and we now present an
updated simultaneous solution for the inner and outer spectroscopic orbits that
incorporates the RV measurements from \cite{McClure:1982},
\cite{Mayor:1987}, \cite{Mermilliod:2009}, \cite{Griffin:2012}, and
\cite{Brogaard:2021}, in addition to our own.
Our velocities from {\tt TODCOR} used templates
with temperatures of 5000 and 4500~K for the primary and secondary,
and $v \sin i$ values of 8 and 6~\kms, respectively.
The secondary-to-primary flux ratio we measured
at the mean wavelength of our observations is
$\ell_{\rm Ab}/\ell_{\rm Aa} = 0.072 \pm 0.012$.
We assumed the
inner and outer orbits are well represented by unperturbed Keplerian
models, and light travel time corrections have been
accounted for.  The elements and auxiliary quantities are listed in
Table~\ref{tab:VB22mcmc1}, from a solution that also includes
astrometry, as described below. Because the eccentricity of the inner
orbit is very small, it was more convenient to solve for a time of
nodal passage, $T_{\rm node, A}$, which is better defined than a time
of periastron passage in such cases, although we also list the latter
among the derived quantities.  As done for the previous system,
we solved for separate multiplicative scale factors for the
nominal RV errors of the primary and secondary from each data set, and
for velocity zero-point offsets for each source relative to our own.
Some of the external sources contribute little to the outer orbit
because they do not cover a sufficient interval in phase, and allowing
their zero-point to vary freely would negate any benefit they could
otherwise provide to that orbit.  Nevertheless, we included them because they do
help to improve the precision of the inner orbit, and therefore of the
absolute masses we report below for the eclipsing components.

\setlength{\tabcolsep}{1pt}
\begin{deluxetable}{l@{\hskip -1ex}cc}
\tablewidth{0pc}
\tablecaption{Spectroscopic Orbital Parameters for vB\,22 \label{tab:VB22mcmc1}}
\tablehead{
\colhead{Parameter} &
\colhead{Value} &
\colhead{Prior}
}
\startdata
\multicolumn{3}{c}{Inner Orbit} \\ [1ex]
\hline \\ [-1.5ex]
 $P_{\rm A}$ (day)                         & $5.60921410 \pm 0.00000058$    & [5.5, 5.7]      \\
 $T_{\rm node, A}$ (BJD)                   & $52190.13836 \pm 0.00049$\phm{2222}  & [52188, 52192]  \\
 $\sqrt{e_{\rm A}}\cos\omega_{\rm Aa}$     & $+0.017 \pm 0.011$\phs         & [$-$1, 1]       \\
 $\sqrt{e_{\rm A}}\sin\omega_{\rm Aa}$     & $+0.0229 \pm 0.0097$\phs       & [$-$1, 1]       \\
 $K_{\rm Aa}$ (\kms)                       & $60.798 \pm 0.030$\phn         & [50, 100]       \\
 $K_{\rm Ab}$ (\kms)                       & $83.790 \pm 0.051$\phn         & [50, 100]       \\
 $\gamma$ (\kms)                           & $+38.475 \pm 0.036$\phn\phs    & [10, 50]        \\
 $\Delta$RV$_{\rm McClure}$ (\kms)             & $-0.10 \pm 0.24$\phs           & [$-$5, 5]       \\
 $\Delta$RV$_{\rm Griffin}$ (\kms)             & $-0.760 \pm 0.056$\phs         & [$-$5, 5]       \\
 $\Delta$RV$_{\rm Mayor}$ (\kms)               & $-0.08 \pm 0.19$\phs           & [$-$5, 5]       \\
 $\Delta$RV$_{\rm Brogaard}$ (\kms)            & $+0.038 \pm 0.092$\phs         & [$-$5, 5]       \\[0.5ex]
\hline \\ [-1.5ex]
\multicolumn{3}{c}{Outer Orbit} \\ [1ex]
\hline \\ [-1.5ex]
 $P_{\rm AB}$ (day)                        & $3077.8 \pm 8.6$\phn\phn\phn   & [1000, 5000]    \\
 $T_{\rm peri, AB}$ (BJD)                  & $49570 \pm 20$\phm{222}        & [48000, 51000]  \\
 $\sqrt{e_{\rm AB}}\cos\omega_{\rm B}$     & $+0.540 \pm 0.019$\phs         & [$-$1, 1]       \\
 $\sqrt{e_{\rm AB}}\sin\omega_{\rm B}$     & $+0.248 \pm 0.027$\phs         & [$-$1, 1]       \\
 $K_{\rm A}$ (\kms)                        & $2.306 \pm 0.056$              & [1, 5]          \\[0.5ex]
\hline \\ [-1.5ex]
\multicolumn{3}{c}{Error Scaling Factors} \\ [1ex]
\hline \\ [-1.5ex]
 $f_{\rm Aa, DS}$                          & $1.017 \pm 0.059$              & [$-$5, 5]       \\
 $f_{\rm Ab, DS}$                          & $0.996 \pm 0.055$              & [$-$5, 5]       \\
 $f_{\rm Aa, TRES}$                        & $0.96 \pm 0.31$                & [$-$5, 5]       \\
 $f_{\rm Ab, TRES}$                        & $0.40 \pm 0.10$                & [$-$5, 5]       \\
 $f_{\rm Aa, McClure}$                     & $0.93 \pm 0.15$                & [$-$5, 5]       \\
 $f_{\rm Ab, McClure}$                     & $1.23 \pm 0.24$                & [$-$5, 5]       \\
 $f_{\rm Aa, Griffin}$                     & $1.080 \pm 0.077$              & [$-$5, 5]       \\
 $f_{\rm Ab, Griffin}$                     & $1.12 \pm 0.19$                & [$-$5, 5]       \\
 $f_{\rm Aa, Mayor}$                       & $1.59 \pm 0.31$                & [$-$5, 5]       \\
 $f_{\rm Ab, Mayor}$                       & $1.33 \pm 0.48$                & [$-$5, 5]       \\
 $f_{\rm Aa, Brogaard}$                    & $1.13 \pm 0.30$                & [$-$5, 5]       \\
 $f_{\rm Ab, Brogaard}$                    & $1.20 \pm 0.28$                & [$-$5, 5]       \\[0.5ex]
\hline \\ [-1.5ex]
\multicolumn{3}{c}{Derived Quantities} \\ [1ex]
\hline \\ [-1.5ex]
 $e_{\rm A}$                               & $0.00094 \pm 0.00041$          & \nodata         \\
 $\omega_{\rm Aa}$ (deg)                   & $55 \pm 27$                    & \nodata         \\
 $e_{\rm AB}$                              & $0.354 \pm 0.017$              & \nodata         \\
 $\omega_{\rm A}$ (deg)                    & $24.7 \pm 2.9$\phn             & \nodata         \\
 $T_{\rm peri, A}$ (BJD)                   & $52191.00 \pm 0.42$\phm{2222}  & \nodata         \\
 $M_{\rm Aa}\sin^3 i$ ($M_{\sun}$)         & $1.0181 \pm 0.0014$            & \nodata         \\
 $M_{\rm Ab}\sin^3 i$ ($M_{\sun}$)         & $0.73874 \pm 0.00085$          & \nodata         \\
 $q_{\rm A} \equiv M_{\rm Ab}/M_{\rm Aa}$  & $0.72560 \pm 0.00058$          & \nodata         \\
 $a_{\rm Aa} \sin i$ ($10^6$ km)           & $4.6895 \pm 0.0023$            & \nodata         \\
 $a_{\rm Ab} \sin i$ ($10^6$ km)           & $6.4629 \pm 0.0040$            & \nodata         \\
 $a_{\rm A} \sin i$ ($R_{\sun}$)           & $16.0314 \pm 0.0033$\phn       & \nodata         \\
 $M_{\rm B} \sin i /(M_{\rm A} + M_{\rm B})^{2/3}$ & $0.1474 \pm 0.0030$    & \nodata         \\
 $a_{\rm A} \sin i_{\rm AB}$ ($10^6$ km)   & $91.3 \pm 1.9$\phn             & \nodata         
\enddata

\tablecomments{$T_{\rm node,A}$, $T_{\rm peri,AB}$, and $T_{\rm peri,A}$
are referred to BJD 2,400,000. The values listed correspond to the mode of the
  posterior distributions from our MCMC analysis, with uncertainties
  representing the 68.3\% credible intervals. Priors in square
  brackets in the last column are uniform over the ranges specified,
  except the error inflation factors $f$, which are log-uniform.
  Velocity offsets $\Delta$ are in the sense ``CfA minus other''.}

\end{deluxetable}
\setlength{\tabcolsep}{6pt}

\begin{figure*}
  \epsscale{1.17}
  \hspace*{-1mm}\includegraphics[width=\textwidth]{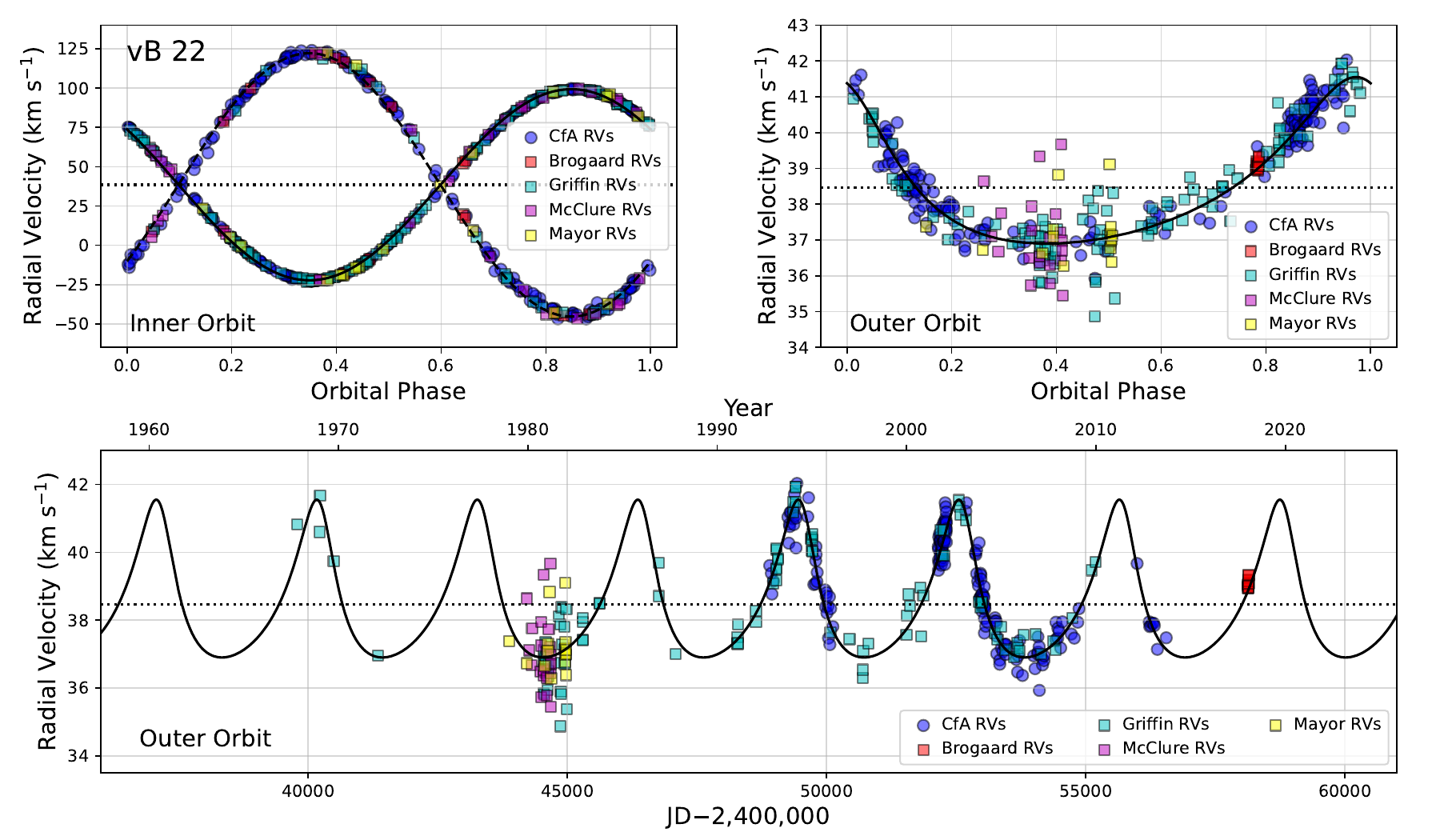}

  \figcaption{RV measurements for vB\,22 together with our models for
    the inner and outer orbits (phase-folded plots in the top panels;
    see the text). Each panel has the motion from the other orbit
    subtracted. The bottom panel shows the outer orbit as a function
    of time. The center-of-mass velocity of the triple system is
    indicated with the dotted lines. \label{fig:VB22}}

\end{figure*}

The results from our orbital solution are similar to those reported by
others, only more precise. We note that the Gaia mission also detected
vB\,22 as a complex system, and reported both an SB2 solution as well
as astrometric acceleration due to the third star (star `B'). However, while the
RV semiamplitude $K_{\rm Aa}$ of the more massive component from Gaia
is similar to ours, their $K_{\rm Ab}$ is 8~\kms\ larger than what we and others
derive, and must be biased. Figure~\ref{fig:VB22} shows our models for
the inner and outer spectroscopic orbits, along with the observations.

It is also possible to constrain the motion of the tertiary of vB\,22
on the plane of the sky, using available astrometric information.  For
this, we combined the RVs with measurements from the Hipparcos mission
taken over a period of 2.5~yr, and with the highly precise parallax,
position, and proper motion from Gaia~DR3. The use of the Hipparcos
intermediate data (abscissa residuals) introduces five adjustable
parameters in the form of corrections $\Delta\alpha^*$ and
$\Delta\delta$ to the catalog values of the position of the barycenter
at the mean catalog epoch of 1991.25, corrections
$\Delta\mu_{\alpha}^*$ and $\Delta\mu_{\delta}$ to the barycenter
proper motion components\footnote{We follow here the practice in the
  Hipparcos catalog in defining $\Delta\alpha^* \equiv \Delta\alpha
  \cos\delta$ and $\Delta\mu_{\alpha}^* = \Delta\mu_{\alpha}
  \cos\delta$.}, and a correction to the catalog parallax,
$\Delta\pi$.  To reduce the number of free parameters, we constrained
$\Delta\pi$ using the improved parallax reported as part of the
astrometric acceleration solution from Gaia, referenced above
($\pi_{\rm Gaia} = 21.458$~mas), adjusted for a zero-point offset of
+0.035~mas, as recommended by \cite{Lindegren:2021}. Because Hipparcos
did not spatially resolve the tertiary (nor did Gaia), the
measurements track the motion of the center of light, so an additional
parameter $a^{\prime\prime}_{\rm phot}$ must be solved for.  This
represents the semimajor axis of the photocenter orbit, which is a
scaled-down version of the relative orbit, reflected about the
origin. Checks based on a model isochrone for the Hyades show that,
despite the difference in wavelengths between the two space missions,
the semimajor axis of the photocenter is not expected to be
significantly different for Hipparcos and Gaia, so we have assumed
here that they are the same. The final two parameters of the
astrometric orbit are the inclination angle, expressed as $\cos i_{\rm
  AB}$, and the position angle of the ascending node, $\Omega$.

The formalism for incorporating Hipparcos abscissa residuals has been
described in the Hipparcos documentation \citep[see also,
  e.g.,][]{Torres:2007}.  The predicted position at the Gaia epoch
($\alpha_{\rm G}^*$, $\delta_{\rm G}$) was computed as follows, from
the other parameters of the model:
\begin{equation}
  \label{eq:positions}
  \begin{aligned}
  \alpha_{\rm G}^* &= \alpha_{\rm H}^* + \Delta\alpha^* +
  (\mu_{\alpha, {\rm H}}^* + \Delta\mu_{\alpha}^*)(t_{\rm G} - t_{\rm
    H}) + \Delta\alpha_{\rm G}^* \\
  \delta_{\rm G} &= \delta_{\rm H} + \Delta\delta + (\mu_{\delta, {\rm
      H}} + \Delta\mu_{\delta})(t_{\rm G} - t_{\rm H}) +
  \Delta\delta_{\rm G}~.
  \end{aligned}
\end{equation}
Here $\alpha_{\rm H}^*$ and $\delta_{\rm H}$ are the published
Hipparcos positions at the catalog epoch, $\mu_{\alpha, {\rm H}}^*$
and $\mu_{\delta, {\rm H}}$ are the published proper motion
components, and $t_{\rm G}$ and $t_{\rm H}$ are the mean Hipparcos and
Gaia catalog epochs taken from \cite{Brandt:2021}. The quantities
$\Delta\alpha_{\rm G}^*$ and $\Delta\delta_{\rm G}$ represent the
contributions to the position of the photocenter from orbital motion
at the Gaia epoch.  They are expressed in terms of the classical
Thiele-Innes constants as follows,
\begin{equation}
  \label{eq:thieleinnes}
  \begin{aligned}
  \Delta\alpha^* & = a^{\prime\prime}_{\rm phot} (Bx+Gy) \\
  \Delta\delta & = a^{\prime\prime}_{\rm phot} (Ax+Fy)~,
  \end{aligned}
\end{equation}
in which $x = \cos E - e_{\rm AB}$ and $y = \sqrt{1-e_{\rm AB}^2}\sin
E$, with $E$ being the eccentric anomaly. Because Gaia only observed
vB\,22 for a fraction of the 8.4~yr period of the tertiary, and the
path on the sky is curved, computing the orbital position at the mean
Gaia catalog epoch is not the same as taking the average of the
positions over each individual epoch \citep[see an illustration of
  this in Figure~3 of][]{Brandt:2024}. We therefore calculated
$\Delta\alpha_{\rm G}^*$ and $\Delta\delta_{\rm G}$ at each individual
Gaia epoch, as predicted by the Gaia Observation Scheduling Tool
\citep[GOST;][]{Fernandez:2019}, and then took the average.

The predicted proper motion components at the Gaia epoch,
$\mu_{\alpha,{\rm G}}^*$, $\mu_{\delta,{\rm G}}$, were calculated as
\begin{equation}
  \label{eq:propermotions}
  \begin{aligned}
  \mu_{\alpha,{\rm G}}^* &= \mu_{\alpha,{\rm H}}^* +
  \Delta\mu_{\alpha}^* + \Delta\mu_{\alpha,{\rm G}}^* \\
  \mu_{\delta,{\rm G}} &= \mu_{\delta,{\rm H}} +
  \Delta\mu_{\delta}^* + \Delta\mu_{\delta,{\rm G}}~,
  \end{aligned}
\end{equation}
where the last terms on the right are the derivatives of the last
terms on the right-hand side of eqs.[\ref{eq:positions}].

We used the Hipparcos abscissa residuals from the original catalog
\citep{ESA:1997} as well as those from the revised version
\citep{vanLeeuwen:2007}, which are different.  \cite{Brandt:2018,
  Brandt:2021} has shown that a combination of the original and
revised catalogs with relative weights of 0.4 and 0.6, respectively,
leads to an improvement in the positions and proper motions over
either version.  Here we have chosen to apply the same recipe to
independent astrometric-spectroscopic solutions using the original and
revised abscissae, which give very consistent results. The astrometric
parameters from this procedure are listed in
Table~\ref{tab:VB22mcmc2}, and the spectroscopic ones were given
earlier.

\setlength{\tabcolsep}{8pt}
\begin{deluxetable}{lcc}
\tablewidth{0pc}
\tablecaption{Astrometric Parameters for vB\,22, and Derived Properties \label{tab:VB22mcmc2}}
\tablehead{
\colhead{Parameter} &
\colhead{Value} &
\colhead{Prior}
}
\startdata
 $a^{\prime\prime}_{\rm phot}$ (mas)     &  $14.95 \pm 0.47$\phn            & [0, 50]         \\
 $\cos i_{\rm AB}$                       &  $0.424 \pm 0.017$               & [$-1$, 1]       \\
 $\Omega$ (deg)                          &  $70.7 \pm  3.9$\phn             & [0, 360]        \\
 $\Delta\alpha^*$ (mas)                  &  $+14.7 \pm 1.1$\phn\phs         & [$-100$, 100]   \\
 $\Delta\delta$ (mas)                    &  $-1.6 \pm 1.1$\phs              & [$-100$, 100]   \\
 $\Delta\mu_{\alpha}^*$ (mas yr$^{-1}$)  &  $+1.824 \pm 0.035$\phs          & [$-100$, 100]   \\
 $\Delta\mu_{\delta}^*$ (mas yr$^{-1}$)  &  $-5.307 \pm 0.026$\phs          & [$-100$, 100]   \\[0.5ex]
\hline \\ [-1.5ex]
\multicolumn{3}{c}{Derived Properties} \\ [1ex]
\hline \\ [-1.5ex]
 $i_{\rm AB}$ (deg)                      &  $64.9 \pm 1.1$\phn              & \nodata         \\
 $a^{\prime\prime}$ (mas)                &  $112.76 \pm 0.24$\phn\phn       & \nodata         \\
 $M_{\rm B}$ ($M_{\sun}$)                &  $0.2611 \pm 0.0056$             & \nodata         \\
 $\mu_{\alpha,{\rm bary}}^*$ (mas yr$^{-1}$) & $+114.894 \pm 0.035$\phn\phn\phs & \nodata     \\
 $\mu_{\delta,{\rm bary}}^*$ (mas yr$^{-1}$) & $-26.697 \pm 0.026$\phn\phs  & \nodata         \\
 $M_{\rm Aa}$ ($M_{\sun}$)               &  $1.0269 \pm 0.0014$             & \nodata         \\
 $M_{\rm Ab}$ ($M_{\sun}$)               &  $0.74515 \pm 0.00086$              & \nodata         \\
 $R_{\rm Aa}$ ($R_{\sun}$)               &  $0.927 \pm 0.012$               & \nodata         \\
 $R_{\rm Ab}$ ($R_{\sun}$)               &  $0.737 \pm 0.010$               & \nodata         \\
 $\log g_{\rm Aa}$ (cgs)                 &  $4.516 \pm 0.011$               & \nodata         \\
 $\log g_{\rm Ab}$ (cgs)                 &  $4.575 \pm 0.012$               & \nodata         \\
 $\beta$                                 &  $0.0034 \pm 0.0042$             & \nodata         \\
 $\Delta$Hp (mag)                        &  $6.2 \pm 1.3$                   & \nodata         
\enddata

\tablecomments{The values listed correspond to the mode of the
  posterior distributions from our MCMC analysis, with uncertainties
  representing the 68.3\% credible intervals. Priors in square
  brackets in the last column are uniform over the ranges specified.
  Distributions for the Gaia~DR3 parallax and for the photometric
  parameters taken from Table~5 of \citep{Brogaard:2021} were assumed
  to be Gaussian.}

\end{deluxetable}
\setlength{\tabcolsep}{6pt}

The most recent analysis of the light curve of vB\,22 by
\cite{Brogaard:2021} provided accurate measurements of the photometric
parameters of the system. The higher precision of the spectroscopic
minimum masses and projected semimajor axis of the eclipsing pair in
Table~\ref{tab:VB22mcmc1} now allows us to improve on the absolute
masses and radii for the components. From Table~5 of
\cite{Brogaard:2021}, we adopted weighted mean values for the
inclination angle of the inner orbit, the radius ratio, and the sum of
the relative radii of $i = 85\fdg657 \pm 0\fdg045$, $r_2/r_1 = 0.796 \pm 0.013$, and
$r_1 + r_2 = 0.1035 \pm 0.0011$. Revised masses, radii, and surface
gravity determinations are included in the bottom section of
Table~\ref{tab:VB22mcmc2}, along with the inferred inclination angle
and semimajor axis of the relative orbit of the tertiary,
$a^{\prime\prime}$. With this, we were able to infer a rather precise
value for the mass of the tertiary as well, formally good to about
2\%.  This adds to the interest of the system, as all three components
now have well determined masses. Those for components Aa and Ab are
known to better than 0.13\%.
Even though the third star is not visible in the spectra, it is possible
to infer its brightness in the Hipparcos band, relative to the combined
light of the inner binary, $\Delta$Hp, via the classical relation
\begin{equation}
  a^{\prime\prime}_{\rm phot} = a^{\prime\prime} \left(\frac{M_{\rm B}}{M_{\rm A}+M_{\rm B}} - \beta\right)~,
\end{equation}
where $\beta = (1+10^{0.4\Delta{\rm Hp}})^{-1}$ is the fractional
brightness of the third star. In practice, however, $\beta$ is poorly
determined. The magnitude difference is $\Delta{\rm Hp} = 6.2 \pm
1.3$~mag.

Confirmation or improvement of the astrometric orbit of the tertiary
and its brightness could come from the direct detection of the star by
ground-based near-infrared imaging techniques. The angular separation
and position angle can now be estimated to be
[0\farcs050, 155\arcdeg] in mid 2026 and [0\farcs068,
  229\arcdeg] in mid 2027.

\subsection*{(246) vB\,259}  
\label{sec:L33}

This cluster member was announced as a spectroscopic binary by
\cite{Griffin:1988}, who estimated the period to be about 3~yr. It was
only seen as single-lined at the time, but subsequent observations
allowed both components to be measured.  A double-lined orbit was
published by \cite{Griffin:2012}, based on about 20 RV measurements of
the two components made between 1990 and 2010.

The observations at the CfA began in 1979 and continued until 2003.
All 44 spectra taken with the Digital Speedometers are double-lined.
The velocities for both stars were measured with {\tt TODCOR},
using best-fitting
templates with temperatures of 4750~K and 4500~K for the primary and
secondary, and no rotational broadening.
We measured a flux ratio of
$\ell_2/\ell_1 = 0.528 \pm 0.045$ at the mean wavelength of our observations.

The companion of vB\,259 was detected in $K$-band speckle observations
by \cite{Patience:1998} at a separation of just 43~mas, and a
brightness difference of 0.14~mag at that wavelength. A handful of
speckle measurements by others since have now covered most of the
orbit, constraining the astrometric elements fairly well. A solution
of the speckle orbit was published recently by \cite{Tokovinin:2024}.

\begin{figure}
  \epsscale{1.17}
  \hspace*{-1mm}\includegraphics[width=0.48\textwidth]{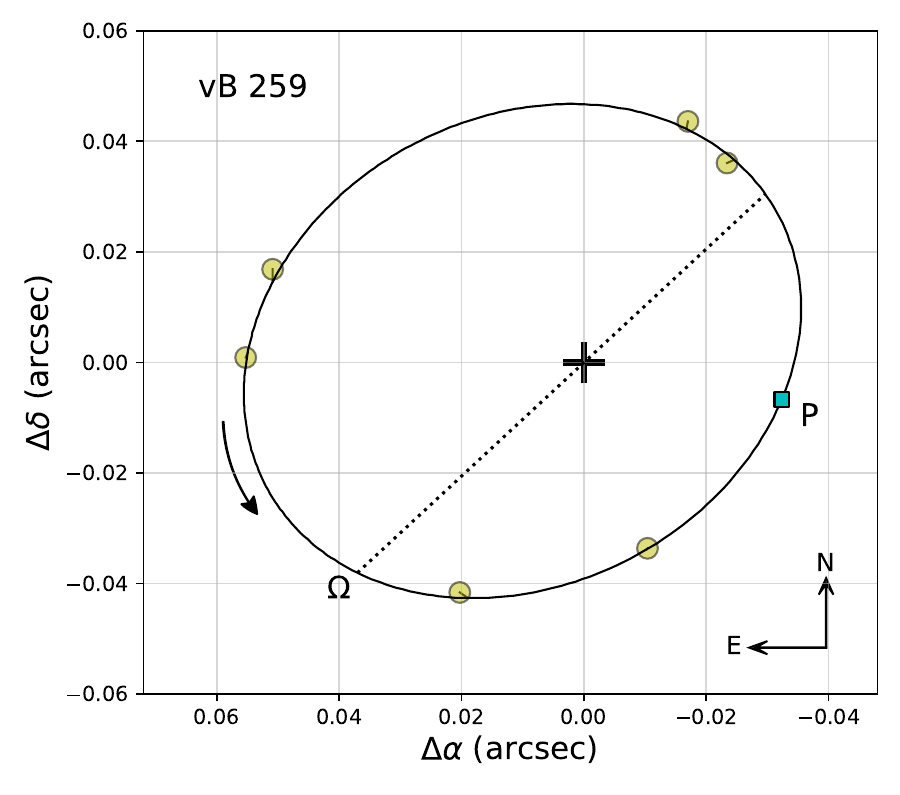}
  \hspace*{-1mm}\includegraphics[width=0.48\textwidth]{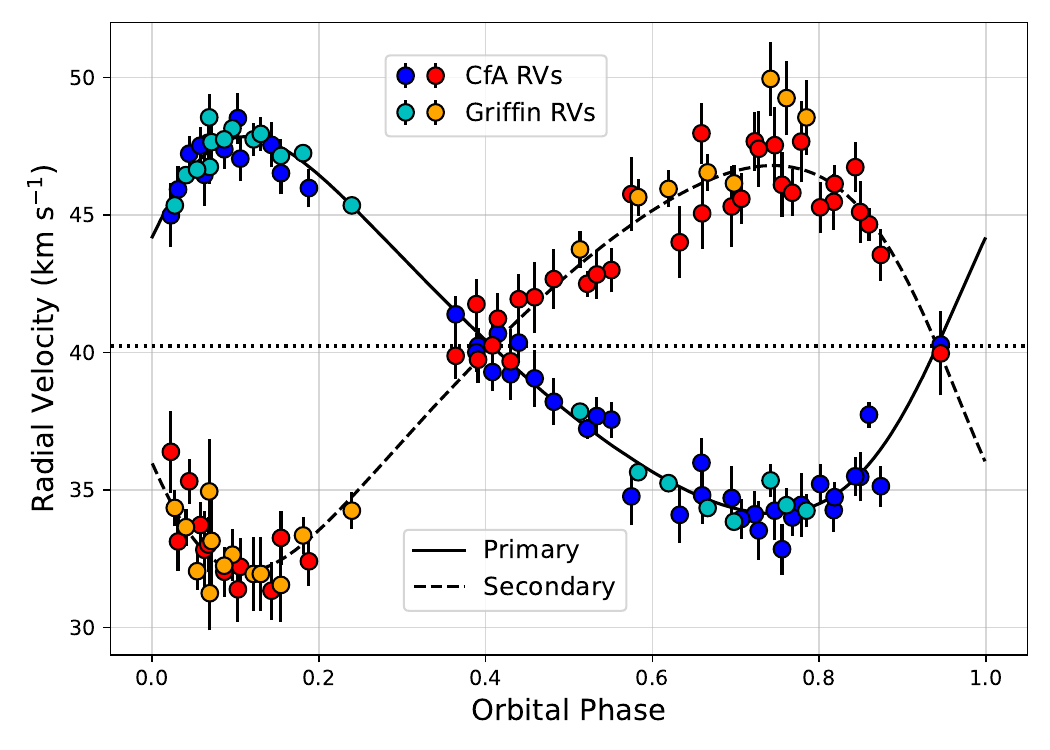}

  \figcaption{\emph{Top}: Astrometric observations of vB\,259 together with our
    model for the orbit. Periastron is marked with a square and
    labeled `P'. The dotted
    line represents the line of nodes, with the ascending node labeled
    $\Omega$.
    \emph{Bottom}: RV measurements together with our model for
    the spectroscopic orbit. The center-of-mass velocity is indicated with the
    dotted line. \label{fig:L33skyrv}}

\end{figure}

For this work we performed an MCMC analysis combining these
astrometric measurements with our own RVs and those of Griffin. We
incorporated also the abscissa residuals from the Hipparcos mission,
weighted as explained above for vB\,22, which show hints that the
motion of the photocenter is detectable. They therefore help to
constrain the astrometric elements. The results are listed in
Table~\ref{tab:L33} in a similar format as employed previously.
Plots of the visual and spectroscopic orbit models and
observations are shown in Figure~\ref{fig:L33skyrv}.
Unfortunately, the dynamical masses are not
well constrained in this case because the inclination of the orbit is
rather low, and its uncertainty is significant.

\setlength{\tabcolsep}{6pt}
\begin{deluxetable}{lcc}
\tablewidth{0pc}
\tablecaption{Orbital Parameters for vB\,259 \label{tab:L33}}
\tablehead{
\colhead{Parameter} &
\colhead{Value} &
\colhead{Prior}
}
\startdata
 $P$ (days)                             & $1042.14 \pm 0.45$\phm{222}  & [500, 2000]     \\
 $T_{\rm peri}$ (BJD$-$2,400,000)       & $49947.6 \pm 5.3$\phm{2222}  & [49700, 50500]  \\
 $a^{\prime\prime}$ (\arcsec)           & $0.05003 \pm 0.00072$        & [0.02, 0.10]    \\
 $\sqrt{e}\cos\omega_1$                 & $+0.227 \pm 0.014$\phs       & [$-1$, 1]       \\
 $\sqrt{e}\sin\omega_1$                 & $-0.430 \pm 0.011$\phs       & [$-1$, 1]       \\
 $\cos i$                               & $+0.824 \pm 0.021$\phs       & [$-1$, 1]       \\
 $\Omega$ (deg)                         & $135.8 \pm 1.4$\phn\phn      & [0, 360]        \\
 $K_1$ (\kms)                           & $6.852 \pm 0.092$            & [1, 20]         \\
 $K_2$ (\kms)                           & $7.38 \pm 0.17$              & [1, 20]         \\
 $\gamma$ (\kms)                        & $+40.239 \pm 0.090$\phn\phs  & [20, 60]        \\
 $\Delta$RV$_{\rm Griffin}$ (\kms)      & $-0.55 \pm 0.13$\phs         & [$-5$, 5]       \\
 $a^{\prime\prime}_{\rm phot}$ (mas)    & $7.0 \pm 2.9$                & [0, 50]         \\
 $\Delta\alpha^*$ (mas)                 & $+3.6 \pm 2.1$\phs           & [$-100$, 100]   \\
 $\Delta\delta$ (mas)                   & $-1.4 \pm 1.2$\phs           & [$-100$, 100]   \\
 $\Delta\mu_{\alpha}^*$ (mas yr$^{-1}$) & $+3.6 \pm 2.9$\phs           & [$-100$, 100]   \\
 $\Delta\mu_{\delta}$ (mas yr$^{-1}$)   & $-6.0 \pm 2.3$\phs           & [$-100$, 100]   \\[0.5ex]
\hline \\ [-1.5ex]
\multicolumn{3}{c}{Error Scaling Factors} \\ [1ex]
\hline \\ [-1.5ex]
 $f_{\rm CfA,1}$                        & $1.08 \pm 0.12$              & [$-5$, 5]       \\
 $f_{\rm CfA,2}$                        & $0.93 \pm 0.11$              & [$-5$, 5]       \\
 $f_{\rm Griffin,1}$                    & $1.08 \pm 0.20$              & [$-5$, 5]       \\
 $f_{\rm Griffin,2}$                    & $1.10 \pm 0.20$              & [$-5$, 5]       \\
 $f_{\theta}$                           & $2.3 \pm 1.3$                & [$-5$, 5]       \\
 $f_{\rho}$                             & $1.26 \pm 0.66$              & [$-5$, 5]       \\[0.5ex]
\hline \\ [-1.5ex]
\multicolumn{3}{c}{Derived Properties} \\ [1ex]
\hline \\ [-1.5ex]
 $i$ (deg)                              & $34.5 \pm 2.2$\phn           & \nodata         \\
 $e$                                    & $0.2376 \pm 0.0078$          & \nodata         \\
 $\omega_1$ (deg)                       & $297.9 \pm 1.9$\phn\phn      & \nodata         \\
 $a$ (au)                               & $2.33 \pm 0.14$              & \nodata         \\
 $M_1$ ($M_{\sun}$)                     & $0.78 \pm 0.15$              & \nodata         \\
 $M_2$ ($M_{\sun}$)                     & $0.72 \pm 0.13$              & \nodata         \\
 $M_{\rm tot}$ ($M_{\sun}$)             & $1.52 \pm 0.28$              & \nodata         \\
 $q \equiv M_2/M_1$                     & $0.928 \pm 0.025$            & \nodata         \\
 $\pi_{\rm orb}$ (mas)                  & $21.3 \pm 1.5$\phn           & \nodata         \\
 Distance (pc)                          & $46.3 \pm 3.3$\phn           & \nodata         
\enddata

\tablecomments{The
  values listed in the table correspond to the mode of the posterior
  distributions from our MCMC analysis, with uncertainties
  representing the 68.3\% credible intervals. $\Delta_{\rm Griffin}$
  is the offset to be applied to the Griffin RVs to place them on the
  same system as the CfA velocities. The $f$ symbols represent
  multiplicative scaling factors for the nominal errors of the radial
  velocity and astrometric measurements, solved for simultaneously
  with the other elements. Priors in square brackets are uniform over
  the ranges specified, except those for the error inflation factors
  $f$, which are log-uniform. }

\end{deluxetable}
\setlength{\tabcolsep}{6pt}

\subsection*{(374) vB 80}  
\label{sec:VB80}

This bright visual binary member, with a current separation of about 1\farcs5
and a brightness difference of 2.5~mag, was observed many times in
combined light with the Digital Speedometers at the CfA between 1989
and 1999, including several attempts to split the components on nights
of good seeing.  However, the limited signal-to-noise ratios with
those instruments and especially the rapid rotational velocity of the primary
star ($v \sin i \approx 180~\kms$) have prevented us from extracting
useful RV information. A handful of spectra of the combined light were
also gathered with TRES more recently, between 2011 and 2013.

The latest solution for the 172~yr, highly eccentric astrometric orbit
of vB\,80 ($e = 0.915$) was reported by \cite{Torres:2019a}. That study
made use of ground-based visual and speckle observations, as well as
proper motions for the individual components from the Gaia and
Hipparcos missions, but did not use RVs because they were not
considered to be reliable at the time. Nevertheless, the astrometry
alone allowed the determination of the mass ratio as well as the
dynamical masses for both stars.

\begin{figure}
  \epsscale{1.17}
  \hspace*{-1mm}\includegraphics[width=0.48\textwidth]{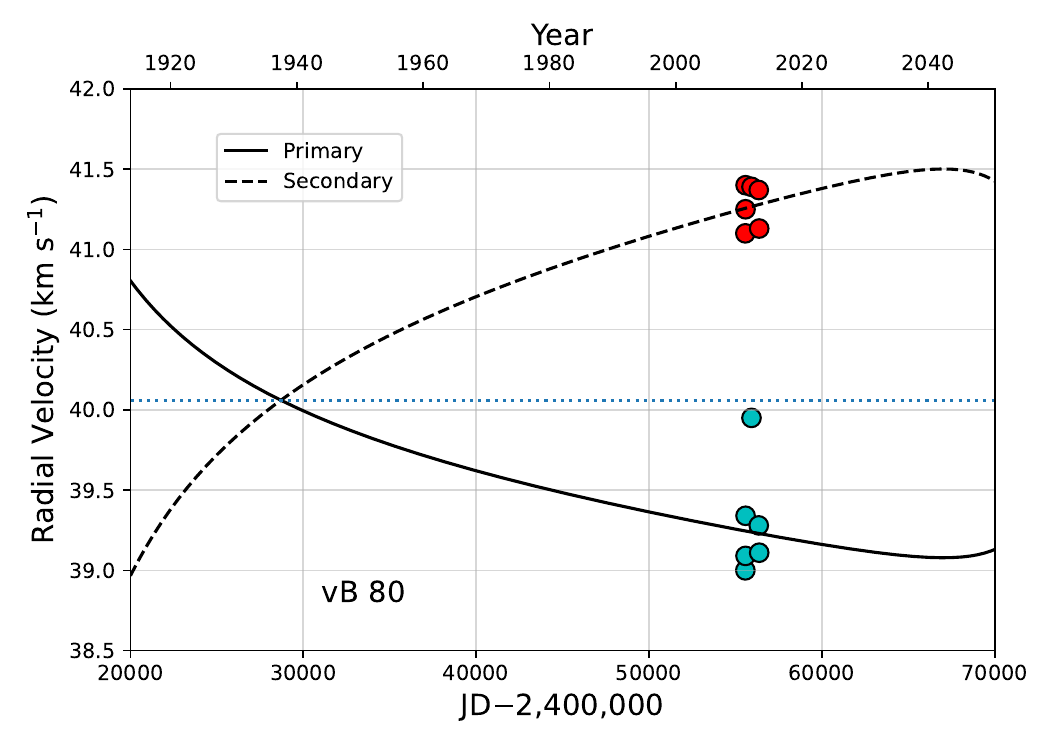}

  \figcaption{RV measurements for vB\,80, compared with a section of the
    spectroscopic orbit calculated from the visual elements of
    \cite{Torres:2019a} and the velocity semiamplitudes inferred from
    the individual masses in that work. The only free parameter we
    have adjusted is the center-of-mass velocity (dotted line),
    $\gamma = 40.06~\kms$. The one discrepant primary measurement
    happens to be from a spectrum of much lower signal-to-noise ratio
    compared to the others.\label{fig:VB80orbit}}

\end{figure}

We have reanalyzed the six TRES spectra of the combined light using
{\tt TODCOR}, and we are now able to extract meaningful velocities for the two
components using templates with effective temperatures of 7750 and 5750~K, along
with $v \sin i$ values of 180 and 6~\kms.
Given the angular separation and the 2\farcs3 diameter of
the spectrograph fiber, the amount of flux from each star entering the
instrument under typical sky conditions is highly dependent on seeing
and guiding, so for this case we left the flux ratio between the
primary and secondary as a free parameter for each exposure. The
velocities we obtained
are shown in Figure~\ref{fig:VB80orbit} together with the
spectroscopic orbit of vB\,80, all of whose parameters are known or can
be inferred from the visual
elements by \cite{Torres:2019a} and the masses, except for the
center-of-mass velocity. With all other orbital parameters held fixed,
the determination of this missing parameter using the measured
velocities yields $\gamma \approx 40.06~\kms$, which is in very good
agreement with the expected astrometric velocity for the system
(39.93~\kms). The difference of only 0.13~\kms\ is well within the
internal velocity dispersion of the cluster (see Section~\ref{sec:dispersion}).

\subsection*{(417) vB 96}
\label{sec:H578}

This is a double-lined binary with a period just under 14~yr, and is a
long-recognized member of the Hyades. The spectroscopic orbit
published by \cite{Griffin:2012} rests on spectrometer scans he collected
between 2002 and 2012, although only the ones after 2007 allowed him
to measure RVs for both components. As this represents less than one cycle,
Griffin was forced to hold the period fixed at a value determined by
including earlier measurements of lower quality that did not split
the components, on the assumption that they represented the primary.

Observations at the CfA were
carried out earlier with the Digital Speedometers between 1980 and 2001, and
both stars can be seen in most of the 58 spectra, which cover 1.5
cycles of the binary. The templates for
our {\tt TODCOR} analysis had temperatures of 5500 and 4750~K for the
primary and secondary, and zero rotational broadening.
The flux
ratio we determined at the mean wavelength of our observations is
$\ell_2/\ell_1 = 0.200 \pm 0.017$.

The companion of vB\,96 was spatially resolved in 1985 with the speckle
interferometry technique by \cite{McAlister:1987}, at a separation of
0\farcs15.  The double star designation is WDS~04340+1510.  About four
dozen astrometric observations have been obtained since, and trace the
orbit over 2.5 cycles quite well \citep[e.g.,][]{Cvetkovic:2008},
although measurements near
periastron are sparse because the angular separation there is less
than 0\farcs05. The Gaia mission reported the binary was partially
resolved in 9\% of the observations.

\begin{figure}
  \epsscale{1.17}
  \hspace*{-1mm}\includegraphics[width=0.48\textwidth]{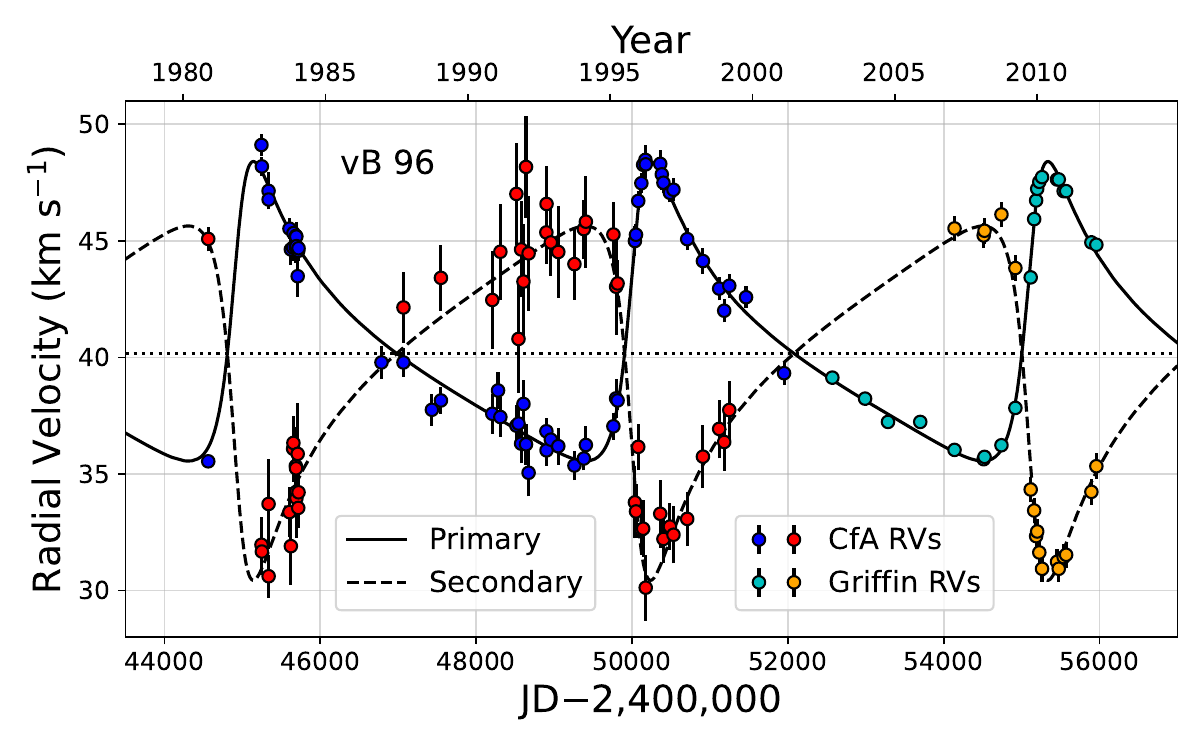}

  \figcaption{RV measurements for vB\,96, along with our best-fit model.
  The dotted line marks the center-of-mass velocity.\label{fig:H578rv}}

\end{figure}

\begin{figure}
  \epsscale{1.17}
  \hspace*{-1mm}\includegraphics[width=0.48\textwidth]{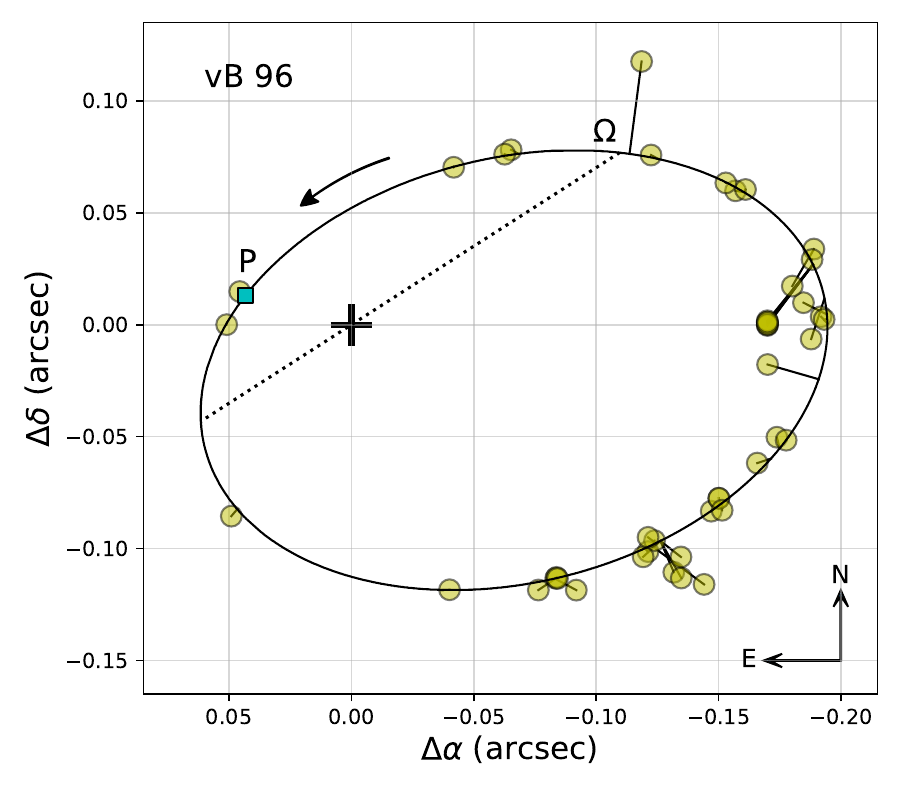}
  \hspace*{-1mm}\includegraphics[width=0.48\textwidth]{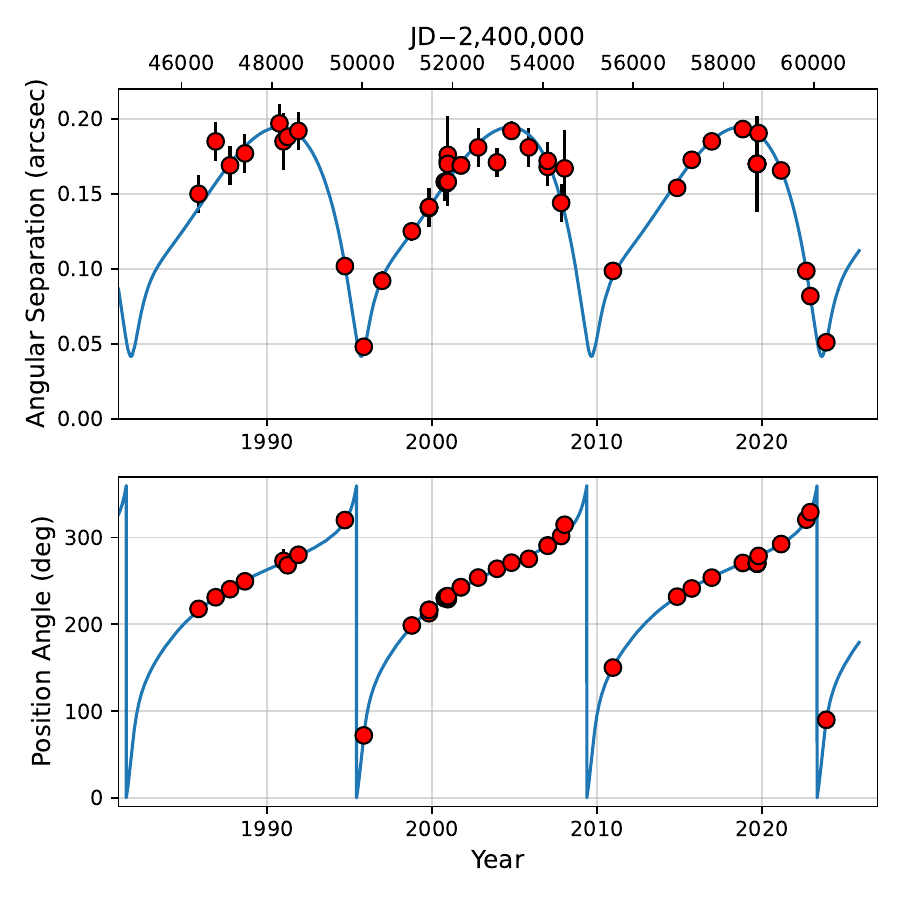}

  \figcaption{Astrometric orbit of vB~96; see
    Figures~\ref{fig:L33skyrv} and \ref{fig:+28624pasep}.
    \label{fig:H578sky}}

\end{figure}

Here we have combined the RV and speckle measurements in order to derive
model-independent masses for the components. The results of our MCMC
analysis are given in Table~\ref{tab:H578} in a similar format as for
other systems. The dynamical masses for the primary and secondary of vB\,96 are
determined to about 4\% and 3\%, respectively, and the uncertainty of
the orbital parallax is 1.3\%, a factor of two smaller than the
value reported in the Gaia\,DR3 catalog. The Gaia parallax is also
about 1~mas smaller than ours: $\pi_{\rm DR3} = 20.68 \pm 0.61$~mas,
including a correction for the known zero-point offset
\citep{Lindegren:2021}.  The spectroscopic orbit is shown in
Figure~\ref{fig:H578rv}, and the astrometric measurements together
with our model are seen in Figure~\ref{fig:H578sky}.

\begin{figure}
  \epsscale{1.17}
  \hspace*{-1mm}\includegraphics[width=0.48\textwidth]{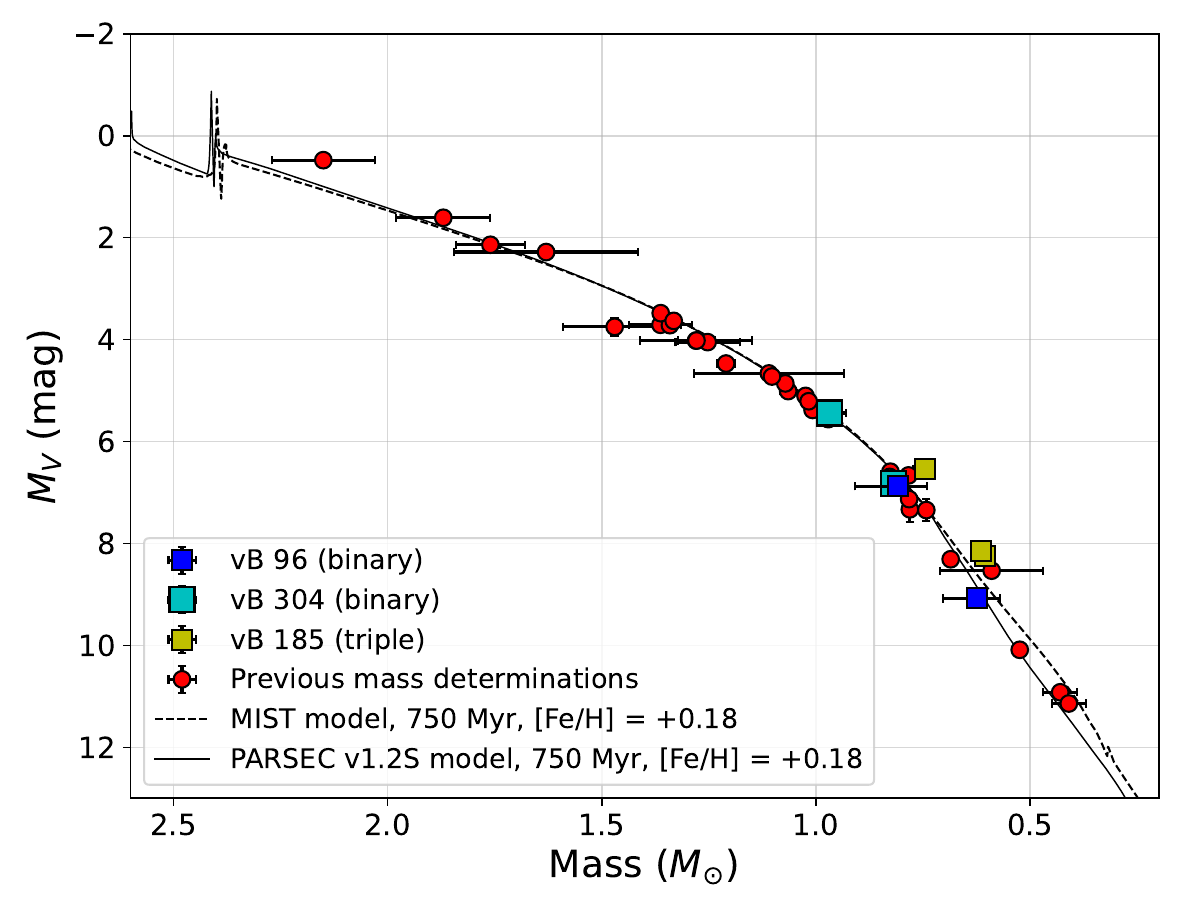}

  \caption{Mass-luminosity relation for the Hyades in the visual
  band. Empirical determinations are shown for vB\,96, vB\,304,
  and vB\,185, as labeled. Circles represent other previously
  published dynamical mass determinations in the cluster. The
  lines show model isochrones from the PARSEC~v1.2S series
  \citep{Chen:2014} and the MIST series \citep{Choi:2016}.
  \label{fig:mlrelation}}
\end{figure}

The $V$-band brightness difference between the components, from an
average of eight independent measures contained in the WDS, is $1.38
\pm 0.03$~mag. By adopting a consensus apparent brightness from the
literature of $V = 8.495 \pm 0.020$~mag, along with the distance we
obtain, we infer absolute visual magnitudes for the components of
$5.440 \pm 0.035$ and $6.815 \pm 0.042$~mag. These values along with
the masses are compared against evolutionary models for the Hyades in
Figure~\ref{fig:mlrelation}, and show good agreement. The secondary falls
one or two tenths of a magnitude below the predictions, and seems to
follow an emerging trend found for other cluster stars \citep[see,
  e.g.,][]{Torres:2024a}.

\setlength{\tabcolsep}{6pt}
\begin{deluxetable}{lcc}
\tablewidth{0pc}
\tablecaption{Orbital Parameters for vB\,96 \label{tab:H578}}
\tablehead{
\colhead{Parameter} &
\colhead{Value} &
\colhead{Prior}
}
\startdata
 $P$ (days)                        & $5097.5 \pm 2.8$\phm{222}    & [3000, 10000]   \\
 $T_{\rm peri}$ (BJD$-$2,400,000)  & $50036.9 \pm 4.7$\phm{2222}  & [48000, 52000]  \\
 $a^{\prime\prime}$ (\arcsec)      & $0.15220 \pm 0.00042$        & [0.02, 0.40]      \\
 $\sqrt{e}\cos\omega_1$            & $+0.3607 \pm 0.0045$\phs     & [$-1$, 1]         \\
 $\sqrt{e}\sin\omega_1$            & $-0.6899 \pm 0.0030$\phs     & [$-1$, 1]         \\
 $\cos i$                          & $+0.6699 \pm 0.0037$\phs     & [$-1$, 1]         \\
 $\Omega$ (deg)                    & $305.03 \pm 0.38$\phn\phn    & [0, 360]        \\
 $K_1$ (\kms)                      & $6.426 \pm 0.069$            & [1, 30]         \\
 $K_2$ (\kms)                      & $7.60 \pm 0.14$              & [1, 30]         \\
 $\gamma$ (\kms)                   & $+40.171 \pm 0.072$\phn\phs  & [20, 60]        \\
 $\Delta$RV$_{\rm Griffin}$ (\kms) & $-1.068 \pm 0.091$\phs   & [$-5$, 5]         \\[0.5ex]
\hline \\ [-1.5ex]
\multicolumn{3}{c}{Error Scaling Factors} \\ [1ex]
\hline \\ [-1.5ex]
 $f_{\rm CfA,1}$                   & $1.04 \pm 0.10$              & [$-5$, 5]         \\
 $f_{\rm CfA,2}$                   & $1.01 \pm 0.10$              & [$-5$, 5]         \\
 $f_{\rm Griffin,1}$               & $1.15 \pm 0.21$              & [$-5$, 5]         \\
 $f_{\rm Griffin,2}$               & $1.06 \pm 0.22$              & [$-5$, 5]         \\
 $f_{\theta}$                      & $1.93 \pm 0.22$              & [$-5$, 5]         \\
 $f_{\rho}$                        & $3.13 \pm 0.34$              & [$-5$, 5]         \\[0.5ex]
\hline \\ [-1.5ex]
\multicolumn{3}{c}{Derived Properties} \\ [1ex]
\hline \\ [-1.5ex]
 $i$ (deg)                         & $47.94 \pm 0.29$\phn         & \nodata         \\
 $e$                               & $0.6059 \pm 0.0020$          & \nodata         \\
 $\omega_1$ (deg)                  & $297.62 \pm 0.39$\phn\phn    & \nodata         \\
 $a$ (au)                          & $7.046 \pm 0.084$            & \nodata         \\
 $M_1$ ($M_{\sun}$)                & $0.969 \pm 0.040$            & \nodata         \\
 $M_2$ ($M_{\sun}$)                & $0.820 \pm 0.026$            & \nodata         \\
 $M_{\rm tot}$ ($M_{\sun}$)        & $1.795 \pm 0.064$            & \nodata         \\
 $q \equiv M_2/M_1$                & $0.846 \pm 0.019$            & \nodata         \\
 $\pi_{\rm orb}$ (mas)             & $21.63 \pm 0.28$\phn         & \nodata         \\
 Distance (pc)                     & $46.22 \pm 0.60$\phn         & \nodata         
\enddata

\tablecomments{When expressed in Julian years, the period is $13.9562 \pm 0.0077$~yr,
  and the time of periastron passage is $1995.871 \pm 0.013$.  The
  values listed in the table correspond to the mode of the posterior
  distributions from our MCMC analysis, with uncertainties
  representing the 68.3\% credible intervals. $\Delta_{\rm Griffin}$
  is the offset to be applied to the Griffin RVs to place them on the
  same system as the CfA velocities. The $f$ symbols represent
  multiplicative scaling factors for the nominal errors of the radial
  velocity and astrometric measurements, solved for simultaneously
  with the other elements. Priors in square brackets are uniform over
  the ranges specified, except those for the error inflation factors
  $f$, which are log-uniform. }

\end{deluxetable}
\setlength{\tabcolsep}{6pt}

\vskip 30pt
\subsection*{(450) vB 304}  
\label{sec:L83}

This is a hierarchical triple system (HD\,285947), and a member of the
cluster. The inner binary has a period of 60\,d and a moderate
eccentricity \citep{Griffin:1985a}, but only the primary was detected
in those authors' spectroscopic observations. The third star in the
system was discovered in a $K$-band speckle imaging survey of the
Hyades by \cite{Patience:1998}, at a separation of about 0\farcs2. The
system has received the double star designation WDS~04384+1733.
Subsequently, \cite{Tomkin:2002} were able to detect the secondary of
the inner binary in their higher quality spectra, as well as a drift
of the center-of-mass velocity in response to the third star. The
tertiary, which they estimated to be a much fainter mid-M dwarf,
remained undetected. \cite{Tomkin:2002} combined their RVs with those
of Griffin, and estimated the period of the tertiary to be about
8000\,d, although they could not rule out a longer period of roughly
10,000\,d.

Our own observations of vB\,304 at the CfA began in 1980. We measured
the velocities of both components of the inner pair (listed in
Table~\ref{tab:RVsb2}) using templates with $T_{\rm eff} = 4750$ and 4000~K
for the primary and secondary, and $v \sin i = 2~\kms$ for both.
The measured flux ratio is
$\ell_{\rm Ab}/\ell_{\rm Aa}= 0.108 \pm 0.006$ at 5187~\AA.
Although our spectra show no sign of the
third star, its dynamical influence on the velocities of the primary
and secondary is obvious. A combined spectroscopic solution of the
inner and outer orbits using our RVs, similar to the one performed by
\cite{Tomkin:2002}, supports the period of $\sim$8000\,d proposed by
those authors, and our velocities now cover two cycles of that orbit.
Furthermore, astrometric measurements of the relative position of the
tertiary since its discovery have been obtained by several observers,
and those observations independently confirm the outer period.  An
astrometric-only orbital solution by \cite{Tokovinin:2021a} gives the
period as $22.92 \pm 0.15$~yr ($8383 \pm 55$\,d), and other orbital
elements consistent with ours.

Here we have performed a joint MCMC astrometric-spectroscopic orbital
solution combining the RVs of Griffin, Tomkin, and our own, along with
the astrometric measurements of the tertiary from the WDS. As pointed
out by \cite{Tomkin:2002}, however, in principle the Griffin
measurements of the primary could be biased due to blending with the
much fainter secondary star, which was not seen in those observations.
Because they go back to 1973, the Griffin measurements are potentially
valuable for constraining the outer orbit. Therefore, as a test, we
incorporated them and allowed them to contribute to the determination
of their own value of the velocity semiamplitude $K_{\rm Aa}$,
separate from the one constrained by the other observations. The
remaining spectroscopic elements of the inner binary were constrained
by all observations. While the Griffin
semiamplitude did come out marginally smaller than the value from the
other observations, the difference was much smaller than the
uncertainties, so we chose to ignore it and to solve for a single
common parameter $K_{\rm Aa}$. A similar concern could be raised about
the even fainter tertiary, which neither we nor Tomkin or Griffin
detected, but its effect will be completely negligible.

Table~\ref{tab:L83} presents the results and other derived properties
from our combined astrometric-spectroscopic solution, compared to
orbital solutions by others. For the inner binary, the times of
observation were adjusted iteratively in our solution for the light
travel time resulting from motion in the outer orbit.

\setlength{\tabcolsep}{4pt}
\begin{deluxetable*}{lccccc}
\tablewidth{0pc}
\tablecaption{Orbital Parameters for vB\,304, Compared with Results by Others \label{tab:L83}}
\tablehead{
\colhead{Parameter} &
\colhead{\cite{Griffin:1985a}} &
\colhead{\cite{Tomkin:2002}} &
\colhead{\cite{Tokovinin:2021a}} &
\colhead{This paper} &
\colhead{Prior}
}
\startdata
 $P_{\rm A}$ (day)                        & $60.821 \pm 0.013$\phn       & $60.8081 \pm 0.0005$\phn       &  \nodata                  & $60.80936 \pm 0.00017$\phn      & [40, 80]       \\
 $T_{\rm peri,A}$ (BJD)                   & $44397.6 \pm 0.4$\phm{2222}  & $50417.70 \pm 0.06$\phm{2222}  &  \nodata                  & $50174.555 \pm 0.041$\phm{2222} & [51140, 51200] \\
 $\sqrt{e_{\rm A}}\cos\omega_{\rm Ab}$    & \nodata                      & \nodata                        &  \nodata                  & $-0.3670 \pm 0.0014$\phs        & [$-$1, 1]        \\
 $\sqrt{e_{\rm A}}\sin\omega_{\rm Ab}$    & \nodata                      & \nodata                        &  \nodata                  & $+0.2723 \pm 0.0019$\phs        & [$-$1, 1]        \\
 $K_{\rm Aa}$ (\kms)                      & $25.27 \pm 0.22$\phn         & $25.85 \pm 0.04$\phn           &  \nodata                  & $25.818 \pm 0.019$\phn          & [5, 50]        \\
 $K_{\rm Ab}$ (\kms)                      & \nodata                      & $33.44 \pm 0.10$\phn           &  \nodata                  & $33.439 \pm 0.089$\phn          & [5, 50]        \\
 $P_{\rm AB}$ (day)                       & \nodata                      & $8074 \pm 81$\phn\phn          &  $8386 \pm 55$\phn\phn    & $8310 \pm 17$\phn\phn           & [5000, 12000]  \\
 $T_{\rm peri,AB}$ (BJD)                  & \nodata                      & $51548 \pm 26$\phm{222}        &  $51538 \pm 26$\phm{222}  & $51561 \pm 14$\phm{222}         & [50000, 53000] \\
 $a_{\rm AB}^{\prime\prime}$ (\arcsec)    & \nodata                      & \nodata                        &  $0.210 \pm 0.005$        & $0.2117 \pm 0.0056$             & [0.1, 0.5]     \\
 $\sqrt{e_{\rm AB}}\cos\omega_{\rm A}$    & \nodata                      & \nodata                        &  \nodata                  & $-0.376 \pm 0.016$\phs          & [$-$1, 1]        \\
 $\sqrt{e_{\rm AB}}\sin\omega_{\rm A}$    & \nodata                      & \nodata                        &  \nodata                  & $-0.6943 \pm 0.0070$\phs        & [$-$1, 1]        \\
 $\cos i_{\rm AB}$                        & \nodata                      & \nodata                        &  \nodata                  & $0.803 \pm 0.024$               & [$-$1, 1]        \\
 $\Omega_{\rm AB}$ (deg)                  & \nodata                      & \nodata                        &  $161.7 \pm 3.1$\phn\phn  & $341.3 \pm 2.1$\phn\phn         & [0, 360]       \\
 $K_{\rm A}$ (\kms)                       & \nodata                      & $2.18 \pm 0.06$                &  \nodata                  & $2.146 \pm 0.051$           & [0.1, 10]      \\
 $\gamma$ (\kms)                          & $+42.43 \pm 0.14$\phn\phs    & $+40.17 \pm 0.07$\phn\phs          &  \nodata                  & $+40.165 \pm 0.040$\phn\phs     & [30, 60]       \\
 $\Delta$RV (\kms)                        & \nodata                      & \nodata                        &  \nodata                  & $-0.412 \pm 0.051$\phs          & [$-$5, 5]        \\[0.5ex]
\hline \\ [-1.5ex]
\multicolumn{6}{c}{Error Scaling Factors} \\ [1ex]
\hline \\ [-1.5ex]
 $f_{\rm Aa,Tom}$                        & \nodata                      & \nodata                        &  \nodata                  & $1.128 \pm 0.091$               & [$-$5, 5]        \\
 $f_{\rm Ab,Tom}$                        & \nodata                      & \nodata                        &  \nodata                  & $1.06 \pm 0.15$                 & [$-$5, 5]        \\
 $f_{\rm Aa,CfA}$                        & \nodata                      & \nodata                        &  \nodata                  & $1.004 \pm 0.093$               & [$-$5, 5]        \\
 $f_{\rm Ab,CfA}$                        & \nodata                      & \nodata                        &  \nodata                  & $0.998 \pm 0.088$               & [$-$5, 5]        \\
 $f_{\theta}$                             & \nodata                      & \nodata                        &  \nodata                  & $3.0 \pm 1.0$                   & [$-$5, 5]        \\
 $f_{\rho}$                               & \nodata                      & \nodata                        &  \nodata                  & $3.00 \pm 0.81$                 & [$-$5, 5]        \\[0.5ex]
\hline \\ [-1.5ex]
\multicolumn{6}{c}{Derived Properties} \\ [1ex]
\hline \\ [-1.5ex]
 $e_{\rm A}$                              & $0.218 \pm 0.008$            & $0.2102 \pm 0.0016$            &  \nodata                  & $0.20886 \pm 0.00093$           & \nodata        \\
 $\omega_{\rm Aa}$ (deg)                  & $143.0 \pm 2.1$\phn\phn      & $142.9 \pm 0.4$\phn\phn        &  \nodata                  & $143.43 \pm 0.27$\phn\phn       & \nodata        \\
 $i_{\rm A}$ (deg)                        & \nodata                      & \nodata                        &  \nodata                  & $70.9 \pm 6.1$\phn              & \nodata        \\
 $i_{\rm AB}$ (deg)                       & \nodata                      & \nodata                        &  $36.2 \pm 2.5$\phn       & $36.6 \pm 2.3$\phn              & \nodata        \\
 $e_{\rm AB}$                             & \nodata                      & $0.604 \pm 0.017$              &  $0.611 \pm 0.015$        & $0.6232 \pm 0.0090$             & \nodata        \\
 $\omega_{\rm A}$ (deg)                   & \nodata                      & $242.9 \pm 2.5$\phn\phn        &  $241.4 \pm 2.3$\phn\phn          & $241.6 \pm 1.2$\phn\phn         & \nodata        \\
 $K_{\rm B}$ (\kms)                       & \nodata                      & \nodata                        &  \nodata                  & $7.65 \pm 0.84$                 & \nodata        \\
 $a_{\rm AB}$ (au)                        & \nodata                      & \nodata                        &  \nodata                  & $9.83 \pm 0.26$                 & \nodata        \\
 $M_{\rm AB}$ ($M_{\sun}$)                & \nodata                      & \nodata                        &  \nodata                  & $1.83 \pm 0.15$                 & \nodata        \\
 $M_{\rm A}$ ($M_{\sun}$)                 & \nodata                      & \nodata                        &  \nodata                  & $1.43 \pm 0.15$                 & \nodata        \\
 $M_{\rm Aa}$ ($M_{\sun}$)                & \nodata                      & \nodata                        &  \nodata                  & $0.809 \pm 0.085$               & \nodata        \\
 $M_{\rm Ab}$ ($M_{\sun}$)                & \nodata                      & \nodata                        &  \nodata                  & $0.624 \pm 0.066$               & \nodata        \\
 $M_{\rm B}$ ($M_{\sun}$)                 & \nodata                      & \nodata                        &  \nodata                  & $0.403 \pm 0.014$               & \nodata        \\
 $q \equiv M_{\rm Ab}/M_{\rm Aa}$         & \nodata                      & $0.773 \pm 0.002$              &  \nodata                  & $0.7721 \pm 0.0022$             & \nodata        \\
 $q \equiv M_{\rm B}/M_{\rm A}$           & \nodata                      & \nodata                        &  \nodata                  & $0.278 \pm 0.031$               & \nodata        
\enddata

\tablecomments{The values listed correspond to the mode of the
  posterior distributions from our MCMC analysis, with uncertainties
  representing the 68.3\% credible intervals. Times of periastron
  passage are referred to BJD~2,400,000. The values of $T_{\rm
    peri,A}$ by other authors differ from ours by very nearly an
  integer number of cycles. $\Delta$RV represents a systematic
  velocity offset added to our velocities to translate them to the
  reference frame of Tomkin. The $f$ symbols represent multiplicative
  scaling factors for the nominal errors of the radial-velocity and
  astrometric measurements, solved for simultaneously with the other
  elements. Priors in square brackets in the last column are uniform
  over the ranges specified, except those for the error inflation
  factors, which are log-uniform. }

\end{deluxetable*}
\setlength{\tabcolsep}{6pt}

\begin{figure}
  \epsscale{1.17}
  \hspace*{-1mm}\includegraphics[width=0.48\textwidth]{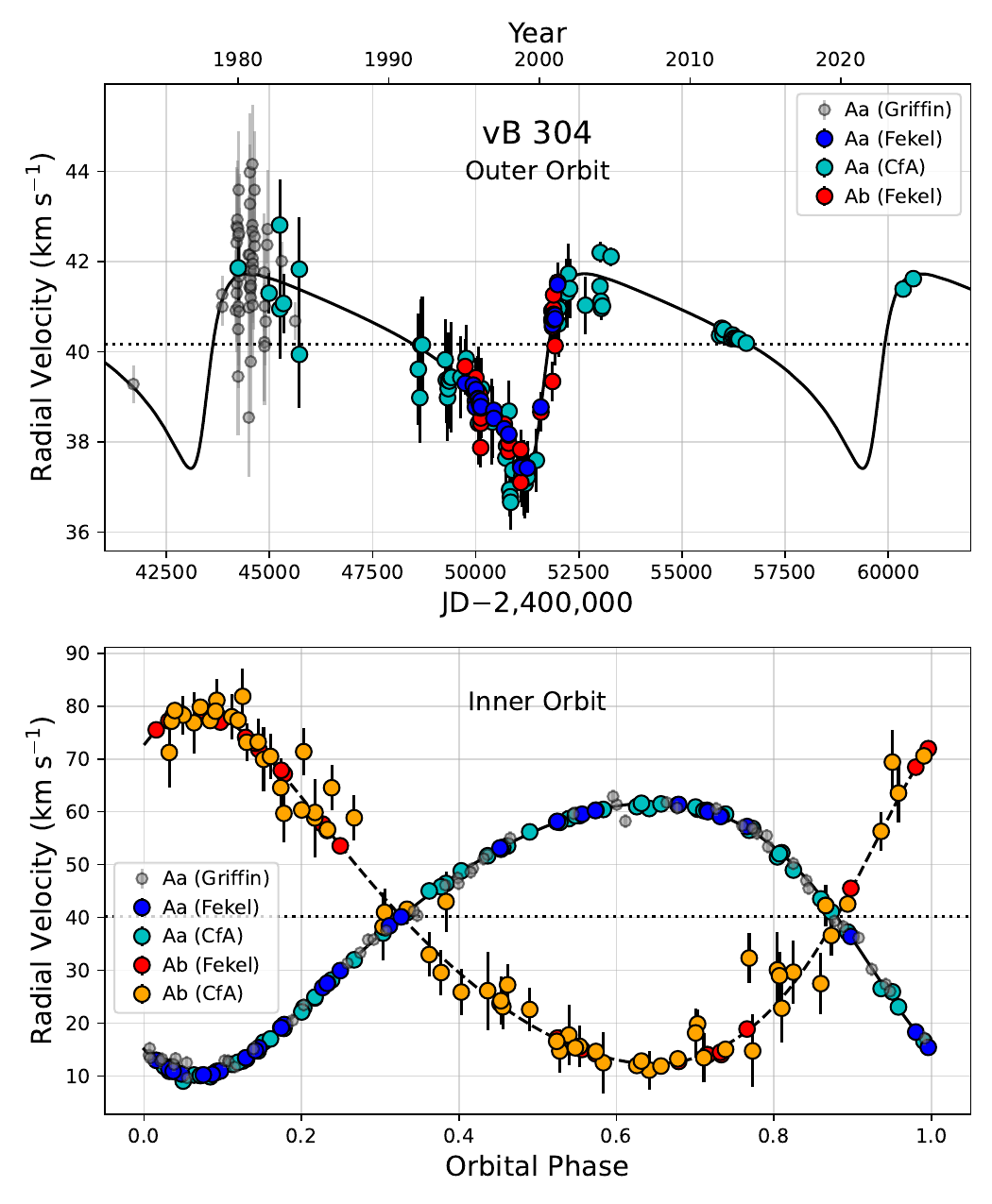}

  \figcaption{\emph{Top:} Our model for the outer spectroscopic orbit
    of vB\,304, together with the RV measurements of the primary and
    secondary from which the inner motion has been subtracted. The
    dotted line represents the center of mass of the triple system.
    \emph{Bottom:} Velocities in the inner spectroscopic orbit, after
    removing the motion of the inner binary in the outer
    orbit.  \label{fig:L83rvs}}

\end{figure}

\begin{figure}
  \epsscale{1.17}
  \hspace*{-1mm}\includegraphics[width=0.48\textwidth]{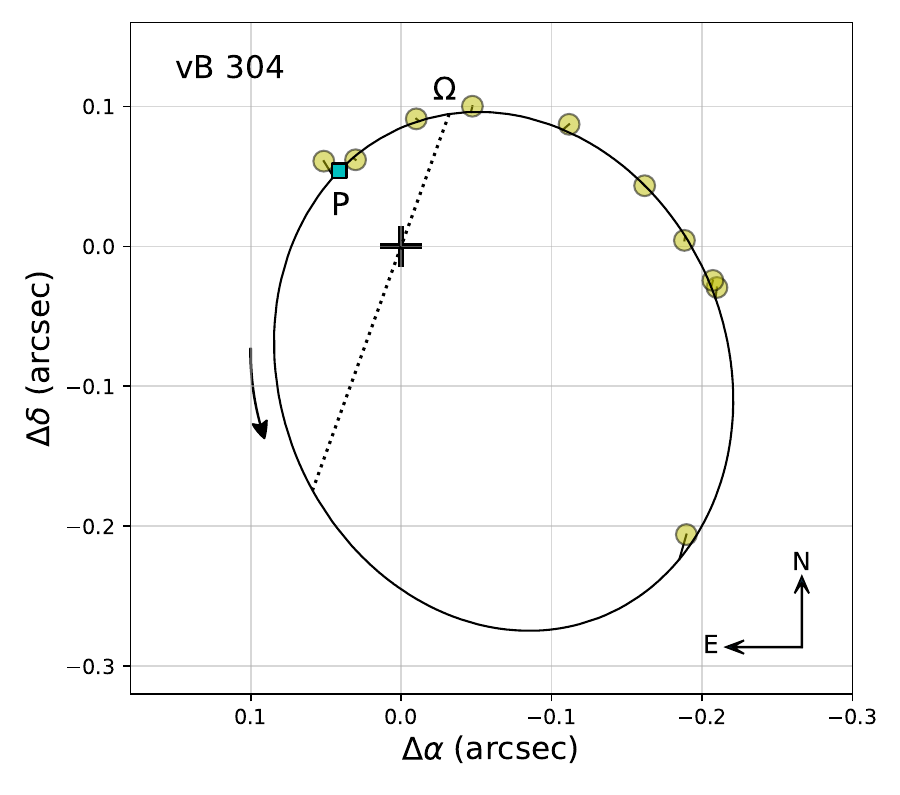}
  \hspace*{-1mm}\includegraphics[width=0.48\textwidth]{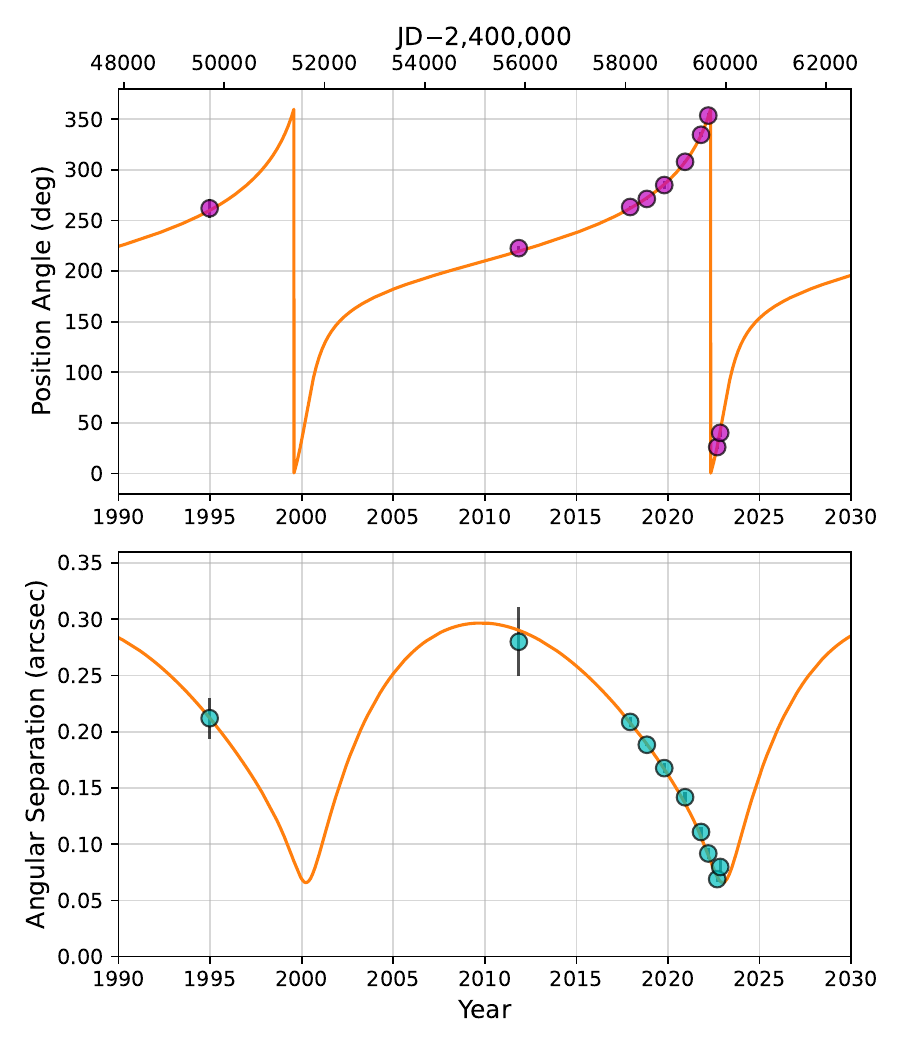}

  \figcaption{Astrometric orbit of vB\,304, similar to
    Figure~\ref{fig:H578sky}.\label{fig:L83sky}}

\end{figure}

The inner and outer spectroscopic orbits are shown in
Figure~\ref{fig:L83rvs}.  The orbit of the tertiary on the plane of
the sky is shown in the top panel of Figure~\ref{fig:L83sky}.  The
lower panels display the position angle and angular separation as a
function of time, along with our best-fit model. The WDS measurements
cover slightly more than one orbital cycle.

While the astrometry provides additional properties of the outer
orbit, it is not sufficient to make a determination of the masses of
the three stars in vB\,304 because the tertiary has not been detected
spectroscopically. For this, we made use of the parallax provided in
the Gaia~DR3 catalog, $\pi_{\rm DR3} = 21.54 \pm
0.11$~mas\footnote{This value includes a parallax zero-point correction
  of +0.036~mas \citep{Lindegren:2021}.}. Because of the complicated
nature of the source, the reported parallax uncertainty is several
times larger than is typical for stars of this brightness. If taken at
face value, this parallax leads to absolute masses for the three stars
as listed in the bottom section of Table~\ref{tab:L83}. Other derived
properties are listed there as well. The masses we inferred for the
primary and secondary are somewhat larger than those suggested by
\cite{Tomkin:2002}, which relied on an adopted spectral type for the
primary.

With an estimate of the absolute magnitudes of the primary and
secondary, a comparison is possible against predictions from stellar
evolution models for the Hyades cluster. Our measured spectroscopic
flux ratio, converted to the $V$ band using PHOENIX model spectra from
\cite{Husser:2013}, leads to a brightness difference of $\Delta V =
2.20 \pm 0.13$~mag. Apparent magnitude measurements for vB\,304 in the
literature vary by a little more than 0.1~mag, and $B-V$ estimates by
0.3~mag, possibly due to intrinsic variability caused by spots. We
therefore adopted the brighter of the magnitude estimates, $V = 10.077
\pm 0.069$ \citep[from the Tycho-2 Catalog, transformed to the Johnson
  system;][]{Hog:2000}, which may be expected to more closely
represent the unspotted stellar surfaces. After a small correction of
+0.01~mag to remove the light contribution from the tertiary (based on
its mass), we obtained absolute magnitudes for the primary and
secondary of $M_V = 6.878 \pm 0.072$ and $9.08 \pm 0.13$,
respectively, ignoring extinction. The comparison against model
isochrones for the Hyades is shown in Figure~\ref{fig:mlrelation}, together
with other empirical mass determinations in the cluster.  The
properties of vB\,304 derived here should be considered to be
preliminary, given the incomplete astrometric coverage and the need
to rely on the Gaia parallax to constrain the solution.

\subsection*{(466) vB 185}   
\label{sec:L82}

This is another hierarchical triple system, and a member of the
cluster, also known as HD~29608. As in the case of vB\,304, the
initial report of a spectroscopic orbit for the inner binary is due to
\cite{Griffin:1985a}, and was based on the detection of the primary
star only. The period is 276 days. The outer companion was first
recorded by the Hipparcos mission at a separation of 0\farcs46, and
carries the designation WDS~04404+1631.\footnote{A claimed detection
  in 1991 by speckle interferometry \citep{Mason:1993}, along with
  several subsequent ones, turns out to be of a different binary, as
  indicated in the WDS.} In a more detailed spectroscopic study by
\cite{Tomkin:2003}, all three stars in the system were detected, and a
drift in the center-of-mass velocity of the inner binary was measured
for the first time, resulting from the action of the tertiary. Masses
for the components were estimated from assumed spectral types of K0,
K5, and K5 for components Aa, Ab, and B.

Spectroscopic observations at the CfA began on New Years Day, 1980.
The spectra from the Digital Speedometers only show the primary
clearly; the velocities may be found in Table~\ref{tab:RVsb1}.
The more recent spectra from TRES show all three stars, and their radial
velocities were measured with {\tt TRICOR}. They are listed in
Table~\ref{tab:L82rvs}. For the primary we adopted a synthetic
template with $T_{\rm eff} = 5000$~K, and for the secondary and
tertiary we used $T_{\rm eff} = 4250$~K. Surface gravities of $\log g
= 4.5$ were assumed, along with no rotational broadening. The measured
flux ratios at the mean wavelength of our observations
($\sim$5187~\AA) are $\ell_2/\ell_1 = 0.160 \pm 0.011$ and
$\ell_3/\ell_1 = 0.175 \pm 0.013$.

\setlength{\tabcolsep}{6pt}
\begin{deluxetable}{lccc}
\tablewidth{0pc}
\tablecaption{Radial Velocity measurements of vB\,185 from TRES \label{tab:L82rvs}}
\tablehead{
\colhead{BJD} &
\colhead{$RV_{\rm Aa}$} &
\colhead{$RV_{\rm Ab}$} &
\colhead{$RV_{\rm B}$}
\\
\colhead{(2,400,000+)} &
\colhead{(\kms)} &
\colhead{(\kms)} &
\colhead{(\kms)}
}
\startdata
  55910.9633  &  57.24  &  23.27  &  38.20 \\
  55965.6634  &  51.09  &  31.31  &  38.20 \\
  56016.5925  &  35.97  &  50.40  &  38.21 \\
  56196.0179  &  57.60  &  23.83  &  38.19 \\
  56228.0360  &  54.23  &  28.39  &  38.34 \\
  56290.8993  &  36.80  &  49.63  &  38.63 \\
  56309.8581  &  31.60  &  56.82  &  37.80 \\
  56348.7480  &  25.74  &  63.11  &  37.14 \\
  56385.6217  &  33.11  &  55.19  &  37.98 \\
  56559.0011  &  39.74  &  46.98  &  38.61 \\
  60354.6585  &  58.37  &  25.03  &  36.20 \\
  60364.7123  &  57.55  &  26.03  &  36.41 \\
  60386.6782  &  53.95  &  29.86  &  36.11 \\
  60591.0211  &  54.78  &  29.05  &  36.38 \\
  60600.8943  &  56.80  &  26.84  &  36.19 \\
  60612.9908  &  58.12  &  26.36  &  36.39 \\
  60614.0255  &  58.30  &  24.77  &  36.49 
\enddata
\tablecomments{Initial uncertainties for our orbital analysis were
  assumed to be 0.15, 0.60, and 0.60~\kms\ for stars Aa, Ab, and B,
  respectively.}

\end{deluxetable}
\setlength{\tabcolsep}{6pt}

More than two dozen additional measurements of the relative position
of the wide companion of vB\,304 have been made since its discovery,
and are listed in the WDS, although they only cover a bit less than
half of the outer orbit. An astrometric solution with a period of
68.2~yr was reported by \cite{Tokovinin:2021b}.  The spectroscopic
observations of \cite{Griffin:1985a} predate the astrometric ones by
about 20 years, and the CfA observations started about 10 years before
the astrometry. Both could help with the period. We therefore
carried out a combined MCMC astrometric-spectroscopic analysis, which
included the RVs of \cite{Griffin:1985a}, \cite{Tomkin:2003}, our own,
and the relative positions from the WDS. In principle this allows the
masses of all three stars to be derived, along with the orbital
parallax. A caveat is that the spectroscopic observations of
\cite{Griffin:1985a} and the CfA spectra obtained with the Digital
Speedometers only reveal the brighter primary star, so its radial
velocities may be influenced by the presence of the other two components.
This will generally add scatter, and the measured primary velocities
will tend to underestimate the true RV range for that star, resulting
in a bias of the semiamplitude $K_{\rm Aa}$ toward smaller values.
For this reason, in addition to the standard orbital elements, our
orbital analysis allowed both the Griffin velocities and the Digital
Speedometer velocities to contribute to their own primary velocity
semiamplitude for the inner binary, as done for vB\,304.  We also
allowed for systematic offsets between the velocities of Griffin and
our own relative to those of Tomkin. As a precaution, we also solved
for a systematic offset between the primary and secondary velocities
from TRES, to guard against the possibility of template mismatch.

Initial tests gave unrealistically small masses, along with a small
distance.  More reasonable results were obtained by imposing a
Gaussian prior on the orbital parallax, based on the parallax
$\pi_{\rm DR3}$ as listed in the Gaia~DR3 catalog, along with its
corresponding uncertainty.  Because the system is complex, the
parallax error from Gaia is many times larger than it would otherwise
be. Furthermore, a bias in $\pi_{\rm DR3}$ cannot be ruled out.  We
present the results of our solution in Table~\ref{tab:L82}, although
we consider them to be preliminary at the moment given the incomplete
observational coverage of the outer orbit. It is also possible that
some of the uncertainties are underestimated.

\setlength{\tabcolsep}{4pt}
\begin{deluxetable}{lcc}
\tablewidth{0pc}
\tablecaption{Orbital Parameters for vB\,185 \label{tab:L82}}
\tablehead{
\colhead{Parameter} &
\colhead{Value} &
\colhead{Prior}
}
\startdata
\multicolumn{3}{c}{Outer Orbit} \\ [1ex]
\hline \\ [-1.5ex]
 $P_{\rm AB}$ (yr)                         & $79.2 \pm 1.8$\phn             & [30, 100]       \\
 $T_{\rm peri, AB}$ (yr)                   & $2007.48 \pm 0.21$\phm{222}    & [1980, 2020]    \\
 $a_{\rm AB}^{\prime\prime}$ (\arcsec)     & $0.4878 \pm 0.0041$            & [0.1, 0.7]      \\
 $\sqrt{e_{\rm AB}}\cos\omega_{\rm B}$     & $+0.040 \pm 0.015$\phs         & [$-$1, 1]         \\
 $\sqrt{e_{\rm AB}}\sin\omega_{\rm B}$     & $+0.507 \pm 0.015$\phs         & [$-$1, 1]         \\
 $\cos i_{\rm AB}$                         & $+0.4837 \pm 0.0097$\phs       & [$-$1, 1]         \\
 $\Omega_{\rm AB}$ (deg)                   & $358.95 \pm 0.42$\phn\phn      & [0, 360]        \\
 $K_{\rm A}$ (\kms)                        & $2.416 \pm 0.066$              & [0.5, 20]       \\
 $K_{\rm B}$ (\kms)                        & $5.31 \pm 0.13$                & [0.5, 20]       \\
 $\gamma$ (\kms)                           & $+40.610 \pm 0.069$\phn\phs    & [30, 50]        \\[0.5ex]
\hline \\ [-1.5ex]
\multicolumn{3}{c}{Inner Orbit} \\ [1ex]
\hline \\ [-1.5ex]
 $P_{\rm A}$ (day)                         & $276.7678 \pm 0.0085$\phn\phn  & [200, 400]      \\
 $T_{\rm peri, A}$ (BJD$-$2,400,000)       & $50021.19 \pm 0.66$\phm{2222}  & [49900, 50100]  \\
 $\sqrt{e_{\rm A}}\cos\omega_{\rm Aa}$     & $-0.1546 \pm 0.0042$\phs       & [$-$1, 1]         \\
 $\sqrt{e_{\rm A}}\sin\omega_{\rm Aa}$     & $-0.2698 \pm 0.0035$\phs       & [$-$1, 1]         \\
 $K_{\rm Aa}$ (\kms)                       & $15.950 \pm 0.027$\phn         & [5, 50]         \\
 $K_{\rm Ab}$ (\kms)                       & $19.625 \pm 0.043$\phn         & [5, 50]         \\
 $K_{\rm Aa, Griffin}$ (\kms)              & $14.19 \pm 0.22$\phn           & [5, 50]         \\
 $K_{\rm Aa, DS}$ (\kms)                   & $15.00 \pm 0.16$\phn           & [5, 50]         \\
 $\Delta$RV$_{\rm CfA}$ (\kms)                 & $-0.45 \pm 0.11$\phs           & [$-$5, 5]         \\
 $\Delta$RV$_{\rm Griffin}$ (\kms)             & $-1.21 \pm 0.18$\phs           & [$-$5, 5]         \\
 $\Delta$RV$_{\rm Aa,Ab}$ (\kms)               & $-0.23 \pm 0.17$\phs           & [$-$5, 5]         \\[0.5ex]
\hline \\ [-1.5ex]
\multicolumn{3}{c}{Error Scaling Factors} \\ [1ex]
\hline \\ [-1.5ex]
 $f_{\rm Aa, Tomkin}$                      & $1.41 \pm 0.26$                & [$-$5, 5]         \\
 $f_{\rm Ab, Tomkin}$                      & $0.78 \pm 0.14$                & [$-$5, 5]         \\
 $f_{\rm B, Tomkin}$                       & $1.36 \pm 0.24$                & [$-$5, 5]         \\
 $f_{\rm Aa, TRES}$                        & $1.01 \pm 0.25$                & [$-$5, 5]         \\
 $f_{\rm Ab, TRES}$                        & $1.04 \pm 0.20$                & [$-$5, 5]         \\
 $f_{\rm B, TRES}$                         & $0.76 \pm 0.17$                & [$-$5, 5]         \\
 $f_{\rm Aa, Griffin}$                     & $1.05 \pm 0.12$                & [$-$5, 5]         \\
 $f_{\rm Aa, DS}$                          & $1.55 \pm 0.17$                & [$-$5, 5]         \\
 $f_{\theta}$                              & $1.90 \pm 0.37$                & [$-$5, 5]         \\
 $f_{\rho}$                                & $1.71 \pm 0.36$                & [$-$5, 5]         \\[0.5ex]
\hline \\ [-1.5ex]
\multicolumn{3}{c}{Derived Properties} \\ [1ex]
\hline \\ [-1.5ex]
 $i_{\rm AB}$ (deg)                        & $61.08 \pm 0.64$\phn           & \nodata         \\
 $e_{\rm AB}$                              & $0.259 \pm 0.015$              & \nodata         \\
 $\omega_{\rm B}$ (deg)                    & $85.5 \pm 1.7$\phn             & \nodata         \\
 $a$ (au)                                  & $22.72 \pm 0.36$\phn           & \nodata         \\
 $e_{\rm A}$                               & $0.0967 \pm 0.0019$            & \nodata         \\
 $\omega_{\rm Aa}$ (deg)                   & $240.18 \pm 0.83$\phn\phn      & \nodata         \\
 $M_{\rm A}$ ($M_{\sun}$)                  & $1.351 \pm 0.046$              & \nodata         \\
 $M_{\rm B}$ ($M_{\sun}$)                  & $0.614 \pm 0.021$              & \nodata         \\
 $q_{\rm AB} \equiv M_{\rm B}/M_{\rm A}$   & $0.454 \pm 0.020$              & \nodata         \\
 $M_{\rm Aa}$ ($M_{\sun}$)                 & $0.745 \pm 0.025$              & \nodata         \\
 $M_{\rm Ab}$ ($M_{\sun}$)                 & $0.606 \pm 0.021$              & \nodata         \\
 $q_{\rm A} \equiv M_{\rm Ab}/M_{\rm Aa}$  & $0.8129 \pm 0.0022$            & \nodata         \\
 $\pi_{\rm orb}$ (mas)                     & $21.47 \pm 0.20$\phn           & \nodata         \\
 Distance (pc)                             & $46.58 \pm 0.44$\phn           & \nodata         
\enddata

\tablecomments{The solution accounts for light travel time in the
  outer orbit.  The values listed correspond to the mode of the
  posterior distributions from our MCMC analysis, with uncertainties
  representing the 68.3\% credible intervals. The $f$ symbols
  represent multiplicative scaling factors for the nominal errors of
  the radial velocity and astrometric measurements, solved for
  simultaneously with the other elements. Priors in square brackets
  are uniform over the ranges specified, except those for the error
  inflation factors $f$, which are log-uniform. }

\end{deluxetable}
\setlength{\tabcolsep}{6pt}

\begin{figure}
  \epsscale{1.17}
  \hspace*{-1mm}\includegraphics[width=0.48\textwidth]{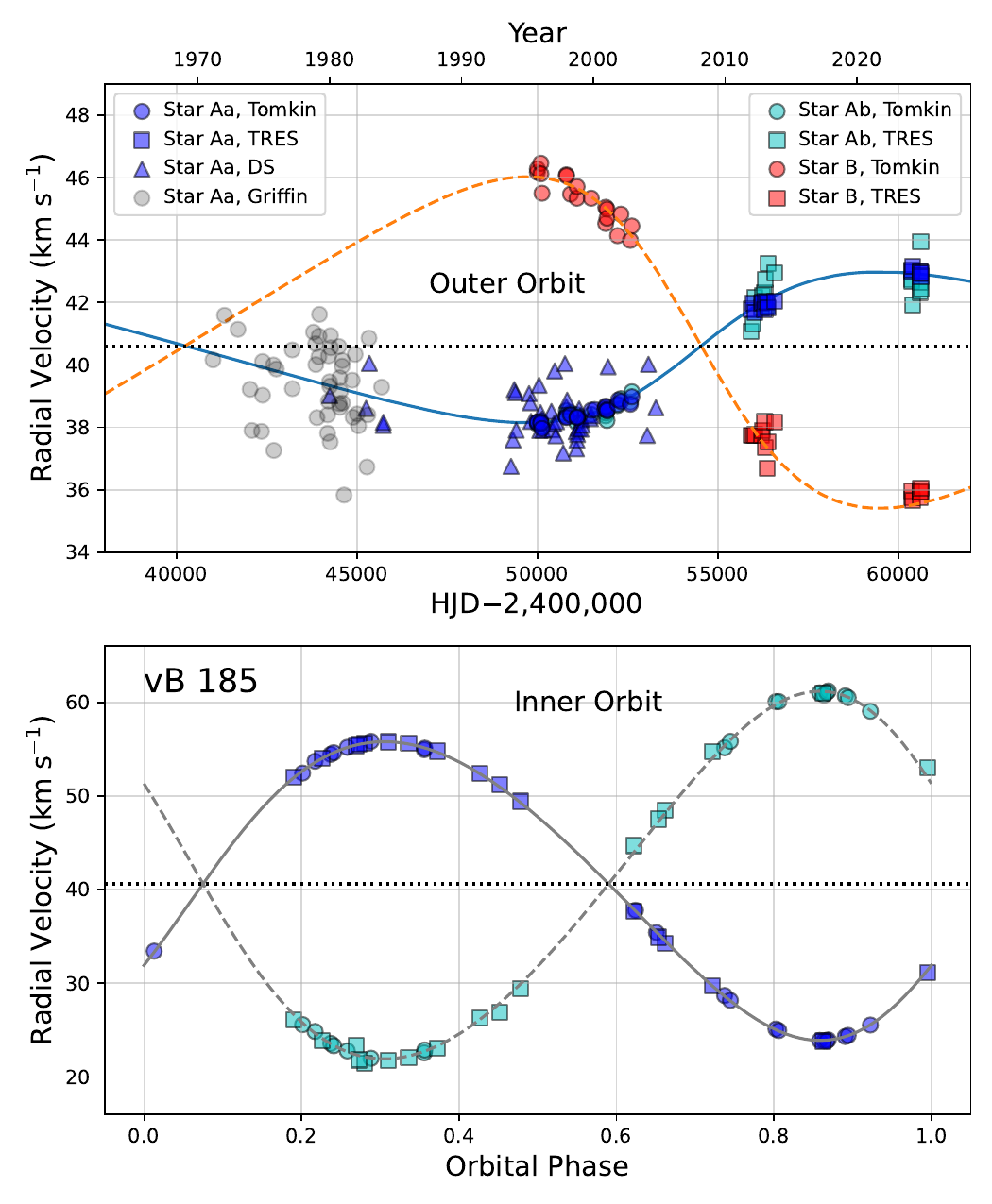}

  \figcaption{\emph{Top:} Outer spectroscopic orbit for vB\,185, with
    the RV measurements from \cite{Tomkin:2003}, \cite{Griffin:1985a},
    and CfA (Digital Speedometers and TRES). Motion in the inner
    binary has been subtracted from the measurements of Aa and Ab. For
    the Digital Speedometer and Griffin velocities, this was done
    using the corresponding semiamplitudes $K_{\rm Aa,DS}$ and $K_{\rm
      Aa,Griffin}$; for those from TRES and \cite{Tomkin:2003}, we
    used $K_{\rm Aa}$ (see Table~\ref{tab:L82}). The dotted line
    represents the center-of-mass velocity of the
    triple. \emph{Bottom:} Inner orbit (symbols as in the top
    panel). \label{fig:L82rvs}}

\end{figure}

\begin{figure}
  \epsscale{1.17}
  \hspace*{-1mm}\includegraphics[width=0.48\textwidth]{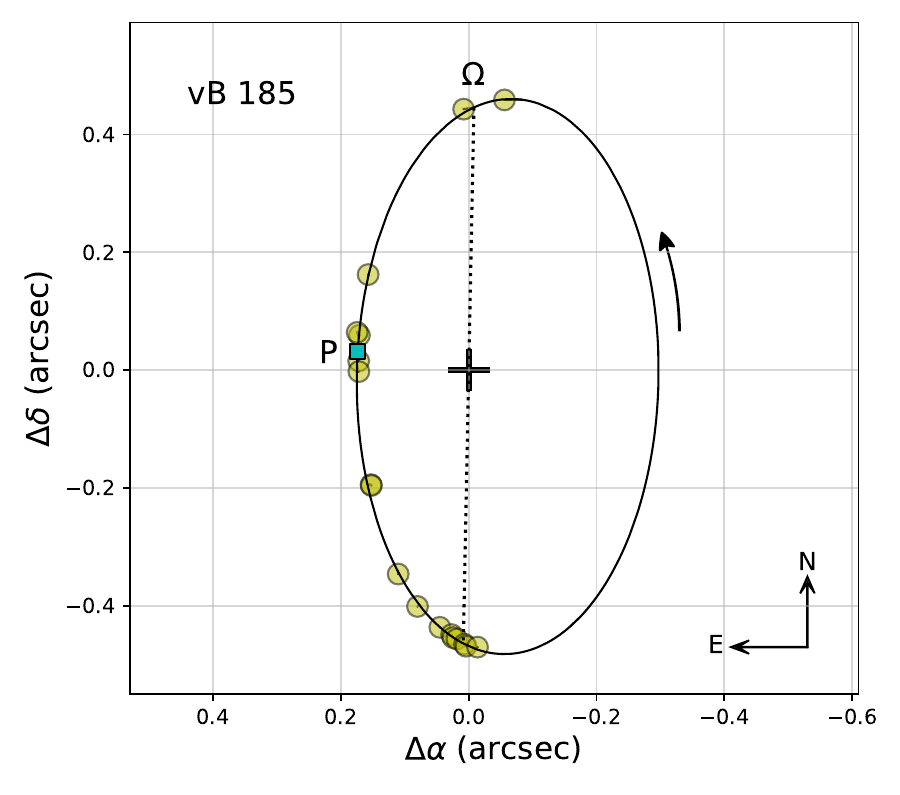}

  \figcaption{Astrometric orbit of vB\,185, with the measurements from
    the WDS. Two lunar occultation observations, which are
    one-dimensional, were not used. The dotted line represents the
    line of nodes, and ``$\Omega$'' marks the ascending node.
    Periastron is indicated with the square labeled
    ``P''. \label{fig:L82sky}}

\end{figure}

The period of the outer orbit (79.2~yr) is 1.5\% longer than the one
reported by \cite{Tokovinin:2021b}. Our position angle for the
ascending node in the outer orbit ($\Omega_{\rm AB}$) and the argument
of periastron for the tertiary ($\omega_{\rm B}$) are in the opposite
quadrant compared to his, as his solution did not make use of radial
velocities to remove the 180\arcdeg\ ambiguity. Other elements show
minor differences. And as anticipated, the primary velocity
semiamplitudes from the Griffin and Digital Speedometer velocities
($14.19 \pm 0.22$ and $15.00 \pm 0.16~\kms$, respectively) are
significantly smaller than the value constrained by the Tomkin and
TRES observations, which resolve all three stars ($K_{\rm Aa} = 15.950
\pm 0.027~\kms$).  The inclination angle for the inner orbit is
inferred to be $79\arcdeg \pm 3\arcdeg$.  We show the inner and outer
spectroscopic orbits in Figure~\ref{fig:L82rvs}, along with our model,
and the astrometric orbit in Figure~\ref{fig:L82sky}.

Our dynamical mass measurements for vB\,185, coupled with estimates of
the intrinsic brightness of each component, offer a chance to compare
them against other similar measurements in the Hyades, and with models
of stellar evolution. We estimated the apparent visual magnitude of
each of the three stars by converting our spectroscopic flux ratios at
$\sim$5187~\AA\ to the visual band, using PHOENIX spectra from
\cite{Husser:2013}, and adopting a total apparent brightness for the
system of $V = 9.472 \pm
0.004$.
The distance modulus was based on the distance in Table~\ref{tab:L82}.
We obtained absolute visual magnitudes for stars Aa, Ab, and B of
$6.524 \pm 0.027$, $8.229 \pm 0.068$, and $8.129 \pm 0.073$,
respectively.\footnote{These values imply a magnitude difference
  between the tertiary and the inner binary of $\Delta V = 1.81 \pm
  0.08$. This is close to the average of the three available direct
  measurements of this difference from the WDS, at wavelengths near
  the $V$ band \citep[$1.84 \pm 0.07$;][]{ESA:1997, Horch:2010}.}  As
seen in Figure~\ref{fig:mlrelation}, although the slope from the
measurements for vB\,185 is consistent with the trend from other
binaries in the Hyades and from the models, all three components fall
slightly above the isochrones. This is likely due to a bias in our
masses, caused by the challenges noted earlier.
A more definitive comparison with theory must await a more complete
coverage of the outer orbit from both astrometric and spectroscopic
observations.

\subsection*{(545) vB 149}  
\label{sec:VB149}

This is a close visual pair of nearly equally bright components
\citep[WDS~04574+0100, HD~31622, $\rho \approx
  0\farcs3$;][]{Aitken:1914}.  The binary nature of the object has
compromised the astrometric solution from Gaia~DR3 (2/3 of the image
profiles were found to be double-peaked), and no parallax or proper
motion were reported in the catalog. However, the parallax and proper
motion from the Hipparcos catalog (source HIP\,23044) indicate it is
not a member of the Hyades, and the distance places it well below the
cluster main sequence.

A handful of early RV measurements of a blend of the two components
made between 1973 and 1992 (\citealt{Griffin:1988};
CORAVEL\footnote{\url{https://webda.physics.muni.cz/}}) showed no
change with time, and confirmed it as a non-member. Our own
observations between 1992 and 1999 supported those findings. In 2004
the spectra we gathered changed character, and appeared clearly
double-lined for a period of about two months.  Subsequent
observations after 2005 were single-lined again, or showed slight
broadening. This suggested we caught the periastron passage in a
possibly very eccentric binary with a long orbital period, most likely
containing the same companion discovered by the visual observers. 
Analysis with {\tt TODCOR}
allowed us to measure the velocities of both components in all our
spectra, with adopted template parameters of $T_{\rm eff} = 5750$~K for both
stars, along with $v \sin i$ values of 5 and 4~\kms\ for the primary
and secondary components, respectively. The flux ratio we determined is
$\ell_2/\ell_1 = 0.780 \pm 0.036$.
However, the orbital period cannot be determined from the velocities alone.

Fortunately, the measurements of the relative position of vB\,149 going
back to 1913 complement the spectroscopy very well, and allow the
period to be determined reliably. We used a listing of all such
measurements from the WDS, together
with our velocities, to carry out a joint MCMC
astrometric-spectroscopic orbital solution. The resulting elements and
derived properties are presented in Table~\ref{tab:VB149}. 
The orbit is indeed very eccentric ($e = 0.9000 \pm
0.0050$), quite close to edge-on ($i = 98\fdg2 \pm 1\fdg9$), and has a
period of 31.8~yr.

\setlength{\tabcolsep}{6pt}
\begin{deluxetable}{lcc}
\tablewidth{0pc}
\tablecaption{Orbital Parameters for vB\,149 \label{tab:VB149}}
\tablehead{
\colhead{Parameter} &
\colhead{Value} &
\colhead{Prior}
}
\startdata
 $P$ (day)                         & $11597 \pm 61$\phn\phn\phn   & [6000, 20000]   \\
 $T_{\rm peri}$ (BJD$-$2,400,000)  & $53071.9 \pm 4.4$\phm{2222}  & [52000, 54000]  \\
 $a^{\prime\prime}$ (\arcsec)      & $0.1732 \pm 0.0016$          & [0.02, 0.20]    \\
 $\sqrt{e}\cos\omega_1$            & $+0.9437 \pm 0.0037$\phs     & [$-$1, 1]         \\
 $\sqrt{e}\sin\omega_1$            & $-0.096 \pm 0.021$\phs       & [$-$1, 1]         \\
 $\cos i$                          & $-0.142 \pm 0.034$\phs       & [$-$1, 1]         \\
 $\Omega$ (deg)                    & $124.96 \pm 0.55$\phn\phn    & [0, 360]        \\
 $K_1$ (\kms)                      & $13.59 \pm 0.42$\phn         & [1, 30]         \\
 $K_2$ (\kms)                      & $13.93 \pm 0.47$\phn         & [1, 30]         \\
 $\gamma$ (\kms)                   & $+36.907 \pm 0.073$\phn\phs  & [0, 60]         \\[0.5ex]
\hline \\ [-1.5ex]
\multicolumn{3}{c}{Error Scaling Factors} \\ [1ex]
\hline \\ [-1.5ex]
 $f_{\theta}$                      & $1.84 \pm 0.42$              & [$-$5, 5]         \\
 $f_{\rho}$                        & $1.73 \pm 0.40$              & [$-$5, 5]         \\
 $f_1$                             & $1.00 \pm 0.17$              & [$-$5, 5]         \\
 $f_2$                             & $1.01 \pm 0.17$              & [$-$5, 5]         \\[0.5ex]
\hline \\ [-1.5ex]
\multicolumn{3}{c}{Derived Properties} \\ [1ex]
\hline \\ [-1.5ex]
 $i$ (deg)                         & $98.2 \pm 1.9$\phn           & \nodata         \\
 $e$                               & $0.9000 \pm 0.0050$          & \nodata         \\
 $\omega_1$ (deg)                  & $354.2 \pm 1.3$\phn\phn      & \nodata         \\
 $a$ (au)                          & $12.93 \pm 0.36$\phn         & \nodata         \\
 $M_1$ ($M_{\sun}$)                & $1.083 \pm 0.096$            & \nodata         \\
 $M_2$ ($M_{\sun}$)                & $1.058 \pm 0.089$            & \nodata         \\
 $q \equiv M_2/M_1$                & $0.977 \pm 0.037$            & \nodata         \\
 $\pi_{\rm orb}$ (mas)             & $13.41 \pm 0.38$\phn         & \nodata         \\
 Distance (pc)                     & $74.6 \pm 2.2$\phn           & \nodata         
\enddata

\tablecomments{The values listed correspond to the mode of the
  posterior distributions from our MCMC analysis, with uncertainties
  representing the 68.3\% credible intervals. The argument of
  periastron listed ($\omega_1$) is for the primary. The symbols
  $f_{\theta}$ and $f_{\rho}$ represent multiplicative scaling factors
  for the nominal errors of the astrometric measurements, solved for
  simultaneously with the other elements, and $f_1$ and $f_2$ are the
  corresponding factors for the primary and secondary radial
  velocities. Priors in square brackets are uniform over the ranges
  specified, except those for the error inflation factors $f$, which
  are log-uniform. }

\end{deluxetable}
\setlength{\tabcolsep}{6pt}

\begin{figure}
  \epsscale{1.17}
  \hspace*{-1mm}\includegraphics[width=0.48\textwidth]{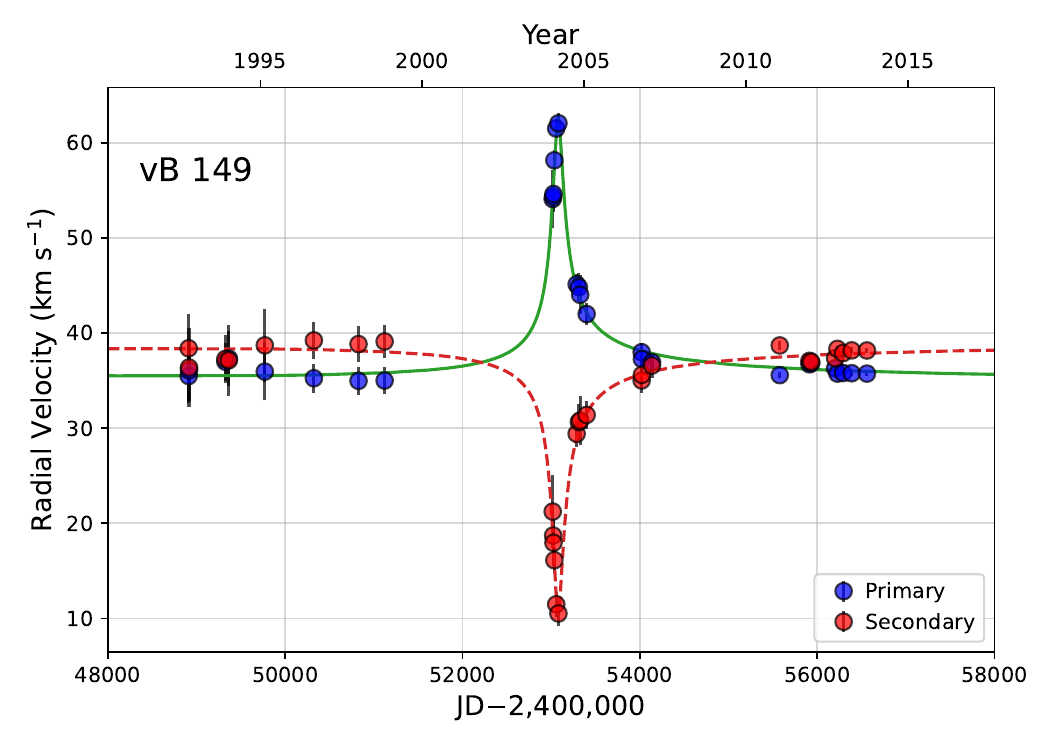}

  \figcaption{Radial velocity measurements of vB\,149, with our model
    for the spectroscopic orbit. \label{fig:VB149rvs}}

\end{figure}

\begin{figure}
  \begin{center}
  \begin{tabular}{c}
  \hspace*{-5ex}\includegraphics[width=0.45\textwidth]{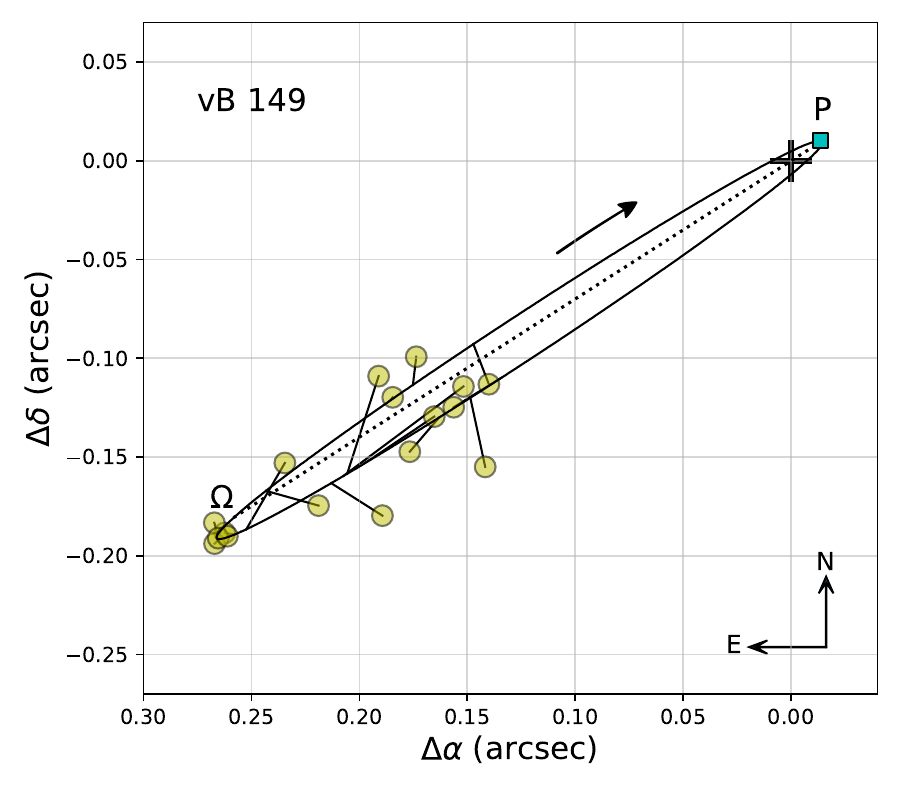} \\
  \includegraphics[width=0.45\textwidth]{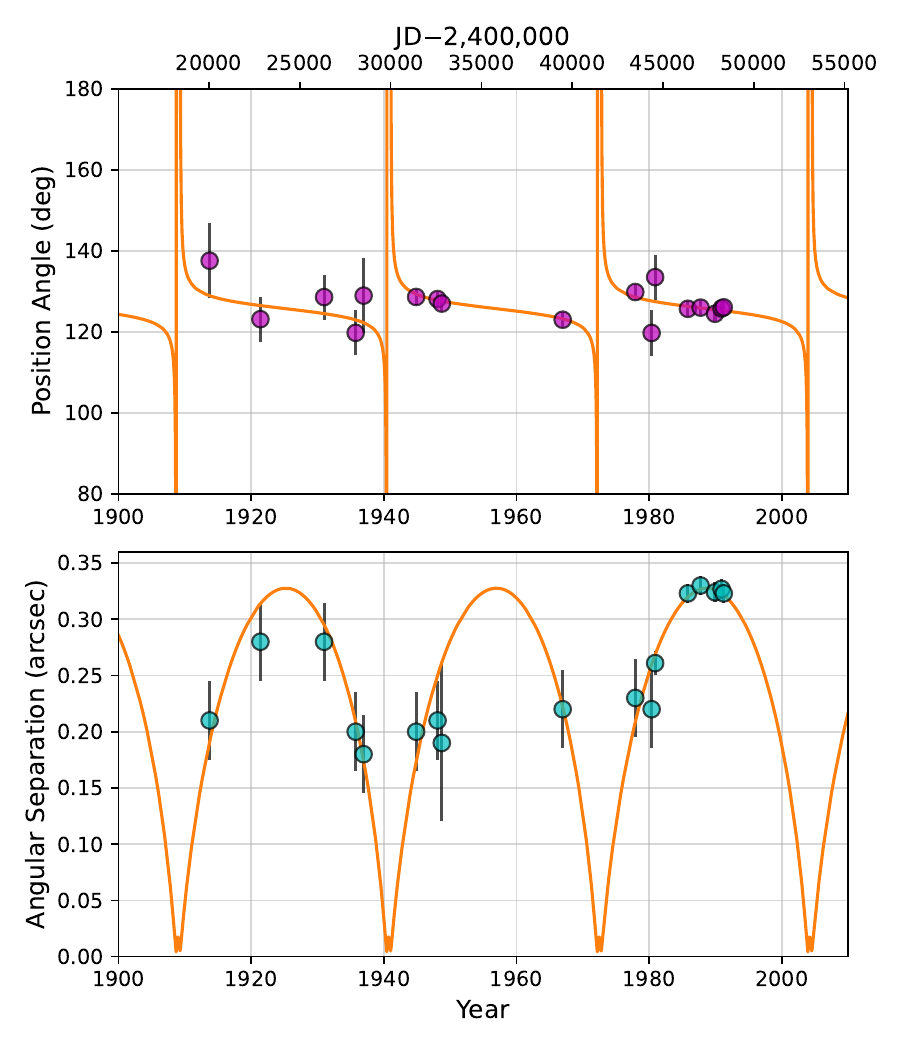}
  \end{tabular}
  \caption{\emph{Top:} Measured positions of the secondary of
    vB\,149 relative to the primary (plus sign) on the plane of the
    sky.  Periastron is indicated with a square labeled ``P'', and the
    dotted line represents the line of nodes. The letter ``$\Omega$''
    marks the ascending node, at which the secondary is receding from
    the observer. \emph{Middle and bottom:} Position angles and
    separations as a function of time, shown with our
    model. \label{fig:VB149sky}}
    \end{center}
\end{figure}

Figure~\ref{fig:VB149rvs} displays the spectroscopic model with our
velocities. The top panel of Figure~\ref{fig:VB149sky} shows that,
because of the high eccentricity, the binary has only
been spatially resolved near its maximum angular separation. The lower
panels display the measurements as a function of time, which cover
about two and a half orbital cycles.

The original Hipparcos catalog \citep{ESA:1997} reported a parallax of
$12.62 \pm 1.89$~mas. The revised version of the catalog
\citep{vanLeeuwen:2007} gave $12.27 \pm 1.76$~mas. The orbital
parallax we obtain here, $13.41 \pm 0.38$~mas, is several times more
precise, but is consistent with the earlier estimates within their
larger uncertainties. The corresponding distance is $74.6 \pm 2.2$~pc,
far behind the cluster. Both components are of solar type, with masses
marginally larger than the Sun.

Because of the very elongated shape of its orbit, vB\,149 is relatively
rare among the known visual binaries, as is the system vB\,80 discussed
earlier ($e = 0.915$). Only about 5.5\% of all systems
in the Sixth Catalog of Orbits of Visual Binary Stars
\citep{Hartkopf:2001} have eccentricities as high as 0.9 or higher.
While the maximum separation between the components of vB\,149 is about
24.6~au, at periastron they come within just 1.3~au of each other.

\vskip 20pt
\subsection*{(585) HD 34031}  
\label{sec:HIP024480}

Our 9 RVs (one from the year 2000, and the others from 2012--2014)
show a clear variation (see Table~\ref{tab:RVsb1}), but are not sufficient to establish an orbit.
By adding 4 measurements from the public Elodie
archive\footnote{\url{http://atlas.obs-hp.fr/elodie/}} (2004--2005) and 7
measurements from the public Sophie
archive\footnote{\url{http://atlas.obs-hp.fr/sophie/}} (2009--2010), we were able
to establish that this is a triple system. We adopted individual RV uncertainties
for these two data sets of 0.20 and 0.12~\kms, respectively. The
inner orbital period is 78.5\,d, and the outer period appears to be
850\,d, although we consider the latter to be somewhat tentative at the
present time, given the small number of observations. The elements of
our preliminary orbital solution that solves for both orbits
simultaneously are given in Table~\ref{tab:HIP024480}. The
center-of-mass velocity clearly indicates HD~34031 is not a member of
the Hyades.

\setlength{\tabcolsep}{2pt}
\begin{deluxetable}{lcc}
\tablewidth{0pc}
\tablecaption{Orbital Parameters for HD\,34031\label{tab:HIP024480}}
\tablehead{
\colhead{Parameter} &
\colhead{Value} &
\colhead{Prior}
}
\startdata
 $P_{\rm A}$ (day)                        & $78.5400 \pm 0.0058$\phn       & [50, 100]       \\
 $T_{\rm peri,A}$ (BJD$-$2,400,000)       & $55114.91 \pm 0.70$\phm{2222}  & [55100, 55170]  \\
 $\sqrt{e_{\rm A}}\cos\omega_{\rm Ab}$    & $+0.234 \pm 0.011$\phs         & [$-1$, 1]         \\
 $\sqrt{e_{\rm A}}\sin\omega_{\rm Ab}$    & $+0.141 \pm 0.013$\phs         & [$-1$, 1]         \\
 $K_{\rm Aa}$ (\kms)                      & $11.867 \pm 0.047$\phn         & [5, 20]         \\
 $P_{\rm AB}$ (day)                       & $848.9 \pm 1.9$\phn\phn        & [300, 1000]     \\
 $T_{\rm peri,AB}$ (BJD$-$2,400,000)      & $54721.9 \pm 4.7$\phm{2222}    & [54500, 55300]  \\
 $\sqrt{e_{\rm AB}}\cos\omega_{\rm A}$    & $+0.476 \pm 0.019$\phs         & [$-1$, 1]         \\
 $\sqrt{e_{\rm AB}}\sin\omega_{\rm A}$    & $+0.400 \pm 0.035$\phs         & [$-$1, 1]         \\
 $K_{\rm A}$ (\kms)                       & $3.721 \pm 0.077$              & [1, 5]          \\
 $\gamma$ (\kms)                          & $+13.474 \pm 0.082$\phn\phs    & [0, 20]         \\
 $\Delta$RV (\kms)                        & $+0.703 \pm 0.065$\phs         & [$-$5, 5]         \\[0.5ex]
\hline \\ [-1.5ex]
\multicolumn{3}{c}{Derived Properties} \\ [1ex]
\hline \\ [-1.5ex]
 $e_{\rm A}$                              & $0.0750 \pm 0.0043$            & \nodata         \\
 $\omega_{\rm Aa}$ (deg)                  & $31.0 \pm 3.2$\phn             & \nodata         \\
 $a_{\rm Aa} \sin i$ ($10^6$ km)          & $12.782 \pm 0.051$\phn         & \nodata         \\
 $M_{\rm Ab} \sin i /(M_{\rm Aa} + M_{\rm Ab})^{2/3}$ & $0.23806 \pm 0.00094$  & \nodata     \\
 $e_{\rm AB}$                             & $0.386 \pm 0.027$              & \nodata         \\
 $\omega_{\rm A}$ (deg)                   & $39.5 \pm 3.1$\phn             & \nodata         \\
 $a_{\rm A} \sin i$ ($10^6$ km)           & $40.21 \pm 0.56$\phn           & \nodata         \\
 $M_{\rm B} \sin i /(M_{\rm A} + M_{\rm B})^{2/3}$ & $0.1531 \pm 0.0021$   & \nodata         
\enddata

\tablecomments{The values listed correspond to the mode of the
  posterior distributions from our MCMC analysis, with uncertainties
  representing the 68.3\% credible intervals. $\Delta$RV represents a
  systematic velocity offset between the Sophie and Elodie
  measurements and our own.  Priors in square brackets in the last
  column are uniform over the ranges specified, except those for the
  error inflation factors, which are log-uniform. Light travel time
  effects have been accounted for in the solution.}

\end{deluxetable}
\setlength{\tabcolsep}{6pt}

\begin{figure}
  \epsscale{1.17}
  \hspace*{-1mm}\includegraphics[width=0.48\textwidth]{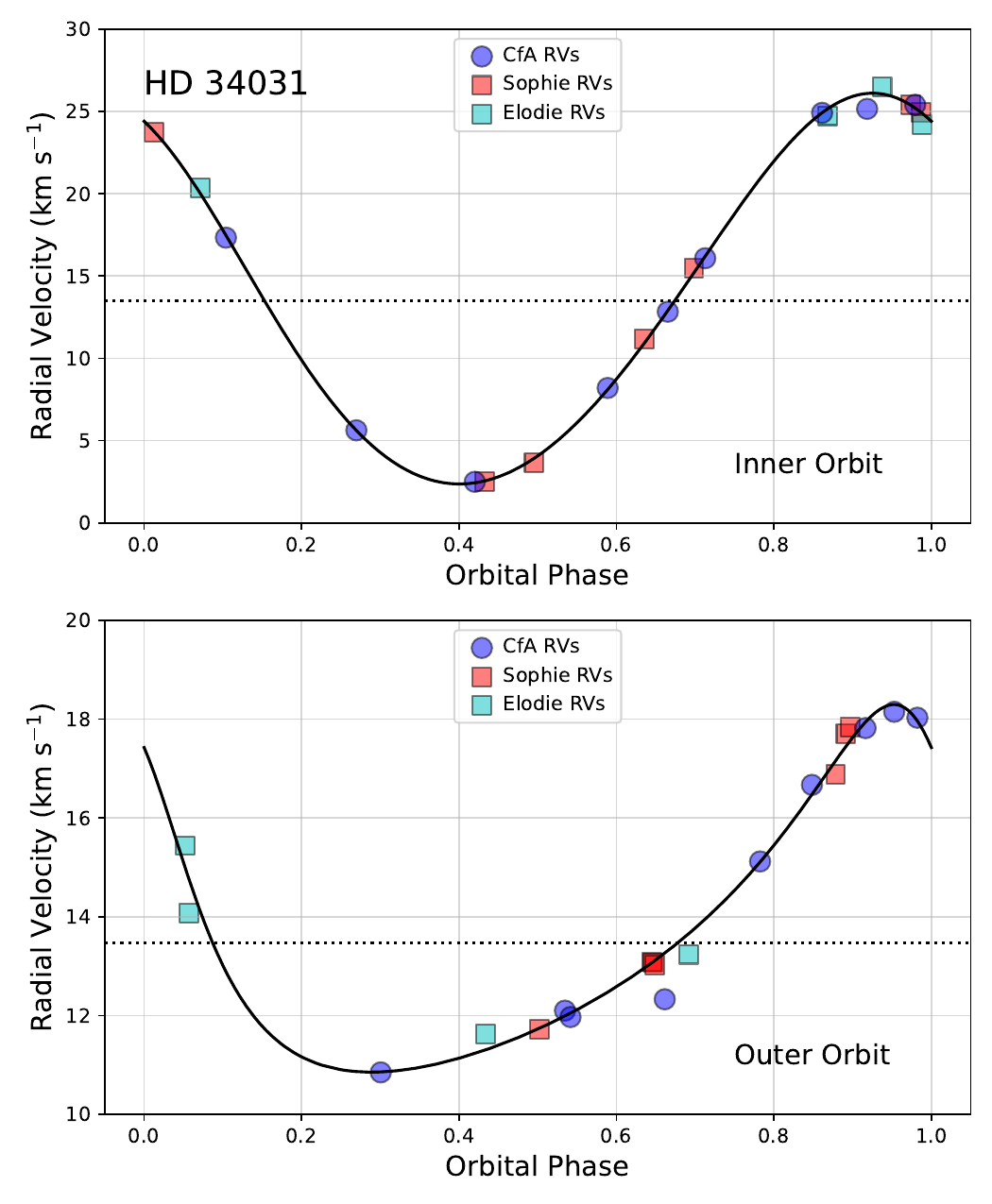}

  \figcaption{RV measurements of HD\,34031, together with our models
    for the inner and outer orbits. In each case, the motion in the
    other orbit has been subtracted for display purposes. The dotted
    line represents the center-of-mass
    velocity. \label{fig:HIP024480rv}}

\end{figure}

The ratio of the outer and inner periods is among the smallest known
($\sim$10.8).  An alternative outer period of about 480\,d is also possible,
and would imply a period ratio of only $\sim$6, although we
consider it less likely.  That solution gives an rms residual
somewhat larger than the model we favor. Plots of our preferred model
with the observations are shown in Figure~\ref{fig:HIP024480rv}.

\section*{Appendix B: Notes}
\label{sec:appendixC}

In the following we include additional information of interest for
selected targets in our sample. In more than two dozen cases, an
object in our survey forms a wide physical pair with another survey target.
We list all such cases with angular separations up to 30\arcsec\ in
Table~\ref{tab:pairs}. About half of these pairs are considered
members of the cluster. In several instances one or the other component,
or both, are themselves binaries, implying a higher multiplicity system.
Other targets have even wider companions listed in the WDS,
which we do not include here.

\setlength{\tabcolsep}{5pt}
\begin{deluxetable*}{llrrrcll}
\tablewidth{0pc}
\tablecaption{Physical Pairs With Separations under 30\arcsec\ Among the Survey Targets\label{tab:pairs}}
\tablehead{
\colhead{Visual Primary} &
\colhead{Visual Secondary} &
\colhead{$G_{\rm prim}$} &
\colhead{$G_{\rm sec}$} &
\colhead{$\rho$} &
\colhead{Memb.} &
\colhead{Prim. Binarity} &
\colhead{Sec. Binarity}
\\
\colhead{} &
\colhead{} &
\colhead{(mag)} &
\colhead{(mag)} &
\colhead{($\arcsec$)} &
\colhead{} &
\colhead{} &
\colhead{}
}
\startdata
(5)     HD 15128A       &   (6)     HD 15128B        &  7.77  &  9.31  &  1.41   &  NM  & DUP                    &  DUP \\
(10)    HD 17383        &   (9)     HIP 13042        &  8.95  &  10.31 &  23.88  &  NM  & DUP                    &  \nodata \\
(11)    G 75-42A        &   (12)    G 75-42B         &  11.39 &  11.99 &  1.30   &  NM  & VAR,DUP,RUWE           &  DUP,RUWE \\
(27)    BD+25 522A      &   (26)    BD+25 522B       &  10.57 &  12.42 &  3.23   &  M   & RUWE                   &  \nodata \\
(32)    HIP 15368A      &   (31)    HIP 15368B       &  11.56 &  12.59 &  0.89   &  NM  & DUP,RUWE               &  DUP \\
(40)    vB 3            &   (41)    vB 207           &  8.13  &  10.12 &  19.91  &  NM  & SB1,DUP,RUWE,VBO       &  \nodata \\
(48)    LP 413-18       &   (49)    LP 413-19        &  11.69 &  12.08 &  16.15  &  M   & SB2,DUP,RUWE           &  SB2,RUWE \\
(73)    HD 24098A       &   (72)    HD 24098B        &  6.40  &  10.08 &  4.59   &  M   & \nodata                &  \nodata \\
(88)    LP 474-428 A    &   (89)    LP 474-428 B     &  13.13 &  13.40 &  2.21   &  M   & DUP,RUWE               &  DUP,RUWE \\
(96)    HD 25068A       &   (97)    HD 25068B        &  9.93  &  12.35 &  4.29   &  NM  & RUWE                   &  \nodata \\
(111)   HG 7-73A        &   (112)   HG 7-73B         &  11.33 &  11.75 &  5.50   &  M   & RUWE                   &  \nodata \\
(134)   G 39-1A         &   (133)   G 39-1B          &  7.04  &  8.95  &  2.82   &  NM  & VAR,ACC                &  VAR,DUP \\
(136)   vB 11           &   (135)   vB 12            &  5.98  &  8.68  &  3.58   &  M   & DUP                    &  LT,RUWE \\
(143)   LP 357-168 A    &   (144)   LP 357-168 B     &  12.55 &  12.65 &  1.76   &  M   & DUP                    &  DUP \\
(154)   REID 88         &   (153)   REID 88B         &  12.36 &  13.77 &  5.55   &  M   & VAR,LT,DUP,RUWE        &  \nodata \\
(162)   PELS 20         &   (161)   HD 284163B       &  9.07  &  13.29 &  5.07   &  M   & SB3p,DUP,RUWE,ACC,vbop &  \nodata \\
(250)   HD 27691        &   (249)   HD 27691B        &  7.14  &  8.37  &  1.25   &  M   & SB2,DUP,ACC,vbo        &  DUP \\
(277)   vB 140          &   (276)   vB 140B          &  8.73  &  12.95 &  7.11   &  NM  & SB2,RUWE,ACC,vbo       &  SB2,RUWE \\
(296)   EPIC 210953703  &   (295)   EPIC 210953791   &  10.96 &  11.09 &  5.48   &  NM  & VAR,DUP,RUWE           &  \nodata \\
(301)   vB 56           &   (300)   HD 27962B        &  4.30  &  8.10  &  1.81   &  M   & ACC                    &  LT \\
(345)   HAN 432         &   (346)   HG 7-232B        &  11.36 &  13.23 &  1.67   &  M   & SB1,DUP,RUWE           &  DUP \\
(440)   HD 29205A       &   (439)   HD 29205B        &  8.79  &  9.08  &  2.02   &  NM  & SB1,RUWE               &  \nodata \\
(501)   LP 416-565      &   (502)   LP 416-566       &  12.54 &  13.17 &  3.54   &  M   & RUWE                   &  \nodata \\
(531)   HD 284908       &   (530)   vB 145           &  9.30  &  10.96 &  3.91   &  NM  & VAR,RUWE               &  \nodata \\
(553)   HD 31734A       &   (552)   HD 31734B        &  8.10  &  10.23 &  1.68   &  NM  & DUP                    &  \nodata \\
(577)   vB 131          &   (578)   vB 132           &  5.97  &  8.94  &  11.42  &  M   & VB                     &  DUP,VB \\
(616)   HD 39251A       &   (617)   HD 39251B        &  7.82  &  9.55  &  2.51   &  NM  & SB1                    &  \nodata 
\enddata
\tablecomments{After the SIMBAD names of the brighter and fainter component of each pair, the columns list the Gaia magnitudes,
  the angular separation based on the Gaia~DR3 positions, membership, and the binarity codes as described in Section~\ref{sec:binarity}.}

\end{deluxetable*}
\setlength{\tabcolsep}{6pt}

In several cases the Hipparcos intermediate astrometric data (abscissa
residuals) have sufficient precision to detect the motion of the
center-of-light of SB1 or SB2 systems, when the phase coverage is
reasonably complete. This allows the inclination
angle $i$ to be determined, along with the semimajor axis of the photocenter
($a^{\prime\prime}_{\rm phot}$) and the position angle of the ascending
node ($\Omega$). We report those results in Table~\ref{tab:hipparcos},
from separate solutions we carried out for each system.

\setlength{\tabcolsep}{5pt}
\begin{deluxetable*}{lcccccc}
\tablewidth{0pc}
\tablecaption{Survey Targets With Photocenter Motion Detectable in the Hipparcos Observations\label{tab:hipparcos}}
\tablehead{
\colhead{Target} &
\colhead{Memb.} &
\colhead{SB1/SB2} &
\colhead{$P$ (days)} &
\colhead{$a_{\rm phot}^{\prime\prime}$ (mas)} &
\colhead{$i$ (deg)} &
\colhead{$\Omega$ (deg)}
}
\startdata
(7) G\,36-30      &  NM  &  SB1  &  1228  & $41.2 \pm 7.1$\phn    &  $76.3  \pm 3.2$\phn      &  $49.1 \pm 6.7$\phn        \\ 
(14) HD\,17922    &  M   &  SB1  &   319  & $9.9 \pm 1.3$         &  $116.8 \pm 7.2$\phn\phn  &  $230.8 \pm 5.0$\phn\phn   \\ 
(103) vB\,8       &  M   &  SB1  &   856  & $8.07 \pm 0.72$       &  $89.8 \pm 7.3$\phn       &  $190.6 \pm 9.7$\phn\phn   \\ 
(257) vB\,43      &  M   &  SB2  &   590  & $7.0 \pm 1.8$         &  $21 \pm 69$              &  \phn$6 \pm 11$            \\ 
(343) vB\,68      &  M   &  SB2  &   330  & $9.2 \pm 1.4$         &  $99 \pm 10$              &  $137 \pm 11$\phn          \\ 
(362) vB\,77\tablenotemark{a}      &  M   &  SB1  &   239  & $10.82 \pm 0.98$\phn  &  $18 \pm 38$              &  $ 308.6 \pm 5.8$\phn\phn  \\ 
(378) vB\,82      &  M   &  SB1  &    91  & $1.98 \pm 0.64$       &  $81 \pm 27$              &  \phn$1 \pm 26$            \\ 
(392) vB\,285\tablenotemark{b}     &  M   &  SB1  &   848  & $12.0 \pm 1.5$\phn    &  $17 \pm 44$              &  $218.3 \pm 6.6$\phn\phn   \\ 
(443) vB\,102     &  M   &  SB1  &   734  & $9.4 \pm 1.3$         &  $112.4 \pm 7.5$\phn\phn  &  $119.7 \pm 9.2$\phn\phn   \\ 
(456) HD\,29479   &  NM  &  SB1  &   817  & $10.4 \pm 1.3$\phn    &  $85.0 \pm 4.8$\phn       &  $32.0 \pm 5.6$\phn        \\ 
(506) vB\,115     &  M   &  SB1  &  1203  & $11.3 \pm 1.6$\phn    &  $145 \pm 27$\phn         &  $115.7 \pm 7.6$\phn\phn   \\ 
(556) HD\,32662\tablenotemark{c}   &  NM  &  SB1  &   998  & $15.4 \pm 1.3$\phn    &  $74.9 \pm 4.9$\phn       &  $335.7 \pm 5.5$\phn\phn   \\ 
(569) HD\,240622  &  NM  &  SB1  &   649  & $8.2 \pm 1.4$         &  $64 \pm 12$              &  $32.4 \pm 8.1$\phn        \\ 
(575) vB\,130     &  M   &  SB1  &   156  & $3.03 \pm 0.83$       &  $148 \pm 73$\phn         &  $357 \pm 20$\phn          
\enddata
\tablenotetext{a}{A similar solution was reported by \cite{Jancart:2005}.}
\tablenotetext{b}{An independent astrometric solution at a different wavelength, based on HST observations, was reported by \cite{McArthur:2011}.}
\tablenotetext{c}{The Hipparcos catalog also reported a solution for the photocenter motion, but adopted an incorrect period (668~d) and
assumed a circular orbit.}
\end{deluxetable*}
\setlength{\tabcolsep}{6pt}

\vni{(39) BD+12\,479.} The 6.88-day SB1 orbit we report is considered to be
preliminary because of the small number of observations.

\vni{(50) vB\,5.} The long-term downward velocity trend we observe
is supported by earlier RVs by \cite{Griffin:1988}, and others, as well
as by the high-precision measurements by \cite{Paulson:2004}, who reported
a slope of $-0.1364$~m~s$^{-1}$ per day over a period of about 6~yr. This is
likely due to the 0\farcs8 speckle companion detected by \cite{Patience:1998}.
No additional companions were found in adaptive optics imaging by
\cite{Guenther:2005}.

\vni{(53) BD+05\,526.} This is listed as an eclipsing binary with
a period of 13.25\,d \citep{Prsa:2022}. Our observations show no
significant change in velocity, and no sign of double lines. It has
a 0\farcs6 visual companion.

\vni{(68) LP\,357-4.} RVs for this object were published previously
by \cite{Torres:2021} as part of a survey of the Pleiades region,
and indicated no significant change. New velocities since then
have now revealed this to be an SB1, and a member of the Hyades,
with elevated chromospheric activity.

\vni{(81) HG\,7-33.} \cite{Stauffer:1997} tentatively identified
this object as being double-lined.

\vni{(102) PELS\,9.} This is listed in the literature as an
exoplanet host (Kepler/K2 designation K2-111\,b).
It is not a member of the cluster.

\vni{(120) HD\,25444.} Our RVs (consistent with a few by
\citealt{Tokovinin:2002}) show a slight increase with time.
Much older (1917--1918) RVs from Mount Wilson strengthen the
trend, which is consistent with what is expected in the 496~yr
astrometric orbit reported for the object (Table~\ref{tab:wdsorbits}).
While the current mean RV is 3~\kms\ lower than expected, it
is trending in the right direction for membership. We tentatively
classify the target as `M?'.

\vni{(128) HG\,7-92.} Although the period is well determined, some of
the other orbital elements are still somewhat uncertain because the
observations do not cover the periastron passage.

\vni{(130) vB\,231.} This cluster member has a substellar companion (a hot Jupiter)
reported by \cite{Quinn:2014}. We do not consider the planet in the
present study, which only deals with stellar companions.

\vni{(134/133) G\,39-1\,A/B.} A premature astrometric orbit has been
reported for this $\sim$3\arcsec\ common proper motion pair, with a period
of 590~yr (Table~\ref{tab:wdsorbits}). They are non-members.

\vni{(170) vB\,17.} \cite{Paulson:2004} reported a downward trend in
their RV measurements of about 180~m~s$^{-1}$ over 2000\,d.

\vni{(250/249) HD\,27691\,A/B.} This is a hierarchical triple system 
and a member of the cluster, in which the visual primary is an SB2 and the
tertiary is about 1\farcs2 away. Our orbit for the SB2 is listed in
Table~\ref{tab:elemSB2a}.
A visual orbit for the tertiary with a period of 247~yr has been reported
by \cite{Josties:2021}. We used {\tt TRICOR} to measure the velocities
of all three stars.
RVs for tertiary were also measured separately
by \cite{Griffin:2012},
who suggested a very slight downward drift over 37~yr of coverage,
consistent with being caused by motion in the 247~yr outer orbit.
The inner pair (250) HD\,27691\,A has been resolved interferometrically using the
Center for High Angular Resolution Astronomy (CHARA) array
\citep{Schaefer:2022}, although the details are not yet available.
The orbital elements for that 4\,d orbit are listed in the Sixth Catalog
of Orbits of Visual Binary Stars \citep{Hartkopf:2001}, and include the
inclination angle ($48\fdg31 \pm 0\fdg30$) and the semimajor axis
($1.266 \pm 0.003$~mas). Combining this information with our
spectroscopic results, we obtain absolute masses of $1.089 \pm 0.065~M_{\sun}$ and
$0.557 \pm 0.020~M_{\sun}$, currently limited by the spectroscopy.
The orbital parallax is $21.75 \pm 0.38$~mas.

\vni{(263) HAN\,296.} We see no change in the velocities.
\cite{Stauffer:1997} reported the detection
of double lines. They used the designation vA\,288.

\vni{(270) vB\,50.} A preliminary astrometric orbit has been reported by
\cite{Zirm:2008} with a period of about 112~yr and a semimajor axis of
$0\farcs73$. This is the same orbit as the spectroscopic one, which has
a large eccentricity ($e = 0.98$) and a still rather poorly determined
period. A solution combining RVs and available astrometry is not entirely
satisfactory. The RV measurements are concentrated around the periastron passage.

\vni{(275) vB\,52.} This object has a close companion at about 1\arcsec\
that we do not resolve. \cite{Paulson:2004} reported a downward trend in
the velocities of 33.3~m~s$^{-1}$ per 1000\,d, likely related to this
companion. Their high-precision RVs show a superimposed variation of unknown origin
with a much shorter period, and a very small amplitude of a few tens
of m~s$^{-1}$.

\vni{(295) EPIC\,210953791.} This target shows double lines. Our 5 
pairs of RV measurements
are listed in Table~\ref{tab:RVsb2},
and used templates with temperatures of 5750~K for both stars
along with $v \sin i$ values of 4 and 2~\kms\ for the primary
and secondary. The measured flux ratio is $\ell_2/\ell_1 = 0.77 \pm 0.05$.
Using the \cite{Wilson:1941} method,
we obtain a mass ratio of $q = 0.98 \pm 0.03$ and a
center-of-mass velocity of $\gamma = 54.4 \pm 0.2~\kms$.
This velocity is consistent with the mean RV of another of
our targets, (296) EPIC\,210953703,
which is 5\farcs5 away. Gaia reports similar parallaxes and proper motions,
suggesting they form a physical pair. Both are non-members.

\vni{(301) vB\,56.} This is a blue straggler. The astrometric acceleration
reported by Gaia is likely due to the 1\farcs8 visual companion, which
shows long-term changes in its RV.

\vni{(316) vB\,62.} Gaia\,DR3 reported a faint companion to this SB1
($\Delta G = 7.8$~mag) 11\farcs1 away with the same parallax and proper motion, suggesting physical association. This would imply the system is triple.
The inner binary has been resolved with the CHARA array, and has an orbital
inclination angle of $30\fdg9 \pm 2\fdg1$.

\vni{(332) TYC\,3333-1738-1.} The 3 spectra we obtained are double-lined;
the RVs from {\tt TODCOR} are listed in Table~\ref{tab:RVsb2}, and used 
templates with a temperature of 5750~K and no rotation for both stars.
The measured flux ratio is $0.785 \pm 0.013$ at $\sim$5187~\AA.
We estimate a mass ratio of $q = 0.971 \pm 0.001$ and a center-of-mass velocity of
$10.28 \pm 0.01~\kms$. This target is not a member of the Hyades.

\vni{(343) vB\,68.} This SB2 has been spatially resolved using the CHARA array,
yielding an inclination angle of $114\fdg2 \pm 3\fdg3$ and a semimajor axis
of $26.9 \pm 1.6$~mas. With this information, we obtained absolute masses for
the components of $1.52 \pm 0.14~M_{\sun}$ and $0.884 \pm 0.087~M_{\sun}$
(limited in this case by the astrometry), and an orbital parallax of
$21.5 \pm 1.4$~mas. The inclination angle and position angle of the ascending
node, as determined using the Hipparcos intermediate data
(Table~\ref{tab:hipparcos}), are consistent
with the interferometric values.

\vni{(345/346) HAN\,432/HG\,7-232B.} The primary of this wide pair listed
in Table~\ref{tab:pairs} (HAN\,432) has a close 0\farcs14 companion half a
magnitude fainter, reported by \cite{Morzinski:2011}.

\vni{(348) vB\,70.} This is $\epsilon$~Tau, one of the four giants in
the Hyades. A planetary companion has been reported by \cite{Sato:2007}.

\vni{(355) HD\,28271.} We have a single RV for this non-member.
\cite{Tokovinin:2001} and \cite{Willmarth:2016} report very similar
SB1 orbits with a period of 461\,d.

\vni{(362) vB\,77.} This SB1 was resolved with the CHARA array, giving
an inclination angle of $16\fdg5 \pm 1\fdg3$. Hipparcos detected the
wobble of the center of light, and the inferred inclination angle and
position angle of the line of nodes (Table~\ref{tab:hipparcos})
are in good agreement with the CHARA values, although they are less precise.

\vni{(369) LP\,358-348.} Three transiting planets have been discovered
around this star, based on photometry from the Kepler/K2 mission
\citep{Mann:2018}. The system is known in the exoplanet field as K2-136.

\vni{(392) vB\,285.} An astrometric orbit corresponding to the motion of
the photocenter of this SB1 was reported by \cite{McArthur:2011}, based
on measurements with the Hubble Space Telescope. They obtained a
semimajor axis of $14.58 \pm 0.24$~mas and an inclination angle of
$134\fdg1 \pm 0\fdg9$. The Hipparcos observations also reveal this
motion, and yield a somewhat similar semimajor axis (at a different
wavelength) of $12.0 \pm 1.5$~mas.

\vni{(401) vB\,288.} \cite{Stauffer:1997} detected and measured the RVs
of the secondary of this SB1 on two occasions. They used the designation
vA\,677.

\vni{(423) HAN\,601.} A 0\farcs08 companion interior to the one
listed by Gaia was reported by \cite{Morzinski:2011}, and is about 1.8~mag
fainter in the $H$ band.

\vni{(429) HAN\,611.} \cite{Tokovinin:2006} reported a period for the
outer orbit of this triple system of 406~yr, although it does not
appear that a full orbital solution is available. The inner binary is an
SB2, and the system is not a member of the Hyades.

\vni{(430) vB\,184.} Membership of this object in the cluster had long been considered
uncertain, until RVs by \cite{Griffin:1985b} that predate ours showed
an average that matched the expected velocity quite precisely, supporting
membership \citep[see also][]{Cudworth:1985}.
However, our measurements since then have shown a consistent
upward trend \emph{away} from the expected velocity (4~\kms\ too positive
as of 2013).
A visual companion that has been followed since 1888 has shown little change in 
position angle, but a significant decrease in angular separation
from 2\farcs3 to 0\farcs1 in 2013. This suggests a highly inclined orbit
possibly approaching periastron around 2013, which might explain the
velocity changes we see. Three measurements between 2015 and 2017 from
Gaia\,DR2 and the GALAH survey are 1~\kms\ lower, and are consistent with
that scenario. On this basis, we have listed the object as `M?'.

\vni{(440/439) HD\,29205\,A/B.} The components of this 2\arcsec\
visual binary were observed separately on nights of good seeing, and
in combined light on other nights. For the latter spectra we used
{\tt TODCOR} to measure the individual velocities. Component A is an SB1,
and both A and B are non-members of the Hyades. A premature visual orbit
based on very little phase coverage has been reported, with a period of 965~yr.

\vni{(443) vB\,102.} An astrometric orbit for the outer companion of this
triple system was reported by \cite{Tokovinin:2021c}. See also
Table~\ref{tab:triples} and Section~\ref{sec:qdistrib}. It is a cluster
member, and the inner pair is an SB1.

\vni{(448) vB\,104.} The secondary in this SB1 has been resolved
interferometrically on two occasions at a separation of about 10~mas,
and the angular diameter of the primary has also been measured \citep{DeFurio:2022}.
We tentatively detect the secondary in our spectra with {\tt TODCOR},
but only in about half of the exposures.

\vni{(490) vB\,142.} It has been claimed \citep[e.g.,][]{Diaz:2012}
that the 989\,d secondary in this SB1 is likely substellar, although this
has not been confirmed.

\vni{(520) vB\,122.} A visual orbit based on Hipparcos data has been
reported by \cite{Soderhjelm:1999}, with a period of 16.28~yr, a semimajor
axis for the relative orbit of 0\farcs188, and $e =  0.46$. The Gaia\,DR3
catalog reports the companion was partially resolved in 21\% of the
scans. Our RVs, and a few earlier ones from \cite{Griffin:1988}, show no
change and are measured from the blended lines of the components.

\vni{(550) HD\,31768.} \cite{Tokovinin:2014} reported a period estimate
of 103~yr for the visual companion.

\vni{(578) vB\,132.} Several astrometric orbits for the close visual
companion of this target have been reported.
The latest one, by \cite{Docobo:2020}, gives an orbital
period of 34.38~yr and a semimajor axis of 0\farcs282. vB\,132 is
a common proper motion companion of (577) vB\,131.

\vni{(615) HD\,39275.} The 4 spectra we collected are double-lined, and yield
a mass ratio of $0.98 \pm 0.16$ and a center-of-mass velocity of
$16.5 \pm 2.2~\kms$, which indicates the system is not a member of the
Hyades. The individual velocities are listed in Table~\ref{tab:RVsb2},
and were obtained with adopted temperatures for the templates of 5500 and 5250~K
for the primary and secondary, and line broadenings of 12 and 10~\kms,
respectively. The measured flux ratio is $0.406 \pm 0.020$.

\vni{(616/617) HD\,39251\,A/B.} The components of this 2\farcs5 visual
binary were observed separately on a few nights, and on other nights
only the combined light could be observed. For the latter we used
{\tt TODCOR} to disentangle them. Star A is an SB1, and the Gaia
parallaxes and proper motions, as well as the mean RVs we measured,
indicate that star B is physically associated.
They are not members of the cluster.

\acknowledgments

The spectra on which this work was based were gathered by a large
number of colleagues, students, and remote observers over the past 45
years. We thank them all. We are also grateful to R.\ J.\ Davis and
J.\ Mink for maintaining the Digital Speedometer and TRES databases,
and for help with the reduction software.
The anonymous referee is thanked for a careful reading of the
manuscript, and for many constructive suggestions.
We acknowledge the
long-term support from the Smithsonian Astrophysical Observatory that
enabled this study, in the form of telescope time and instrumentation.

The research has made extensive use of the SIMBAD and VizieR
databases, operated at the CDS, Strasbourg, France, and of NASA's
Astrophysics Data System Abstract Service. We also acknowledge the use
of the Fourth Catalog of Interferometric Measurements of Binary Stars
\citep[][and online updates]{Hartkopf:2001}, and of the Washington
Double Star Catalog (WDS), maintained at the U.S. Naval Observatory.
R.\ Matson (USNO) was helpful in providing individual astrometric
measurements from the WDS.
The work has used data from the European Space Agency (ESA) mission {\it
  Gaia} (\url{https://www.cosmos.esa.int/gaia}), processed by the {\it
  Gaia} Data Processing and Analysis Consortium (DPAC,
\url{https://www.cosmos.esa.int/web/gaia/dpac/consortium}). Funding
for the DPAC has been provided by national institutions, in particular
the institutions participating in the {\it Gaia} Multilateral
Agreement.
The computational resources used for this research include the
Smithsonian High Performance Cluster (SI/HPC), Smithsonian Institution
(\url{https://doi.org/10.25572/SIHPC}).

\section*{Appendix C: Orbit Figures}
\label{sec:appendixA}

We include here plots of the RV observations and models for 80 single-lined
spectroscopic binaries, of which 30 are new, and 20 double-lined
binaries, of which 9 are new. Additional orbits are presented in
Appendix~B for 5 binaries and 5 triples. Among these, a total of 7
inner or outer spectroscopic orbits are new, along with several of
the astrometric orbits.

In addition to these systems, to our knowledge there are four
binary or multiple systems in the
Hyades with known spectroscopic orbits that host a white dwarf.
Three are included in our survey: the eclipsing binary V471\,Tau, HAN\,346
(a quadruple system),
and vB\,34 (in which the white dwarf is the tertiary, and has a 184~yr
astrometric orbit by \citealt{Zhang:2023}). The fourth white dwarf
system is known as HZ\,9 \citep{Lanning:1981}, and is not in our survey.

\begin{figure*}
  \epsscale{1.12}
  \hspace*{-1mm}\includegraphics[width=\textwidth]{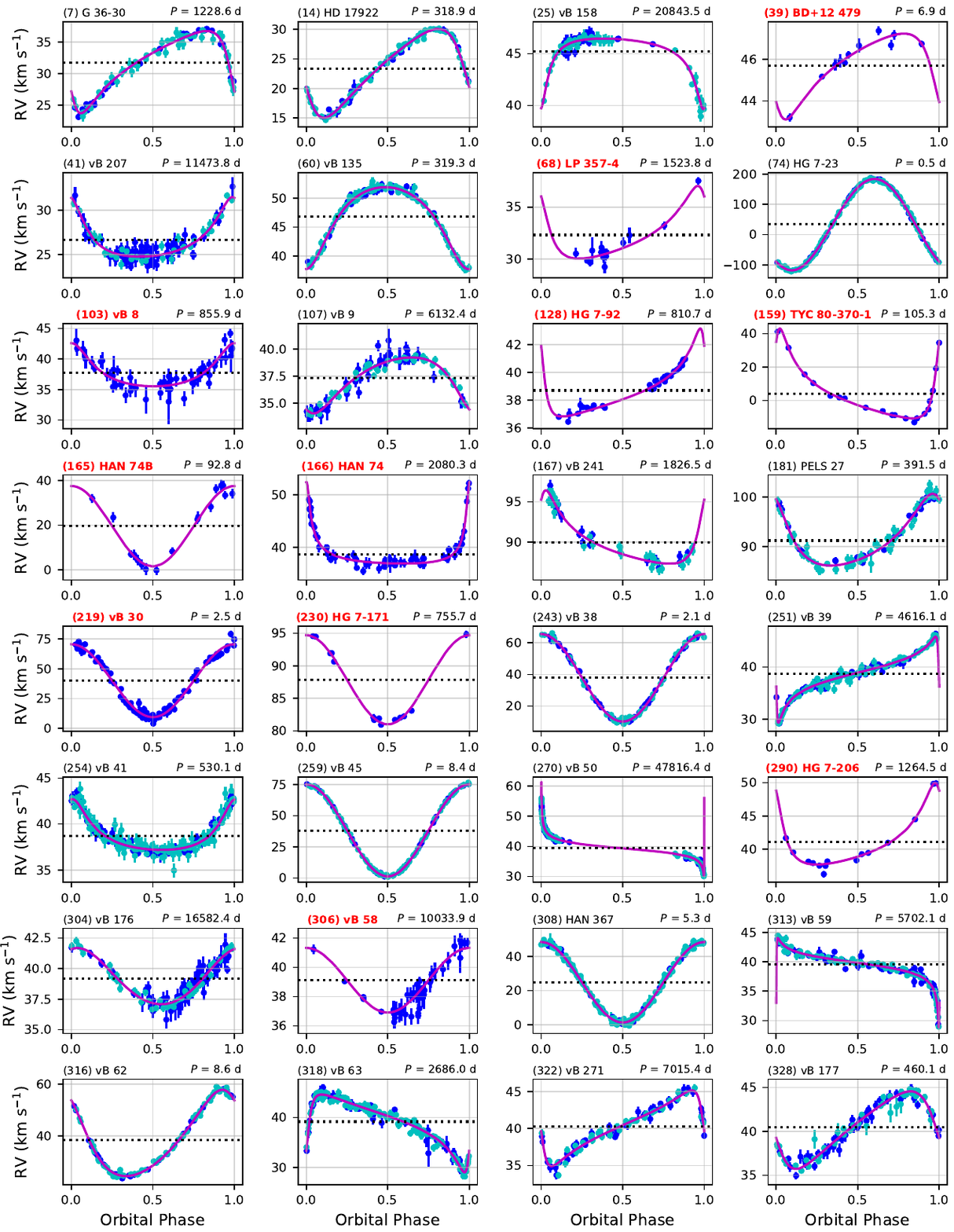}

  \figcaption{Single-lined spectroscopic orbits. Running numbers, SIMBAD names,
  and orbital periods are shown on the title line of each panel. 
  For new orbits, the running numbers and SIMBAD names are shown in red.
  Blue symbols correspond
  to the CfA observations, and cyan symbols to RVs by others (see Table~\ref{tab:elemSB1}). Our best-fit models are shown by the magenta lines, and the center-of-mass velocity of each system is indicated with a dotted line.\label{fig:sb1orbits1}}
\end{figure*}

\addtocounter{figure}{-1}
\begin{figure*}
  \epsscale{1.17}
  \hspace*{-1mm}\includegraphics[width=\textwidth]{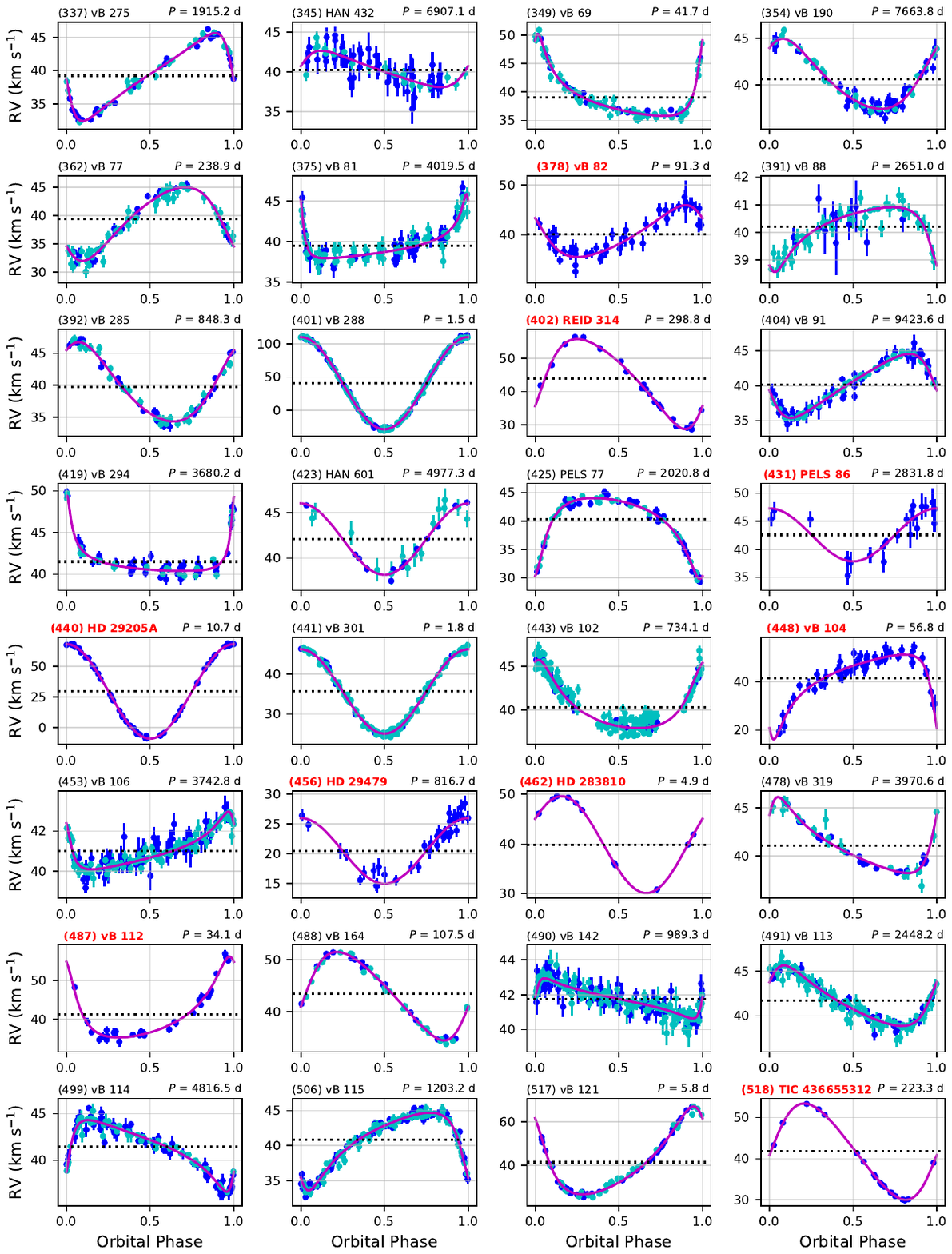}

  \figcaption{Single-lined spectroscopic orbits (continued).}
\end{figure*}

\addtocounter{figure}{-1}
\begin{figure*}
  \epsscale{1.17}
  \hspace*{-1mm}\includegraphics[width=\textwidth]{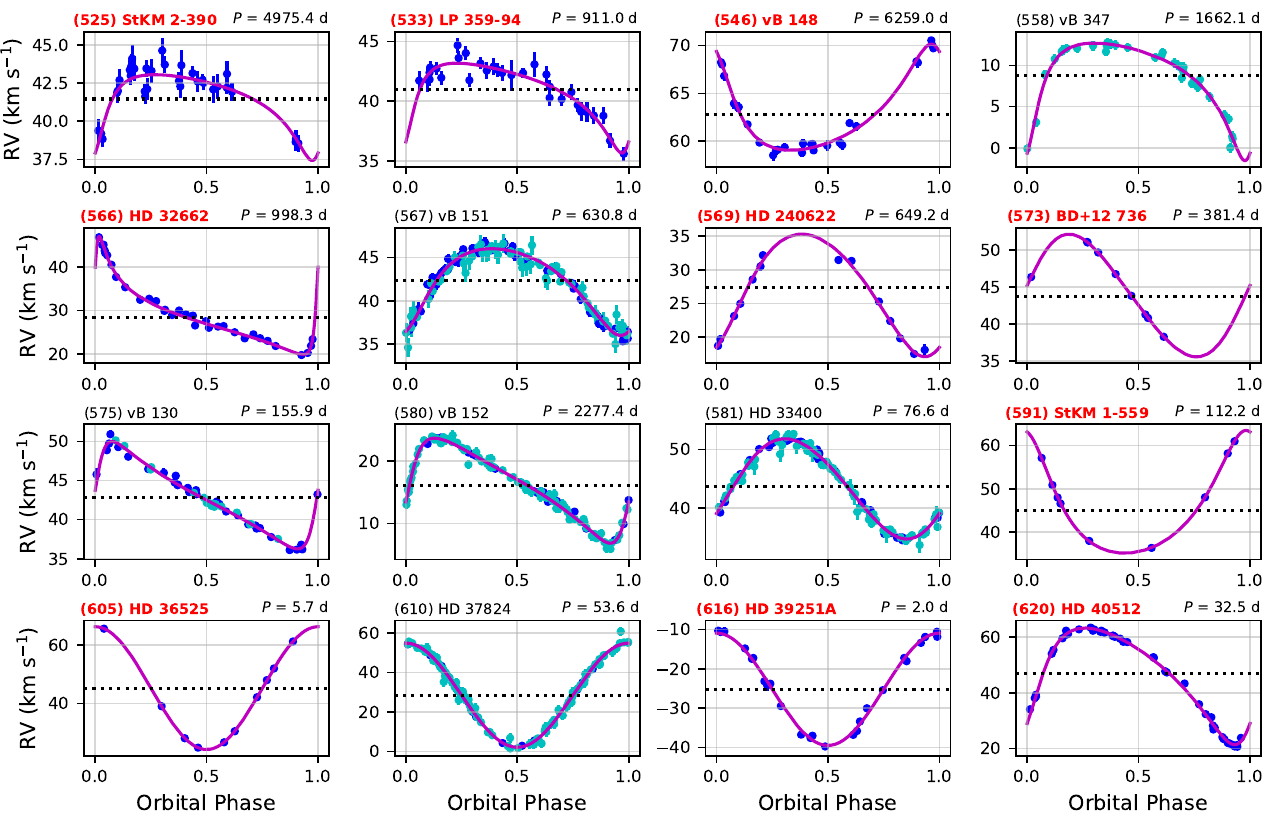}

  \figcaption{Single-lined spectroscopic orbits (continued).}
\end{figure*}

\begin{figure*}
  \epsscale{1.17}
  \hspace*{-1mm}\includegraphics[width=\textwidth]{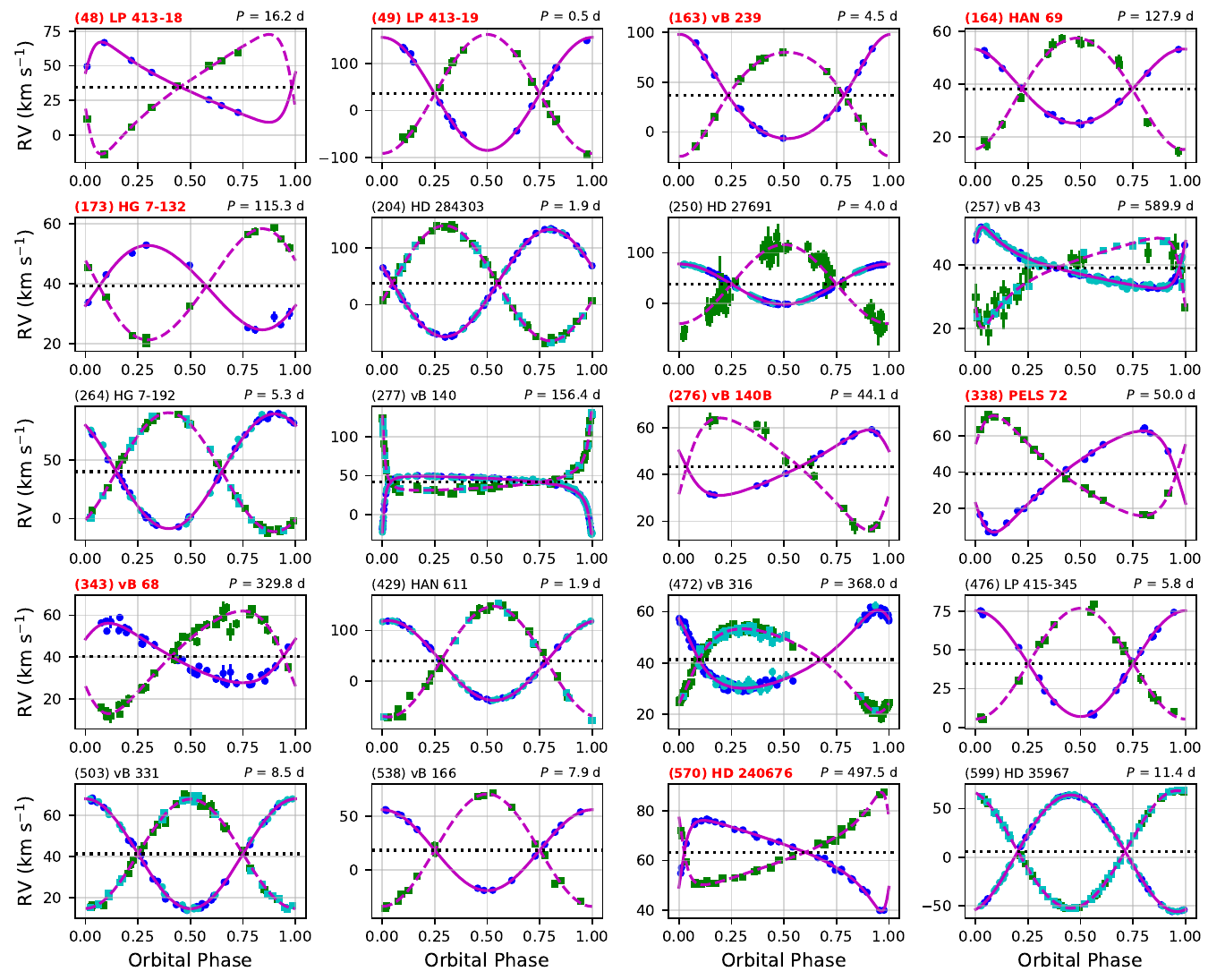}

  \figcaption{Double-lined spectroscopic orbits. Running numbers, SIMBAD names,
  and orbital periods are shown on the title line of each panel.
  For new orbits, the running numbers and SIMBAD names are shown in red.
  Blue and green symbols correspond
  to the CfA observations, and cyan symbols to RVs by others (see Table~\ref{tab:elemSB2a}). Our best-fit model is shown by the red lines (solid for the primary, dashed
  for the secondary), and the center-of-mass velocity of each system is indicated with a dotted line.\label{fig:sb2orbits}}
\end{figure*}

\clearpage



\begin{thebibliography}{}

\bibitem[Aitken(1914)]{Aitken:1914} Aitken, R.\ G.\ 1914, Lick
  Observatory Bulletin, 251, 52

\bibitem[Albrow(2024)]{Albrow:2024} Albrow, M.~D.\ 2024, \mnras, 528, 4, 6211

\bibitem[Altmann et al.(2017)]{Altmann:2017} Altmann, M., Roeser,
  S., Demleitner, M., et al.\ 2017, \aap, 600, L4

\bibitem[Babusiaux et al.(2023)]{Babusiaux:2023} Babusiaux, C.,
  Fabricius, C., Khanna, S., et al.\ 2023, \aap, 674, A32

\bibitem[Balona(1987)]{Balona:1987} Balona, L.~A.\ 1987, South African
Astronomical Observatory Circular, 11

\bibitem[Bashi et al.(2023)]{Bashi:2023} Bashi, D., Mazeh, T., \&
  Faigler, S.\ 2023, \mnras, 522, 1, 1184
  
\bibitem[Bender \& Simon(2008)]{Bender:2008} Bender, C.~F. \& Simon, M.\ 2008, \apj, 689, 416

\bibitem[Benedict et al.(2021)]{Benedict:2021} Benedict, G.~F.,
  Franz, O.~G., Horch, E.~P., et al.\ 2021, \aj, 161, 6, 285

\bibitem[Beyer \& White(2024)]{Beyer:2024} Beyer, A.~C. \& White, R.~J.\ 2024, \apj, 973, 28

\bibitem[Brandt(2018)]{Brandt:2018} Brandt, T.\,d.\ 2018, \apjs, 239,
  31
  
\bibitem[Brandt(2021)]{Brandt:2021} Brandt, T.\ D.\ 2021, \apjs, 254,
  42

\bibitem[Brandt(2024)]{Brandt:2024} Brandt, T.\,d.\ 2024, \pasp, 136,
  073001
  
\bibitem[Brogaard et al.(2021)]{Brogaard:2021} Brogaard, K.,
  Pak{\v{s}}tien{\.{e}}, E., Grundahl, F., et al.\ 2021, \aap, 645,
  A25
  
\bibitem[Buchhave et al.(2012)]{Buchhave:2012} Buchhave, L.\ A.,
  Latham, D.\ W., Johansen, A., et al.\ 2012, \nat, 486, 375
  
\bibitem[Burnham(1894)]{Burnham:1894} Burnham, S.\ W.\ 1894,
  Publications of Lick Observatory, 2, 221

\bibitem[de Bruijne et al.(2001)]{deBruijne:2001} de Bruijne,
  J.\ H.\ J., Hoogerwerf, R., \& de Zeeuw, P.\ T.\ 2001, \aap, 367, 111

\bibitem[Campbell(1910)]{Campbell:1910} Campbell, W.~W.\ 1910, \pasp,
  22, 131, 47

\bibitem[Campbell(1928)]{Campbell:1928} Campbell, W.~W.\ 1928, Publications of Lick Observatory, 16, 1

\bibitem[Chen et al.(2014)]{Chen:2014} Chen, Y., Girardi, L., Bressan,
A., et al.\ 2014, \mnras, 444, 2525

\bibitem[Choi et al.(2016)]{Choi:2016} Choi, J., Dotter, A., Conroy,
  C., et al.\ 2016, \apj, 823, 102
  
\bibitem[Conti(1969)]{Conti:1969} Conti, P.~S.\ 1969, \apj, 156,
661

\bibitem[Cudworth(1985)]{Cudworth:1985} Cudworth, K.~M.\ 1985, \pasp, 97, 348

\bibitem[Cutri et al.(2003)]{Cutri:2003} Cutri, R. M., Skrutskie,
  M. F., van Dyk, S., et al. 2003, The IRSA 2MASS All-Sky Point
  Source Catalog, NASA/IPAC Infrared Science Archive (Washington, DC: NASA)

\bibitem[Cvetkovi{\'c} \& Ninkovi{\'c}(2008)]{Cvetkovic:2008}
  Cvetkovi{\'c}, Z. \& Ninkovi{\'c}, S.\ 2008, \na, 13, 587
  
\bibitem[De Furio et al.(2022)]{DeFurio:2022} De Furio, M., Gardner, T., Monnier, J., et al.\ 2022, \apj, 941, 118

\bibitem[Detweiler et al.(1984)]{Detweiler:1984} Detweiler, H.~L., Yoss, K.~M., Radick, R.~R., et al.\ 1984, \aj, 89, 1038


\bibitem[D{\'\i}az et al.(2012)]{Diaz:2012} D{\'\i}az, R.~F.,
Santerne, A., Sahlmann, J., et al.\ 2012, \aap, 538, A113

\bibitem[Distler et al.(2025)]{Distler:2025} Distler, A., Soares-Furtado, M., Vanderburg, A., et al.\ 2025, \aj, 169, 166

\bibitem[Docobo \& Campo(2020)]{Docobo:2020} Docobo, J.\ A., \& Campo, P.
2020, Double Stars Inf.\ Circ.\ 202, 1

\bibitem[Douglas et al.(2014)]{Douglas:2014} Douglas, S.\ T.,
  Ag{\"u}eros, M.\ A., Covey, K.\ R., et al.\ 2014, \apj, 795, 161

\bibitem[Douglas et al.(2016)]{Douglas:2016} Douglas, S.\ T.,
  Ag{\"u}eros, M.\ A., Covey, K.\ R., et al.\ 2016, \apj, 822, 47

\bibitem[Dravins et al.(1999)]{Dravins:1999} Dravins, D., Lindegren,
  L., \& Madsen, S.\ 1999, \aap, 348, 1040

\bibitem[Duquennoy \& Mayor(1991)]{Duquennoy:1991} Duquennoy, A. \&
  Mayor, M.\ 1991, \aap, 248, 485
  
\bibitem[Dutra-Ferreira et al.(2016)]{Dutra-Ferreira:2016}
  Dutra-Ferreira, L., Pasquini, L., Smiljanic, R., et al.\ 2016, \aap,
  585, A75
  
\bibitem[El-Badry et al.(2019)]{ElBadry:2019} El-Badry, K., Rix, H.-W., Tian, H., et al.\ 2019, \mnras, 489, 5822

\bibitem[ESA(1997)]{ESA:1997} ESA 1997, The Hipparcos and Tycho
  Catalogues, ESA Special Publication, Vol.\ 1200 (Noordwijk: ESA)

\bibitem[Evans \& Oh(2022)]{Evans:2022} Evans, N.~W. \& Oh, S.\ 2022,
 \mnras, 512, 3846

\bibitem[Fabricius et al.(2021)]{Fabricius:2021} Fabricius, C., Luri,
  X., Arenou, F., et al.\ 2021, \aap, 649, A5

\bibitem[Fekel \& Henry(2005)]{Fekel:2005} Fekel, F.~C. \& Henry,
G.~W.\ 2005, \aj, 129, 1669

\bibitem[Fekel et al.(2002)]{Fekel:2002} Fekel, F.~C., Scarfe,
  C.\,d., Barlow, D.~J., et al.\ 2002, \aj, 123, 3, 1723

\bibitem[Fern\'andez-Hern\'andez \& Joliet(2019)]{Fernandez:2019}
  Fern\'andez-Hern\'andez, J., \& Joliet, E. 2019, GOST Software User
  Manual, Gaia~DPAC
  
\bibitem[F\H{u}r\'esz(2008)]{Furesz:2008} F\H{u}r\'esz, G. 2008, PhD
  thesis, Univ.\ Szeged, Hungary

\bibitem[Freund et al.(2020)]{Freund:2020} Freund, S., Robrade, J., Schneider, P.~C., et al.\ 2020, \aap, 640, A66

\bibitem[Friel(1995)]{Friel:1995} Friel, E.\,d.\ 1995, \araa, 33, 381

\bibitem[Gaia Collaboration et al.(2018)]{Gaia:2018} Gaia
  Collaboration, Brown, A.\ G.\ A., Vallenari, A., et al.\ 2018, \aap,
  616, A1

\bibitem[Gaia Collaboration et al.(2021)]{Gaia:2021} Gaia
  Collaboration, Smart, R.\ L., Sarro, L.\ M., et al.\ 2021, \aap, 649,
  A6

\bibitem[Gaia Collaboration et al.(2023a)]{Gaia:2023a} Gaia
  Collaboration, Vallenari, A., Brown, A.\ G.\ A., et al.\ 2023a, \aap,
  674, A1

\bibitem[Gaia Collaboration et al.(2023b)]{Gaia:2023b} Gaia
  Collaboration, Arenou, F., Babusiaux, C., et al.\ 2023b, \aap, 674,
  A34

\bibitem[Geller et al.(2015)]{Geller:2015} Geller, A.\ M., Latham,
  D.\ W., \& Mathieu, R.\ D.\ 2015, \aj, 150, 97

\bibitem[Geller et al.(2013)]{Geller:2013} Geller, A.~M., Hurley, J.~R., \& Mathieu, R.\,d.\ 2013, \aj, 145, 8

\bibitem[Geller et al.(2021)]{Geller:2021} Geller, A.~M., Mathieu,
  R.\,d., Latham, D.~W., et al.\ 2021, \aj, 161, 4, 190
  
\bibitem[Geller \& Mathieu(2012)]{Geller:2012} Geller, A.~M. \&
Mathieu, R.\,d.\ 2012, \aj, 144, 54

\bibitem[Goodman \& Dickson(1998)]{Goodman:1998} Goodman, J. \&
  Dickson, E.~S.\ 1998, \apj, 507, 2, 938

\bibitem[Goos(1908)]{Goos:1908} Goos, F. 1908, Dissertation, Rheinischen
  Friedrich-Wilhelms Universit\"at of Bonn

\bibitem[Gossage et al.(2018)]{Gossage:2018} Gossage, S., Conroy, C., Dotter, A., et al.\ 2018, \apj, 863, 67

\bibitem[Gosset et al.(2025)]{Gosset:2025} Gosset, E., Damerdji, Y.,
  Morel, T.\ et al.\ 2026, \aap, 693, 124

\bibitem[Griffin(1985b)]{Griffin:1985b} Griffin, R.~F.\ 1985b, \pasp, 97, 858

\bibitem[Griffin(2001)]{Griffin:2001} Griffin, R.\ F.\ 2001, The
  Observatory, 121, 244
  
\bibitem[Griffin(2012)]{Griffin:2012} Griffin, R.~F.\ 2012, Journal of
Astrophysics and Astronomy, 33, 29

\bibitem[Griffin(2013)]{Griffin:2013} Griffin, R.~F.\ 2013, The
Observatory, 133, 144

\bibitem[Griffin(2016)]{Griffin:2016} Griffin, R.~F.\ 2016, The
Observatory, 136, 179

\bibitem[Griffin \& Gunn(1974)]{Griffin:1974} Griffin, R.~F. \& Gunn, J.~E.\ 1974, \apj, 191, 545

\bibitem[Griffin \& Gunn(1978)]{Griffin:1978} Griffin, R.~F. \& Gunn,
J.~E.\ 1978, \aj, 83, 1114

\bibitem[Griffin et al.(1982)]{Griffin:1982} Griffin, R.~F., Mayor,
M., \& Gunn, J.~E.\ 1982, \aap, 106, 221

\bibitem[Griffin et al.(1985a)]{Griffin:1985a} Griffin, R.\ F., Gunn,
  J.\ E., Zimmerman, B.\ A., et al.\ 1985a, \aj, 90, 609

\bibitem[Griffin et al.(1988)]{Griffin:1988} Griffin, R.\ F., Gunn,
  J.\ E., Zimmerman, B.\ A., et al.\ 1988, \aj, 96, 172

\bibitem[Guenther et al.(2005)]{Guenther:2005} Guenther, E.~W.,
Paulson, D.~B., Cochran, W.~D., et al.\ 2005, \aap, 442, 3, 1031

\bibitem[Gullikson et al.(2016)]{Gullikson:2016} Gullikson, K., Kraus, A., \& Dodson-Robinson, S.\ 2016, \aj, 152, 40

\bibitem[Gunn et al.(1988)]{Gunn:1988} Gunn, J.~E., Griffin, R.~F.,
  Griffin, R.~E.~M., et al.\ 1988, \aj, 96, 198

\bibitem[Halbwachs et al.(2020)]{Halbwachs:2020} Halbwachs, J.-L.,
Kiefer, F., Lebreton, Y., et al.\ 2020, \mnras, 496, 1355

\bibitem[Halbwachs et al.(2018)]{Halbwachs:2018} Halbwachs, J.-L.,
Mayor, M., \& Udry, S.\ 2018, \aap, 619, A81

\bibitem[Halbwachs et al.(2023)]{Halbwachs:2023} Halbwachs, J.-L.,
  Pourbaix, D., Arenou, F., et al.\ 2023, \aap, 674, A9.

\bibitem[Hanson(1975)]{Hanson:1975} Hanson, R.\ B.\ 1975, \aj, 80, 379

\bibitem[Hao et al.(2024)]{Hao:2024} Hao, C.~J., Xu, Y., Hou,
  L.~G., et al.\ 2024, \apj, 963, 153

\bibitem[Hartkopf et al.(2001)]{Hartkopf:2001} Hartkopf, W.\ I., Mason,
  B.\ D., \& Worley, C.\ E.\ 2001, \aj, 122, 3472

\bibitem[H{\o}g et al.(2000)]{Hog:2000} H{\o}g, E., Fabricius, C.,
  Makarov, V.\ V., et al.\ 2000, \aap, 355, L27

\bibitem[Hole et al.(2009)]{Hole:2009} Hole, K.\ T., Geller, A.\ M.,
  Mathieu, R.\ D., et al.\ 2009, \aj, 138, 159

\bibitem[Holl et al.(2023)]{Holl:2023} Holl, B., Sozzetti, A.,
  Sahlmann, J., et al.\ 2023, \aap, 674, A10

\bibitem[Horch et al.(2010)]{Horch:2010} Horch, E.\ P., Falta, D.,
  Anderson, L.\ M., et al.\ 2010, \aj, 139, 205

\bibitem[Husser et al.(2013)]{Husser:2013} Husser, T.-O., Wende-von
  Berg, S., Dreizler, S., et al.\ 2013, \aap, 553, A6

\bibitem[Hut et al.(1992)]{Hut:1992} Hut, P., McMillan, S., Goodman, J., et al.\ 1992, \pasp, 104, 981

\bibitem[IJspeert et al.(2024)]{IJspeert:2024} IJspeert, L.~W.,
  Tkachenko, A., Johnston, C., et al.\ 2024, \aap, 691, A242

\bibitem[Imbert(2006)]{Imbert:2006} Imbert, M.\ 2006, Romanian
Astronomical Journal, 16, 3

\bibitem[Izmailov(2019)]{Izmailov:2019} Izmailov, I.~S.\ 2019, Astronomy Letters, 45, 1, 30

\bibitem[Jadhav et al.(2024)]{Jadhav:2024} Jadhav, V.~V., Kroupa, P., Wu, W., et al.\ 2024, \aap, 687, A89

\bibitem[Jancart et al.(2005)]{Jancart:2005} Jancart, S., Jorissen, A., Babusiaux, C., et al.\ 2005, \aap, 442, 365

\bibitem[Jantzen(1913)]{Jantzen:1913} Jantzen, K.\ 1913, Astronomische Nachrichten, 196, 117

\bibitem[Jerabkova et al.(2021)]{Jerabkova:2021} Jerabkova, T., Boffin,
  H.\ M.\ J., Beccari, G., et al.\ 2021, \aap, 647, A137

\bibitem[Josties \& Mason(2021)]{Josties:2021} Josties, J., \& Mason, B.\ D. 2021, IAU Double Stars Inf.\ Circ.\ 205, 1

\bibitem[Kraft(1965)]{Kraft:1965} Kraft, R.~P.\ 1965, \apj, 142, 681

\bibitem[Kraft(1967)]{Kraft:1967} Kraft, R.~P.\ 1967, \apj, 150, 551

\bibitem[Kurtz \& Mink(1998)]{Kurtz:1998} Kurtz, M.\ J. \&
  Mink, D.\ J.\ 1998, \pasp, 110, 934
  
\bibitem[Lanning \& Pesch(1981)]{Lanning:1981} Lanning, H.~H. \& Pesch, P.\ 1981, \apj, 244, 280

\bibitem[Laos et al.(2020)]{Laos:2020} Laos, E., Stassun, K.~G., \& Mathieu, R.\,d.\ 2020, \apj, 902, 2, 107

\bibitem[Latham(1985)]{Latham:1985} Latham, D.\ W. 1985, in Proc.\ IAU
  Coll.\ 88, Stellar Radial Velocities, ed.\ A.\ G.\ Philip \&
  D.\ W.\ Latham (Schenectady, NY: L.\ Davis Press), 21

\bibitem[Latham(1992)]{Latham:1992} Latham, D.\ W. 1992, in IAU
  Coll.\ 135, Complementary Approaches to Double and Multiple Star
  Research, ASP Conf.\ Ser.\ 32, eds.\ H.\ A.\ McAlister \&
  W.\ I.\ Hartkopf (San Francisco: ASP), 110

\bibitem[Latham et al.(1996)]{Latham:1996} Latham, D.\ W.,
  Nordstr\"om, B., Andersen, J., Torres, G., Stefanik, R.\ P.,
  Thaller, M., \& Bester, M. 1996, \aap, 314, 864

\bibitem[Latham et al.(2002)]{Latham:2002} Latham, D.~W., Stefanik,
  R.~P., Torres, G., Davis, R.~J., Mazeh, T., Carney, B.~W., Laird,
  J.~B., \& Morse, J.~A.\ 2002, \aj, 124, 1144

\bibitem[Le{\~a}o et al.(2019)]{Leao:2019} Le{\~a}o, I.~C.,
  Pasquini, L., Ludwig, H.-G., et al.\ 2019, \mnras, 483, 5026

\bibitem[Lebreton et al.(2001)]{Lebreton:2001} Lebreton, Y.,
  Fernandes, J., \& Lejeune, T.\ 2001, \aap, 374, 540

\bibitem[Leiner et al.(2015)]{Leiner:2015} Leiner, E.~M., Mathieu,
  R.\,d., Gosnell, N.~M., et al.\ 2015, \aj, 150, 1, 10


\bibitem[Liebing et al.(2023)]{Liebing:2023} Liebing, F., Jeffers,
  S.\ V., Zechmeister, M., et al.\ 2023, \aap, 673, A43

\bibitem[Linck et al.(2024)]{Linck:2024} Linck, E., Mathieu, R.\,d.,
\& Latham, D.~W.\ 2024, \aj, 168, 205

\bibitem[Lindegren(2018)]{Lindegren:2018} Lindegren, L. 2018,
  Re-normalising the Astrometric chi-square in Gaia~DR2,
  GAIA-C3-TN-LU-LL-124-01, (Lund: Gaia~DPAC)

\bibitem[Lindegren et al.(2021)]{Lindegren:2021} Lindegren, L.,
  Bastian, U., Biermann, M., et al.\ 2021, \aap, 649, A4

\bibitem[Lindegren et al.(2000)]{Lindegren:2000} Lindegren, L.,
  Madsen, S., \& Dravins, D.\ 2000, \aap, 356, 1119

\bibitem[Lodieu et al.(2019)]{Lodieu:2019} Lodieu, N., Smart,
  R.~L., P{\'e}rez-Garrido, A., et al.\ 2019, \aap, 623, A35

\bibitem[Lucy \& Ricco(1979)]{Lucy:1979} Lucy, L.~B. \& Ricco,
  E.\ 1979, \aj, 84, 401

\bibitem[Ludendorff(1910)]{Ludendorff:1910} Ludendorff, H.\ 1910,
  Astronomische Nachrichten, 184, 23, 373

\bibitem[Mann et al.(2018)]{Mann:2018} Mann, A.~W., Vanderburg, A., Rizzuto, A.~C., et al.\ 2018, \aj, 155, 4

\bibitem[Mason et al.(1993)]{Mason:1993} Mason, B.\ D., McAlister,
  H.\ A., Hartkopf, W.\ I., et al.\ 1993, \aj, 105, 220

\bibitem[Madsen(2003)]{Madsen:2003} Madsen, S.\ 2003, \aap, 401, 565

\bibitem[Mason et al.(2023)]{Mason:2023} Mason, B.~D., Tokovinin, A., Mendez, R.~A., et al.\ 2023, \aj, 166, 4, 139

\bibitem[Mason et al.(2001)]{Mason:2001} Mason, B.\ D., Wycoff, G.\ L.,
  Hartkopf, W.\ I., et al.\ 2001, \aj, 122, 3466

\bibitem[Mathieu \& Mazeh(1988)]{Mathieu:1988} Mathieu, R.\,d. \&
  Mazeh, T.\ 1988, \apj, 326, 256

\bibitem[Mathieu et al.(2004)]{Mathieu:2004} Mathieu, R.\,d.,
Meibom, S., \& Dolan, C.~J.\ 2004, \apjl, 602, 2, L121

\bibitem[Mayo et al.(2023)]{Mayo:2023} Mayo, A.~W., Dressing, C.\,d., Vanderburg, A., et al.\ 2023, \aj, 165, 235

\bibitem[Mayor \& Mazeh(1987)]{Mayor:1987} Mayor, M. \& Mazeh,
  T.\ 1987, \aap, 171, 157

\bibitem[Mazeh(1990)]{Mazeh:1990} Mazeh, T.\ 1990, \aj, 99, 675

\bibitem[Mazeh(2008)]{Mazeh:2008} Mazeh, T.\ 2008, EAS Publications Series, 29, 1

\bibitem[Mazeh \& Goldberg(1992)]{Mazeh:1992} Mazeh, T. \&
  Goldberg, D.\ 1992, \apj, 394, 592

\bibitem[McAlister et al.(1987)]{McAlister:1987} McAlister, H.~A.,
  Hartkopf, W.~I., Hutter, D.~J., et al.\ 1987, \aj, 93, 688
  
\bibitem[McArthur et al.(2011)]{McArthur:2011} McArthur, B.~E., Benedict, G.~F., Harrison, T.~E., et al.\ 2011, \aj, 141, 172

\bibitem[McClure(1982)]{McClure:1982} McClure, R.\,d.\ 1982, \apj, 254,
  606
  
\bibitem[McNamara \& Sekiguchi(1986)]{McNamara:1986} McNamara,
  B.~J. \& Sekiguchi, K.\ 1986, \apj, 310, 613

\bibitem[Meibom \& Mathieu(2005)]{Meibom:2005} Meibom, S. \& Mathieu,
  R.\,d.\ 2005, \apj, 620, 2, 970

\bibitem[Meibom et al.(2006)]{Meibom:2006} Meibom, S., Mathieu, R.\,d.,
  \& Stassun, K.~G.\ 2006, \apj, 653, 1, 621

\bibitem[Meingast \& Alves(2019)]{Meingast:2019} Meingast, S. \&
  Alves, J.\ 2019, \aap, 621, L3

\bibitem[Mermilliod et al.(2007)]{Mermilliod:2007} Mermilliod, J.-C.,
Andersen, J., Latham, D.~W., et al.\ 2007, \aap, 473, 829

\bibitem[Mermilliod et al.(2008c)]{Mermilliod:2008c} Mermilliod, J.-C.,
Grenon, M., \& Mayor, M.\ 2008c, \aap, 491, 951  

\bibitem[Mermilliod \& Mayor(1999)]{Mermilliod:1999} Mermilliod, J.-C.
\& Mayor, M.\ 1999, \aap, 352, 479  

\bibitem[Mermilliod et al.(2009)]{Mermilliod:2009} Mermilliod, J.-C.,
  Mayor, M., \& Udry, S.\ 2009, \aap, 498, 949

\bibitem[Mermilliod et al.(2008b)]{Mermilliod:2008b} Mermilliod, J.-C.,
Platais, I., James, D.~J., et al.\ 2008b, \aap, 485, 95

\bibitem[Mermilliod et al.(1992)]{Mermilliod:1992} Mermilliod, J.-C.,
  Rosvick, J.\ M., Duquennoy, A., et al.\ 1992, \aap, 265, 513

\bibitem[Mermilliod et al.(2008a)]{Mermilliod:2008a} Mermilliod, J.-C.,
Queloz, D., \& Mayor, M.\ 2008a, \aap, 488, 409  

\bibitem[Meunier et al.(2017)]{Meunier:2017} Meunier, N., Mignon, L.,
  \& Lagrange, A.-M.\ 2017, \aap, 607, A124

\bibitem[Milliman et al.(2014)]{Milliman:2014} Milliman, K.~E.,
  Mathieu, R.\,d., Geller, A.~M., et al.\ 2014, \aj, 148, 2, 38
  

\bibitem[Moe \& Di Stefano(2017)]{Moe:2017} Moe, M. \& Di Stefano, R.\ 2017, \apjs, 230, 15

\bibitem[Morzinski(2011)]{Morzinski:2011} Morzinski, K.\ M. 2011, PhD thesis, Univ.\ of California Santa Cruz

\bibitem[Muller(1978)]{Muller:1978} Muller, P.\ 1978, \aaps, 32, 173

\bibitem[Murphy et al.(2018)]{Murphy:2018} Murphy, S.~J., Moe, M., Kurtz, D.~W., et al.\ 2018, \mnras, 474, 4322

\bibitem[Narayan et al.(2026)]{Narayan:2026} Narayan, R.\ S., Linck, E.,
Mathieu, R.\ D., \& Geller, A.\ M. 2026, \aj, 171, 102

\bibitem[Nine et al.(2020)]{Nine:2020} Nine, A.~C., Milliman, K.~E.,
  Mathieu, R.\,d., et al.\ 2020, \aj, 160, 4, 169

\bibitem[Nordstr\"om et al.(1994)]{Nordstrom:1994} Nordstr\"om, B.,
  Latham, D.\ W., Morse, J.\ A., et al.\ 1994, \aap, 287, 338

\bibitem[North \& Zahn(2003)]{North:2003} North, P. \& Zahn,
  J.-P.\ 2003, \aap, 405, 677

\bibitem[Oh \& Evans(2020)]{Oh:2020} Oh, S. \& Evans, N.~W.\ 2020,
  \mnras, 498, 1920

\bibitem[Olivares et al.(2025)]{Olivares:2025} Olivares, J., Bouy, H., Dorn-Wallenstein, T.~Z., et al.\ 2025, \aap, 693, A12

\bibitem[Patience et al.(1998)]{Patience:1998} Patience, J., Ghez,
  A.\ M., Reid, I.\ N., et al.\ 1998, \aj, 115, 1972

\bibitem[Paulson et al.(2004)]{Paulson:2004} Paulson, D.~B., Cochran,
W.\,d., \& Hatzes, A.~P.\ 2004, \aj, 127, 3579

\bibitem[Pels et al.(1975)]{Pels:1975} Pels, G., Oort, J.\ H., \&
  Pels-Kluyver, H.\ A.\ 1975, \aap, 43, 423
  
\bibitem[Penev \& Schussler(2022)]{Penev:2022} Penev, K.~M. \& Schussler, J.~A.\ 2022, \mnras, 516, 6145

\bibitem[Perryman et al.(1998)]{Perryman:1998} Perryman, M.\ A.\ C.,
  Brown, A.\ G.\ A., Lebreton, Y., et al.\ 1998, \aap, 331, 81

\bibitem[Pinsonneault et al.(2003)]{Pinsonneault:2003} Pinsonneault,
  M.~H., Terndrup, D.~M., Hanson, R.~B., et al.\ 2003, \apj, 598, 588

\bibitem[Plaskett(1915)]{Plaskett:1915} Plaskett, J.~S.\ 1915, Publications of the Dominion Observatory Ottawa, 2, 61

\bibitem[Pr{\v{s}}a et al.(2022)]{Prsa:2022} Pr{\v{s}}a, A., Kochoska, A., Conroy, K.~E., et al.\ 2022, \apjs, 258, 16

\bibitem[Quinn et al.(2014)]{Quinn:2014} Quinn, S.~N., White, R.~J., Latham, D.~W., et al.\ 2014, \apj, 787, 27

\bibitem[Raghavan et al.(2010)]{Raghavan:2010} Raghavan, D.,
  McAlister, H.~A., Henry, T.~J., et al.\ 2010, \apjs, 190, 1
  
\bibitem[Rappaport et al.(2013)]{Rappaport:2013} Rappaport, S.,
  Deck, K., Levine, A., et al.\ 2013, \apj, 768, 33

\bibitem[Reid(1992)]{Reid:1992} Reid, N.\ 1992, \mnras, 257, 257

\bibitem[Reid \& Mahoney(2000)]{Reid:2000} Reid, I.\ N. \& Mahoney,
  S.\ 2000, \mnras, 316, 827

\bibitem[Reino et al.(2018)]{Reino:2018} Reino, S., de Bruijne, J.,
  Zari, E., et al.\ 2018, \mnras, 477, 3197

\bibitem[Risbud et al.(2025)]{Risbud:2025} Risbud, D., Jadhav, V.~V.,
\& Kroupa, P.\ 2025, \aap, 694, A258

\bibitem[R{\"o}ser et al.(2019)]{Roser:2019} R{\"o}ser, S., Schilbach,
  E., \& Goldman, B.\ 2019, \aap, 621, L2

\bibitem[R{\"o}ser et al.(2011)]{Roser:2011} R{\"o}ser, S., Schilbach,
  E., Piskunov, A.~E., et al.\ 2011, \aap, 531, A92

\bibitem[Russell(1914)]{Russell:1914} Russell, H.~N.\ 1914, \apj, 40, 282

\bibitem[Sato et al.(2007)]{Sato:2007} Sato, B., Izumiura, H., Toyota, E., et al.\ 2007, \apj, 661, 527

\bibitem[Savanov \& Dmitrienko(2018)]{Savanov:2018} Savanov, I.~S. \& Dmitrienko, E.~S.\ 2018, Astronomy Reports, 62, 238

\bibitem[Schaefer et al.(2022)]{Schaefer:2022} Schaefer, G., Anugu, N., Bender, C., et al.\ 2022, \baas, 240, 305.07

\bibitem[Schiller \& Milone(1987)]{Schiller:1987} Schiller, S.~J. \&
  Milone, E.~F.\ 1987, \aj, 93, 1471

\bibitem[Schlesinger \& Baker(1910)]{Schlesinger:1910} Schlesinger,
  F. \& Baker, R.~H.\ 1910, Publications of the Allegheny Observatory
  of the University of Pittsburgh, 1, 21, 135

\bibitem[Schwan(1990)]{Schwan:1990} Schwan, H.\ 1990, \aap, 228, 69 

\bibitem[Shahaf et al.(2017)]{Shahaf:2017} Shahaf, S., Mazeh, T.,
  \& Faigler, S.\ 2017, \mnras, 472, 4, 4497
  
\bibitem[S{\"o}derhjelm(1999)]{Soderhjelm:1999} S{\"o}derhjelm, S.\ 1999, \aap, 341, 121
  
\bibitem[Stauffer et al.(1997)]{Stauffer:1997} Stauffer, J.~R., Balachandran, S.~C., Krishnamurthi, A., et al.\ 1997, \apj, 475, 604

\bibitem[Stefanik \& Latham(1985)]{Stefanik:1985} Stefanik, R.\ P. \&
  Latham, D.\ W. 1985, in IAU Coll.\ 88, Stellar Radial Velocities,
  eds.\ A.\ G.\ D.\ Philip and D.\ W.\ Latham (Schelectady: L.\ Davis
  Press), 213
  
\bibitem[Stefanik \& Latham(1992)]{Stefanik:1992} Stefanik, R.~P. \&
  Latham, D.~W.\ 1992, IAU Colloquium 135: Complementary Approaches to
  Double and Multiple Star Research, eds. H.\ A.\ McAlister \&
  W.\ I.\ Hartkopf (San Francisco: ASP), 32, 173
  
\bibitem[Stefanik et al.(1999)]{Stefanik:1999} Stefanik, R.\ P., Latham,
D.\ W., \& Torres, G.\ 1999, IAU Colloq. 170: Precise Stellar Radial
Velocities, 185, 354

\bibitem[Szentgyorgyi \& F\H{u}r\'esz(2007)]{Szentgyorgyi:2007}
  Szentgyorgyi, A.\ H., \& F\H{u}r\'esz, G. 2007, {\it Precision
    Radial Velocities for the Kepler Era}, in The 3rd Mexico-Korea
    Conference on Astrophysics: Telescopes of the Future and San Pedro
    M\'artir, ed.\ S.\ Kurtz, RMxAC, 28, 129

\bibitem[Tal-Or et al.(2019)]{Tal-Or:2019} Tal-Or, L., Trifonov, T.,
Zucker, S., et al.\ 2019, \mnras, 484, L8

\bibitem[Terquem et al.(1998)]{Terquem:1998} Terquem, C., Papaloizou,
  J.~C.~B., Nelson, R.~P., et al.\ 1998, \apj, 502, 2, 788

\bibitem[Tokovinin(1990)]{Tokovinin:1990} Tokovinin, A.~A.\ 1990,
Soviet Astronomy Letters, 16, 440

\bibitem[Tokovinin(2014)]{Tokovinin:2014} Tokovinin, A.\ 2014, \aj, 147, 86

\bibitem[Tokovinin(2021a)]{Tokovinin:2021a} Tokovinin, A. 2021a, IAU
  Inf.\ Circ.\ 203, 3

\bibitem[Tokovinin(2021b)]{Tokovinin:2021b} Tokovinin, A. 2021b, IAU
  Inf.\ Circ.\ 204, 1

\bibitem[Tokovinin(2021c)]{Tokovinin:2021c} Tokovinin, A.\ 2021c,
  \aj, 161, 3, 144
  
\bibitem[Tokovinin \& Gorynya(2001)]{Tokovinin:2001} Tokovinin, A.~A. \& Gorynya, N.~A.\ 2001, \aap, 374, 227

\bibitem[Tokovinin et al.(2022)]{Tokovinin:2022} Tokovinin, A., Mason, B.~D., Mendez, R.~A., et al.\ 2022, \aj, 164, 2, 58

\bibitem[Tokovinin et al.(2024)]{Tokovinin:2024} Tokovinin, A., Mason, B.\,d., Mendez, R.~A., et al.\ 2024, \aj, 168, 28

\bibitem[Tokovinin \& Smekhov(2002)]{Tokovinin:2002} Tokovinin, A.~A. \& Smekhov, M.~G.\ 2002, \aap, 382, 118

\bibitem[Tokovinin et al.(2006)]{Tokovinin:2006} Tokovinin, A., Thomas, S., Sterzik, M., et al.\ 2006, \aap, 450, 681

\bibitem[Tomkin(2003)]{Tomkin:2003} Tomkin, J.\ 2003, The Observatory,
123, 372

\bibitem[Tomkin(2005)]{Tomkin:2005} Tomkin, J.\ 2005, The Observatory,
125, 232

\bibitem[Tomkin \& Griffin(2002)]{Tomkin:2002} Tomkin, J. \& Griffin,
  R.\ F.\ 2002, The Observatory, 122, 1

\bibitem[Tomkin et al.(2007)]{Tomkin:2007} Tomkin, J., Griffin, R.~F.,
\& Alzner, A.\ 2007, The Observatory, 127, 87

\bibitem[Torres(2007)]{Torres:2007} Torres, G.\ 2007, \aj, 133, 2684

\bibitem[Torres(2019)]{Torres:2019a} Torres, G.\ 2019a, \apj, 883, 105 

\bibitem[Torres et al.(2019b)]{Torres:2019b} Torres, G., Stefanik, R.~P., \& Latham, D.~W.\ 2019b, \apj, 885, 9  

\bibitem[Torres et al.(2021)]{Torres:2021} Torres, G., Latham, D.\ W.,
  \& Quinn, S.\ N.\ 2021, \apj, 921, 117

\bibitem[Torres et al.(2002)]{Torres:2002} Torres, G., Neuh{\"a}user,
R., \& Guenther, E.\ W.\ 2002, \aj, 123, 1701

\bibitem[Torres et al.(2024a)]{Torres:2024a} Torres, G., Schaefer,
  G.~H., Stefanik, R.~P., et al.\ 2024a, \apj, 971, 31  

\bibitem[Torres et al.(2024b)]{Torres:2024b} Torres, G., Stefanik, R.~P., \& Latham, D.~W.\ 2024b, \apj, 960, 121  

\bibitem[Torres et al.(2024c)]{Torres:2024c} Torres, G., Schaefer, G.~H., Stefanik, R.~P., et al.\ 2024c, \mnras, 527, 8907  
  
\bibitem[Torres et al.(1997d)]{Torres:1997d} Torres, G., Stefanik,
  R.\ P., Andersen, J., N\"ordstrom, B., Latham, D.\ W., \& Clausen,
  J.\ V. 1997d, \aj, 114, 2764
  
\bibitem[Torres et al.(1997a)]{Torres:1997a} Torres, G., Stefanik, R.~P., \& Latham, D.~W.\ 1997a, \apj, 474, 256

\bibitem[Torres et al.(1997b)]{Torres:1997b} Torres, G., Stefanik, R.~P., \& Latham, D.~W.\ 1997b, \apj, 479, 268

\bibitem[Torres et al.(1997c)]{Torres:1997c} Torres, G., Stefanik, R.~P., \& Latham, D.~W.\ 1997c, \apj, 485, 167

\bibitem[Vaccaro et al.(2015)]{Vaccaro:2015} Vaccaro, T.~R., Wilson,
R.~E., Van Hamme, W., et al.\ 2015, \apj, 810, 157

\bibitem[van Bueren(1952)]{vanBueren:1952} van Bueren, H.\ G.\ 1952,
  \bain, 11, 385

\bibitem[van Leeuwen(2007)]{vanLeeuwen:2007} van Leeuwen, F.\ 2007,
  \aap, 474, 653

\bibitem[Vereshchagin et al.(2013)]{Vereshchagin:2013} Vereshchagin,
  S.~V., Reva, V.~G., \& Chupina, N.~V.\ 2013, Astronomy Reports, 57, 52

\bibitem[Wang et al.(2025)]{Wang:2025} Wang, F., Fang, M., Fu, X.,
  et al.\ 2025, \apj, 979, 92

\bibitem[Wayman et al.(1965)]{Wayman:1965} Wayman, P.~A., Symms,
  L.~S.~T., \& Blackwell, K.~C.\ 1965, Royal Greenwich Observatory
  Bulletins, 98, 33. 

\bibitem[Weis(1983)]{Weis:1983} Weis, E.\ W.\ 1983, \pasp, 95,
  29

\bibitem[Weis et al.(1979)]{Weis:1979} Weis, E.\ W., Deluca, E.\ E., \&
  Upgren, A.\ R.\ 1979, \pasp, 91, 766
  
\bibitem[Willmarth et al.(2016)]{Willmarth:2016} Willmarth, D.~W., Fekel, F.~C., Abt, H.~A., et al.\ 2016, \aj, 152, 46

\bibitem[Wilson(1941)]{Wilson:1941} Wilson, O.~C.\ 1941, \apj, 93, 29

\bibitem[Wilson(1948)]{Wilson:1948} Wilson, R.~E.\ 1948, \apj, 107, 119

\bibitem[Worley \& Douglass(1997)]{Worley:1997} Worley, C.\ E. \&
  Douglass, G.\ G.\ 1997, \aaps, 125, 523
  
\bibitem[Wu et al.(2025)]{Wu:2025} Wu, Y., Hadden, S., Dewberry, J., et al.\ 2025, \apjl, 982, L34

\bibitem[Zahn(1966)]{Zahn:1966} Zahn, J.~P.\ 1966, Annales
  d'Astrophysique, 29, 489

\bibitem[Zahn(1975)]{Zahn:1975} Zahn, J.-P.\ 1975, \aap, 41, 329

\bibitem[Zahn(1977)]{Zahn:1977} Zahn, J.-P.\ 1977, \aap, 57, 383
  
\bibitem[Zahn(1989)]{Zahn:1989} Zahn, J.-P.\ 1989, \aap, 220, 1-2, 112

\bibitem[Zahn \& Bouchet(1989)]{ZahnBouchet:1989} Zahn, J.-P. \&
  Bouchet, L.\ 1989, \aap, 223, 112

\bibitem[Zanazzi(2022)]{Zanazzi:2022} Zanazzi, J.\ J. 2022, \apjl,
  929, L27
  
\bibitem[Zhang et al.(2023)]{Zhang:2023} Zhang, H., Brandt, T.\,d., Kiman, R., et al.\ 2023, \mnras, 524, 695

\bibitem[Zirm(2008)]{Zirm:2008} Zirm, H. 2008, IAU Double Stars Inf.\ Circ.\ 166, 1

\bibitem[Zucker \& Mazeh(1994)]{Zucker:1994} Zucker, S., \& Mazeh,
  T. 1994, \apj, 420, 806

\bibitem[Zucker et al.(1995)]{Zucker:1995} Zucker, S., Torres, G., \&
  Mazeh, T.\ 1995, \apj, 452, 863
  
\bibitem[Zurhellen(1907)]{Zurhellen:1907} Zurhellen, W.\ 1907,
  Astronomische Nachrichten, 173, 23, 353

\end{thebibliography}
\end{document}